\documentclass[a4paper,11pt,reqno]{article}
\usepackage{a4wide}

\usepackage{amsmath,amsfonts,amssymb}
\usepackage[T1]{fontenc}
\usepackage[p,osf]{cochineal}
\usepackage[varqu,varl,var0]{inconsolata}
\usepackage[scale=.95,type1]{cabin}
\usepackage[vvarbb]{newtxmath}
\usepackage[cal=boondoxo]{mathalfa}
\usepackage[mathscr]{euscript}
\usepackage{bbold} 
\usepackage[normalem]{ulem}

\usepackage{pict2e}

\makeatletter
\DeclareRobustCommand{\loplus}{\mathbin{\mathpalette\dog@lsemi{+}}}
\DeclareRobustCommand{\lotimes}{\mathbin{\mathpalette\dog@lsemi{\times}}}
\DeclareRobustCommand{\roplus}{\mathbin{\mathpalette\dog@rsemi{+}}}
\DeclareRobustCommand{\rotimes}{\mathbin{\mathpalette\dog@rsemi{\times}}}

\newcommand{\dog@rsemi}[2]{\dog@semi{#1}{#2}{-90,90}}
\newcommand{\dog@lsemi}[2]{\dog@semi{#1}{#2}{270,90}}
\newcommand{\dog@semi}[3]{%
  \begingroup
  \sbox\z@{$\m@th#1#2$}%
  \setlength{\unitlength}{\dimexpr\ht\z@+\dp\z@\relax}%
  \makebox[\wd\z@]{\raisebox{-\dp\z@}{%
    \begin{picture}(1,1)
    \linethickness{\variable@rule{#1}}
    \roundcap
    \put(0.5,0.5){\makebox(0,0){\raisebox{\dp\z@}{$\m@th#1#2$}}}
    \put(0.5,0.5){\arc[#3]{0.5}}
    \end{picture}%
  }}%
  \endgroup
}
\newcommand{\variable@rule}[1]{%
  \fontdimen8  
  \ifx#1\displaystyle\textfont3\else
    \ifx#1\textstyle\textfont3\else
      \ifx#1\scriptstyle\scriptfont3\else
        \scriptscriptfont3\relax
  \fi\fi\fi
}
\makeatother

\usepackage{nicefrac}
\usepackage{setspace}
\usepackage{datetime}
\usepackage[nosort]{cite}
\usepackage{upgreek}

\newcommand{\RomanNumeral}[1]
    {\MakeUppercase{\romannumeral #1}}
\usepackage{comment}

\usepackage[euler]{textgreek}
\usepackage[breaklinks=true]{hyperref}

\numberwithin{equation}{section}

\hypersetup{
			colorlinks=true,
			urlcolor=blue,
			citecolor=magenta,
			linkcolor=blue,
     }

\let\oldsqrt\sqrt
\def\sqrt{\mathpalette\DHLhksqrt}
\def\DHLhksqrt#1#2{%
\setbox0=\hbox{$#1\oldsqrt{#2\,}$}\dimen0=\ht0
\advance\dimen0-0.2\ht0
\setbox2=\hbox{\vrule height\ht0 depth -\dimen0}%
{\box0\lower0.4pt\box2}}

\newcommand{\RNum}[1]{\uppercase\expandafter{\romannumeral #1\relax}}

\author{
  \begin{minipage}{.97\linewidth}
    \vspace{1cm}
       \begin{center}
      \begin{small}
             \textbf{Anastasios C. Petkou},$^{1}$ 
     \textbf{P. Marios Petropoulos},$^2$ 
      \textbf{David Rivera-Betancour}$^2$
      and
      \textbf{Konstantinos Siampos}$^{3}$
              \end{small}
    \end{center}
    \vspace{0.5cm}
    \hspace{2.4cm}\begin{minipage}{.7\linewidth}
\begin{center}     {\it \begin{footnotesize}
\hbox{\kern-1.8cm\vbox{\vskip0cm
 \begin{itemize}
               \item[$^1$]Division of Theoretical Physics\\School of Physics\\
  Aristotle University of Thessaloniki\\ 
  54124 Thessaloniki, Greece
                           \vskip0.25cm
      \end{itemize}}
\kern-3.2cm\vbox{
\begin{itemize}
 \item[$^2$]Centre de Physique Th\'eorique -- CPHT\\ 
        Ecole Polytechnique, CNRS\footnote{\emph{Centre National de la Recherche Scientifique}, Unit\'e Mixte de Recherche UMR 7644.}\\
        Institut Polytechnique de Paris\\
        91128 Palaiseau Cedex, France         
      \end{itemize}
      \vskip0.cm
}}
     \end{footnotesize}}
\end{center}
    \end{minipage}
    \vspace{0.5cm}\begin{minipage}{.7\linewidth}
\begin{center}     
{\it \begin{footnotesize}
\hbox{\kern0.6cm\vbox{\vskip2cm
}
\kern2.9cm\vbox{
\begin{itemize}
 \item[$^3$] Department of Nuclear and Particle Physics\\ 
              Faculty of Physics\\ 
              National and Kapodistrian University of Athens\\
              15784 Athens Greece
      \end{itemize}\vskip0.05cm
}
}
     \end{footnotesize}}
\end{center}
     \end{minipage}
  \end{minipage}
}

\title{\vspace{1.5cm}
 \boldmath \begin{Large}
    \textbf{\textsc{Relativistic Fluids, Hydrodynamic Frames and their Galilean versus Carrollian Avatars}}
  \end{Large} \unboldmath
}

\date{}

\begin{document}

%\blinddocument
%\blindmathpaper

\begin{titlepage}
\maketitle
\thispagestyle{empty}

 \vspace{-14.cm}
  \begin{flushright}
  CPHT-RR021.042022\\
  \end{flushright}
 \vspace{12.cm}

\begin{center}
\textsc{Abstract}\\  
\vspace{0.8 cm}	
\begin{minipage}{1.0\linewidth}
We comprehensively study Galilean and Carrollian hydrodynamics on arbitrary backgrounds, in the presence of a matter/charge conserved current. For this purpose, we follow two distinct and complementary paths. The first is based on local invariance, be it Galilean or Carrollian diffeomorphism invariance, possibly accompanied by Weyl invariance. The second consists in analyzing the relativistic fluid equations at large or small speed of light, after choosing an adapted gauge, Arnowitt--Deser--Misner--Zermelo for the former and Papapetrou--Randers for the latter. Unsurprisingly, the results agree, but the second approach is superior as it effortlessly captures more elaborate situations with multiple degrees of freedom. It furthermore allows to investigate the fate of hydrodynamic-frame invariance in the two limits at hand, and conclude that its breaking (in the Galilean) or its preservation (in the Carrollian)  are fragile consequences of the behaviour of transport attributes at large or small $c$. Both methods do also agree on the doom of N\oe therian currents generated in the relativistic theory by isometries: conserved currents are not always guaranteed in Newton--Cartan or Carroll spacetimes as a consequence of Galilean or Carrollian isometries. Comparison of Galilean and Carrollian fluid equations exhibits a striking but often superficial resemblance, which we comment in relation to black-hole horizon dynamics, awkwardly akin to Navier--Stokes equations. This congruity is authentic in one instance though and turns out then to describe Aristotelian dynamics, which is the last item in our agenda.

\end{minipage}
\end{center}

\vspace{5cm} 
\end{titlepage}

\onehalfspace

%\noindent\rule{\textwidth}{1.2pt}
%\vspace{-1cm}
\begingroup
\hypersetup{linkcolor=black}
\tableofcontents
\endgroup
\noindent\rule{\textwidth}{0.6pt}

\section{Introduction}

Fluid dynamics is 19th century physics par excellence. It has been thoroughly investigated, expanded and applied in various areas, but continues to raise questions and challenges, which are sometimes conceptual. In relativistic fluids for example, the issue of hydrodynamic-frame invariance is rather subtle. It reflects the freedom to choose arbitrarily the velocity of the fluid, and is rooted in the impossibility to distinguish the mass flow from the energy flow in a relativistic system. This invariance was made popular by Landau and Lifshitz in their formulation of dissipative relativistic hydrodynamics without heat current \cite{Landau}, as opposed to the first formalism for relativistic fluids due to Eckart \cite{Eckart}. The freedom to set freely the velocity field has drawbacks recognized long ago, when implemented in the linearized (or, more generally, truncated) constitutive relations, which accompany the fluid equations of motion. A comprehensive presentation of the subject can be found in \cite{Tolman34, Synge57, Israel1981, Israel87}, where  the main difficulties, namely causality, stability and shock structure, are discussed in length.\footnote{Modern textbooks on fluid mechanics are e.g. \cite{TK,RZ}.} More recent progress has been reported in \cite{Romatschke:2009im, Kovtun:2012rj, Bemfica:2017wps, Grozdanov:2019kge, Grozdanov:2019uhi, Kovtun:2019hdm, Bemfica:2019knx, Hoult:2020, Bemfica:2020zjp, Dore:2021xqq, Hoult:2021gnb}, showing that the debate is still ongoing. It is worth stressing that the above phenomenological descriptions of out-of-equilibrium phenomena are enshrined by relativistic kinetic theory. In particular, the various quoted formalisms -- Eckart, Landau--Lifshitz or others -- arise using the relativistic Boltzmann's equation. Further reading on this subject is proposed in \cite{Israel63, CMarle1, CMarle2, deGroot}.

Hydrodynamic-frame invariance has also emerged  in a slightly more formal context. The asymptotic isometries of the gravitational field in general relativity\footnote{See e.g. the lecture notes \cite{Ruzziconi:2019pzd} for a recent review and further references on this subject.} are related to the symmetries of a fictitious fluid defined on the conformal boundary.\footnote{This fluid is often referred to as ``dual'' or ``holographic'' --  see Refs. \cite{Campoleoni:2018ltl, CMPR, CMPRpos} for the precise symmetry interplay and a complete bibliography of fluid/gravity holographic correspondence.}  When the gravitational field is asymptotically anti-de Sitter, the boundary is time-like and the dual fluid is relativistic. A local transformation of the fluid velocity amounts to a diffeomorphism on the gravitational side. For an asymptotically flat gravitational field, the boundary is null and the associated fluid is Carrollian \cite{CMPPS2}. Does the hydrodynamic-frame invariance survive in that case and does it share the above relationship with the asymptotic isometries? Similarly, and irrespective of any bearing to gravity, why is hydrodynamic-frame invariance lost in ordinary Galilean fluids, where the velocity and the mass density are measurable quantities?

The purpose of the present work is to elaborate on properties of Galilean and Carrollian fluids, in the spirit and as a follow up of Ref. \cite{CMPPS1}. This includes the discussion of hydrodynamic-frame invariance, the addition of a conserved current and its associated chemical potential, the interpretation of Galilean and Carrollian fluid equations as conservation laws stemming out of appropriate diffeomorphism invariances, and the dearth of conservation properties ensuing isometries.

The emergence of Carrollian physics goes back to the works of  L\'evy-Leblond \cite{Levy} and Sen Gupta \cite{SenGupta}. The Carroll group is an ultra-relativistic contraction of Poincar\'e group. It is dual to the better-known non-relativistic contraction, the Galilean group.\footnote{On a purely semantic vein, the given names ``relativistic, ultra-relativistic, non-relativistic''  are all unfortunate, as pointed out with brio by Jean-Marc L\'evy-Leblond. Although it is probably too late to give up the first, one should try to replace the others by \emph{Carrollian} and \emph{Galilean}. Incidentally, Niels Obers and Stefan Vandoren rightfully insist on the ultra-local nature of the Carrollian limit.}
Carrollian symmetry has triggered interest in several directions.
On the mathematical side,  new geometric structures were discovered dubbed Carrollian manifolds \cite{Henneaux:1979vn, Duval:1990hj, Duval:2014uoa, Duval:2014uva, Duval:2014lpa, NC, Duv, Bekaert:2014bwa, Bekaert:2015xua, Hartong:2015xda, Festu, Figueroa-OFarrill:2018ilb, Morand:2018tke,Ciambelli:2019lap, Herfray:2021qmp}, following patterns similar to those leading to the Galilean duals i.e. the Newton--Cartan geometries. From a more physical perspective, the connection of Carrollian symmetry with asymptotic isometries of Ricci-flat gravitational backgrounds and in particular its role in the growing subject of flat holography have attracted outmost attention \cite{Ashtekar:2014zsa, Barnich:2009se, Bergshoeff:2014jla, card, ba, Bagchi2010e, Bagchi2, Att1, Att2,   Merbis:2019wgk}.

Ordinary, Galilean fluid dynamics is the non-relativistic limit of relativistic hydrodynamics. It was originally circumscribed to three-dimensional Euclidean space with absolute time i.e. to the strict Newtonian framework with full Galilean isometry. Efforts have been made to evade this restriction \cite{Vinokur, Avis, Aris, Carlson, Luo, Charron, Jensen:2014ama, Karch, Geracie:2015xfa, Debus} and finally reach the general equations describing a non-relativistic viscous fluid moving on a space endowed with a spatial, time-dependent metric, and covariant  under  Galilean diffeomorphisms such as $t'=t'(t)$ and $\mathbf{x}^{\prime}=\mathbf{x}^{\prime}(t, \mathbf{x})$ \cite{CMPPS1,Armas:2019gnb}.\footnote{See also \cite{Hassaine:1999hn, Horvathy:2009kz} for a discussion on symmetries.}

The more exotic Carrollian fluids are ``flowing'' on Carrollian manifolds and their equations of motion are invariant under Carrollian coordinate transformations, $t'=t'(t, \mathbf{x})$ and $\mathbf{x}^{\prime}=\mathbf{x}^{\prime}(\mathbf{x})$. Although particle motion is forbidden due to the shrinking of the light cone, and despite the absence of a microscopic analysis based on thermodynamics or kinetic theory, the dynamics for a continuous medium seems to make formally sense, involving an ``inverse velocity,'' energy density, pressure etc.  The first instance where Carrollian fluids were quoted is Ref.~\cite{Penna1}. There, it was realized that contrary to a forty-year lore, Einstein dynamics on black-hole horizons and the associated membrane paradigm were not related to the Navier--Stokes equations, but instead to their Carrollian duals. This observation was further discussed in  \cite{Donnay:2019jiz}, and aspects of Carrollian hydrodynamics were analyzed in \cite{ Penna2,  Penna3, dutch, Poovuttikul:2019ckt, dutch2, newdutch}. It is fair to repeat that this sort of fluids lack of microscopic settlement and laboratory applications. Nevertheless, their dynamical equations emerge in various instances where null hypersurfaces are at work, and this justifies a thorough investigation.

The hydrodynamic equations for Galilean or Carrollian fluids were obtained in  \cite{CMPPS1} as limits of the fully covariant relativistic equations on general pseudo-Riemannian manifolds. For the Galilean case, the suitable form of the metric was Zermelo, whereas  Papapetrou--Randers was better adapted to the Carrollian limit. These metrics are indeed form-invariant under Galilean or Carrollian diffeomorphisms respectively (see \cite{Gibbons:2008zi} for further properties). Our study is performed here in the presence of a conserved current, which contributes the dynamics, and fosters the attainment of the Galilean continuity equation. The infinite or vanishing velocity limits are accompanied with some assumptions on the behaviour of the physical quantities such as energy density, heat current or stress  tensor, including important sub-leading terms.  We show that the resilience or the failure of the relativistic hydrodynamic-frame invariance in the non-relativistic or ultra-relativistic limits are closely tied to those behaviours. In a nutshell our conclusions can be summarized as follows. For the Carrollian case, the behaviours at vanishing velocity of light are suggested by the experience acquired with holographic fluids, and turn out to be compatible with hydrodynamic-frame invariance. This is no longer true in the Galilean limit, where  the rules are dictated by non-relativistic physics and disrupt this invariance, unless one concedes to give up matter conservation and at the cost of altering the Navier--Stokes equations. 

The relativistic hydrodynamic equations, namely the vanishing of the covariant energy--momentum tensor divergence, translate the invariance of some effective action with respect to  general diffeomorphisms $t'=t'(t, \mathbf{x})$ and $\mathbf{x}^{\prime}=\mathbf{x}^{\prime}(t, \mathbf{x})$. Similarly,  Galilean or Carrollian equations can be reached upon imposing the corresponding diffeomorphism invariance on the effective actions. The energy--momentum tensor is in these cases traded for other generalized momenta adapted to the local symmetries at hand. Our analysis, performed along the lines of \cite{CM1}, reveals subtleties and slightly differ in comparison with the large-$c$ or small-$c$ limits of the relativistic theory. This betrays that when considering these limits, as opposed to working with action principles directly in Newton--Cartan or Carrollian spacetimes, more information is stored in the equations, and more constraints emerge due to the larger original local invariance. Specifically, Galilean mass conservation (continuity equation) is built in (as shown in \cite{CMPPS1}) irrespective of extra matter current conservation. This confirms that the most striving and economical approach for reaching the dynamical equations is indeed the one based on the limiting procedure, originally used in \cite{CMPPS1}, which can be even extended at wish for incorporating naturally more degrees of freedom, which would require more conjugate variables in the Galilean or Carrollian action principles.

In order to deliver a comprehensive picture of the web of dynamics emanating upon contractions of the plain relativistic group and their associate spacetimes, we briefly venture out and explore the realm of Aristotelian geometries. Introduced by Penrose in \cite{Penrose-Battelle}, they became suddenly popular because they do not possess any boost invariance  (see e.g. \cite{dutch, dutch2, Armas:2020mpr, Pinzani-Fokeeva:2021klb, Rajagopal:2021swe}). The absence of boosts\footnote{The associated group is the \emph{static group}, introduced in \cite{bacryLLinfmas}.} features that both time and space are  absolute  in these geometries. Hence motion and light cone are trivialized, and the notion of fluid becomes even more questionable than in the Carrollian framework. Nonetheless, dynamics can be defined from invariance principles --  no limit involving the speed of light exists that would connect Aristotelian spacetimes to relativistic theories -- and is worth investigating as it appears to stand at the intersection of Galilean and Carrollian physics.

When discussing dynamics in general, and fluid dynamics in particular, part of the duty is to exhibit conserved quantities. These are generally the consequence of symmetries, but this concept should be scrutinized on a case-by-case basis. In a relativistic theory, any (conformal) Killing field provides a conserved current upon contraction with a (traceless) conserved energy--momentum tensor. We prove that this is no longer systematically true for Galilean or Carrollian hydrodynamics, and for instance, boosts present in flat space are not spared -- this would be circumvented in Aristotelian spacetimes if boosts were available. Hence N\oe therian currents in Newton--Cartan or Carroll manifolds arise for a restricted subset of isometries. This is an important spin-off of our study, that compromises former attempts to describe hydrodynamics in flat Newton--Cartan or Carroll spacetimes on the ground of N\oe therian conservation laws.

An executive outline of the present work is as follows. We remind the basics on relativistic fluids in the presence of a conserved matter current with emphasis on hydrodynamic-frame invariance. This analysis is further expanded into two distinct frames, the Zermelo and Papapetrou--Randers, appropriate for the subsequent investigation about Galilean and Carrollian fluids. Galilean fluids are first studied from the conservation perspective mirroring the Galilean-diffeomorphism invariance of Newton--Cartan spacetimes with emphasis on the effect of isometries, when present in the background; next as the infinite-$c$ limit of the relativistic hydrodynamics in Zermelo frame. Hydrodynamic-frame invariance is revisited and we move next to the Galilean massless case. The current analysis is repeated along cognate lines for Carrollian fluid dynamics with a paragraph specifically devoted to the possible multiplication of degrees of freedom in the zero-$c$ limit. Finally we describe the case of Aristotelian fluids in a short chapter before concluding. 

The subject of hydrodynamic-frame invariance and Carrollian fluids has been lately in the spotlight. Some debatable statements have been promoted in the literature, and our views are not always in line with those of other authors. Wherever necessary,  we stress it and provide the adequate elements to make the comparison clear and avoid confusion.  Our approach is meant to be a constructive criticism, and we intentionally supply  a wealth of technical details, some appended in four sections,  to back-up our conclusions, sometimes at the expense of considerably increased length.

\section{Relativistic hydrodynamics}\label{recap}

\subsection{Basic concepts and general equations}\label{rreleq}

\subsubsection*{Energy--momentum and matter conservation}

Fluid mechanics is the description of irreversible off-equilibrium thermodynamics under the assumption that the wave lengths of dynamical phenomena are large compared to typical kinetic scales. It is thus legitimate to assume local thermal equilibrium and use the laws of thermostatics (recalled in App. \ref{thermo}), although the definitions of temperature, chemical potential, entropy etc. are possibly questionable, or at least ambiguous. 

Without external forces and springs or sinks of matter, the basic requirements are covariant energy--momentum and matter (rather than mass) macroscopic conservation, encoded in the following $d+2$ equations:\footnote{Matter conservation could be multiple, or even absent as e.g. in a gas of photons, although no principle forbids the existence of conserved currents in fluids made of massless carriers -- see App. \ref{thermo}.}
\begin{eqnarray}
 \label{conT} 
\nabla_\mu T^{\mu\nu}&=&0,
\\
\label{conJ}
\nabla_\mu J^{\mu}&=&0,
\end{eqnarray}
where we assume the spacetime, of dimension $d+1$, be equipped with a metric $g_{\mu\nu}$. The energy--momentum tensor and the matter current can be decomposed along a vector congruence $u^\mu$ playing the role of velocity field normalized as $u^\mu u_\mu = - c^2$:
\begin{eqnarray}
\label{T} 
T^{\mu \nu}&=&(\varepsilon+p) \frac{u^\mu  u^\nu}{c^2} +p  g^{\mu\nu}+   \tau^{\mu \nu}+ \frac{u^\mu  q^\nu}{c^2}   +\frac{u^\nu  q^\mu}{c^2},\\
\label{curdec}
J^\mu&=&\varrho_0 u^\mu +j^\mu.
\end{eqnarray}
The viscous stress tensor $\tau^{\mu \nu}$ and the heat current $q^\mu$ are purely transverse:
\begin{equation}\label{trans}
u^\mu  q_\mu =0, \quad u^\mu    \tau_{\mu \nu}=0, \quad
u^\mu  T_{\mu \nu}=- q_\nu -{\varepsilon} u_\nu, \quad \varepsilon =\frac{1}{c^2}T_{\mu \nu} u^\mu u^\nu.
\end{equation}
They are expressed in terms of $u^i$ and their spatial components $q_i$ and $\tau_{ij}$ with $i=1,2,\ldots,d$. Similarly is the imperfect particle current from its components $j_i$, since
\begin{equation}\label{transcurr}
u^\mu  j_\mu =0, \quad 
\varrho_0 = -\frac{1}{c^2} u^\mu J_\mu.
\end{equation}
In the above expressions  -- see also App. \ref{thermo}
\begin{itemize}

\item $\varepsilon$ and  $\varrho_0$ are the energy and the matter per unit of proper volume, as measured by an observer moving at velocity $u^\mu$ (comoving);

\item $\varepsilon$ and  $\varrho_0$ are related and so are $T^{\mu \nu}$ and $J^\mu$;

\item $p$ is the local-equilibrium thermodynamic pressure obeying an equation of state of the form    $p=p(T, \mu_0)$, where $T$ and $\mu_0$ are the local temperature and chemical potential;

\item the quantities $j_i$, $q_i$ and $\tau_{ij}$ capture the physical properties of the out of equilibrium state, and are usually expressed as expansions in temperature, chemical potential and velocity derivatives: \emph{the constitutive relations in their hydrodynamic expansion}.
\end{itemize}

It is worth recalling that from the perspective of an effective action\footnote{As usual $\frac{1}{c}\text{d}^{d+1}x =
\text{d}t\wedge \text{d}x^1\wedge\ldots\wedge\text{d}x^d
$.}
$S=\frac{1}{c}\int \text{d}^{d+1}x \sqrt{-g}\mathcal{L}$,
the energy--momentum tensor is defined as 
\begin{equation}
\label{varrelT}
T^{\mu\nu}=\frac{2}{\sqrt{-g}}\frac{\delta S}{\delta g_{\mu\nu}},
\end{equation}
whereas for the matter current, a $U(1)$ gauge field with components $B_\mu $ is needed: 
\begin{equation}
\label{varrelJ}
J^\mu=\frac{1}{\sqrt{-g}}\frac{\delta S}{\delta B_{\mu}}.
\end{equation}
On the one hand, invariance under diffeomorphisms  generated by arbitrary vector fields $\upxi=\xi^{\mu}(t, \mathbf{x})\partial_\mu
$ as
\begin{equation}
\delta_\upxi g_{\mu\nu} = - \mathscr{L}_\upxi g_{\mu\nu},
\end{equation}
where 
\begin{equation}
\mathscr{L}_\upxi g_{\mu\nu}= 
\xi^\rho\partial_\rho g_{\mu\nu}+ g_{\mu\rho}\partial_\nu \xi^\rho
+ g_{\nu\rho}\partial_\mu \xi^\rho
= \nabla_\mu \xi_\nu +\nabla_\nu \xi_\mu,
\end{equation}
implies the conservation equation \eqref{conT}. On the other hand, the matter-conservation equation \eqref{conJ} is a consequence of invariance under
\begin{equation}
\delta_\Lambda B_\mu = - \partial_\mu \Lambda
\end{equation}
with $\Lambda=\Lambda(t, \mathbf{x})$.

It is important to stress at this early stage that we do not assume any isometry, neither here, nor in the subsequent limiting geometries. The energy--momentum tensor and the current should not be confused with any sort of N\oe therian currents, and their conservation is a direct consequence  of local invariances. This should be opposed to other approaches presented in the quoted literature.  

\subsubsection*{Isometries, conformal isometries and extra conservations}

If $\upxi=\xi^\mu\partial_\mu$ is a Killing field it obeys
\begin{equation}
\label{relkill}
\mathscr{L}_\upxi g_{\mu\nu}=0.
\end{equation}
Hence, due to \eqref{conT}, the current defined as 
\begin{equation}
\label{relconcur}
I_\mu=\xi^\nu T_{\mu\nu}
\end{equation}
is divergence-free
\begin{equation}
\label{relconcureq}
\nabla_\mu I^\mu=0.
\end{equation}
Using Stokes and Gauss theorems,
\begin{equation}
\int_{\mathscr{W}} \text{d}^{d+1}x \sqrt{-g}\,\nabla_\mu I^\mu= \oint_{\partial\mathscr{W}} \ast\text{I}
\label{gaustok-gen},
\end{equation}
where $ \mathscr{W}$ is a domain inside $\mathscr{M}$ and $\ast\text{I}$ is the $\mathscr{M}$-Hodge dual of $\text{I} = I_\mu \text{d}x^\mu$ ($\epsilon_{01\ldots d}=1$), 
we infer that 
\begin{equation}
\label{relconch}
Q_{I}=\frac{1}{c}\int_{\upSigma_d}\ast\text{I}
\end{equation}
is conserved. Here, ${\upSigma_d}$ is an arbitrary space-like hypersurface of $\mathscr{M}$, and ``conserved'' means that the value of $Q_{I}$ is independent of the choice of $\upSigma_d$.\footnote{If $\upSigma_d$ belongs to a family of hypersurfaces defined as $\tau(t, \mathbf{x})=\text{const.}$, the conservation is expressed as $ \frac{\text{d}Q_{I}}{\text{d}\tau}=0$. This needs not be so though, and the wording ``conservation'' is to some extent reductive, as no reference to time is needed. Care should also be taken with the behaviour of the fields at spatial infinity and if $\upSigma_d$ has itself a boundary.\label{comconservation}
}

When the energy--momentum is trace-free
\begin{equation}
\label{trfree}
 T^\mu_{\hphantom{\mu}\mu}=0,
\end{equation}
a conformal isometry suffices for producing a conservation, along the lines described above. The conformal Killing satisfies 
\begin{equation}
\label{relkillconf}
\mathscr{L}_\upxi g_{\mu\nu}=\frac{2}{d+1} \nabla_\rho \xi^\rho g_{\mu\nu}.
\end{equation}

\subsubsection*{Entropy current and entropy equation}

The variational definitions of the energy--momentum tensor and the matter current as conjugate momenta to some elementary background fields are elegant and general as we will see in the forthcoming sections, but not indispensable. The physics of the fluid relies in fact on the decomposition of these momenta into observable quantities, expressed themselves as derivative expansions, and this requires more information than the knowledge of local symmetries. To this, one should add that the entropy current $S^\mu$ is yet another physical object, which has no variational definition of the sort \eqref{varrelT} or \eqref{varrelJ}. It has actually no microscopic definition such as an expectation value of some observable, but is built order by order in the derivative expansion,\footnote{The first order is often referred to as \emph{classical irreversible thermodynamics}, the second \emph{extended irreversible thermodynamics}, etc.} requiring among others that its divergence (or, alternatively for theories which incorporate memory effects, the integrated divergence  ) be non-negative. The generic form of the entropy current is
\begin{equation}
\label{curs}
S^\mu=\frac{1}{T}\left(p u^\mu-T^{\mu\nu}u_\nu-\mu_0 J^\mu
\right) + R^\mu = \Sigma^\mu  + R^\mu ,
\end{equation} 
where $\Sigma^\mu$ is a kind of universal piece of the current, and  $R^\mu$ depends on the specific  off-equilibrium thermodynamic theory. Using \eqref{trans}, \eqref{transcurr}, \eqref{renth} and \eqref{GD},  $\Sigma^\mu$ is recast as follows: 
\begin{equation}
\label{entrcur}
\Sigma^\mu=\sigma u^\mu-\frac{\mu_0}{T} j^\mu+\frac{1}{T}q^\mu= \frac{\sigma}{\varrho_0} J^\mu- \frac{\cal w}{T\varrho_0}\left(
j^\mu
-\frac{\varrho_0}{\cal w}
q^\mu
\right)
\end{equation}
with $\sigma$ the entropy, ${\cal w}$ the relativistic enthalpy \eqref{renth} and  $\mu_0$ the relativistic chemical potential.

It is convenient, both for the relativistic dynamics and for its Galilean or Carrollian limits, to consider the longitudinal and transverse projections of \eqref{conT}, possibly combined with \eqref{conJ} and the thermodynamic laws \eqref{GD}, \eqref{1stl} and \eqref{1stlLL} in order to trade the energy for the entropy. For the longitudinal projection, we find:\footnote{This is the generalization of Eq. (127,5) in \cite{Landau} for general gravitational backgrounds and in an arbitrary hydrodynamic frame.}
\begin{eqnarray}
\label{en1}
-u_\nu\nabla_\mu T^{\mu\nu}
&=&\text{u}(\varepsilon)+ \left({\cal w}+\frac{\tau}{d}\right) \Theta 
+ \tau^{\mu\nu}\sigma_{\mu\nu}+\nabla_\nu q^\nu+\frac{ a_\nu q^\nu}{c^2},
\\&=&T\nabla_\nu \Sigma^\nu+ \frac{\tau}{d} \Theta + \tau^{\mu\nu}\sigma_{\mu\nu}+T j^\nu  \partial_\nu\frac{\mu_0}{T}+q^\nu \left(\frac{\partial_\nu T}{T} +\frac{a_\nu}{c^2}\right),
\label{en2}
\end{eqnarray}
where $\tau= \tau^{\mu\nu}g_{\mu\nu}$ is the \emph{relativistic non-equilibrium pressure} and 
$\text{u}(f)$ stands for $u^\mu \partial_\mu f$. We have also introduced the following kinematical tensors:\footnote{Our conventions for (anti-) symmetrization are 
$A_{(\mu\nu)}=\frac{1}{2}\left(A_{\mu\nu}+A_{\nu\mu}\right)$ and $ 
A_{[\mu\nu]}=\frac{1}{2}\left(A_{\mu\nu}-A_{\nu\mu}\right)$.}
\begin{eqnarray}
&a_\mu =u^\nu \nabla_\nu u_\mu , \quad
\Theta=\nabla_\mu  u^\mu ,& \label{def21}\\
&\sigma_{\mu \nu}= \nabla_{(\mu } u_{\nu )} + \frac{1}{c^2}u_{(\mu } a_{\nu )} -\frac{1}{d} \Theta\,h_{\mu \nu}  ,&
\label{def23}
\\ &\omega_{\mu \nu}= \nabla_{[\mu } u_{\nu ]} +  \frac{1}{c^2}u_{[\mu }a_{\nu] },&\label{def24}
\end{eqnarray}
which are the acceleration, the expansion, the shear and the vorticity of the velocity field, with
$h_{\mu \nu} $ and  $U_{\mu \nu}$ the projectors onto the space transverse and longitudinal to the velocity field:
\begin{equation}
\label{relproj}
h_{\mu\nu}=\frac{u_{\mu}u_{\nu}}{c^2}+g_{\mu\nu},\quad U_{\mu\nu}=-\frac{u_{\mu}u_{\nu}}{c^2}.
\end{equation}

\subsubsection*{Relativistic hydrodynamic-frame invariance}

The absence of sharp distinction between energy and mass flow in relativistic theories brings some redundancy in the above fluid data such as $q^\mu$ and $j^\mu$. In his seminal theory of relativistic fluids \cite{Eckart}, Eckart shed this redundancy by making  the choice $j^\mu=0$, whereas Landau and Lifshitz \cite{Landau} required instead  $q^\mu=0$. It is more generally admitted that one has the freedom  to redefine  
\begin{equation}
T(x)\to T^{\prime}(x),\quad \mu_0(x)\to\mu_0^{\prime}(x), \quad u^\mu(x)\to u^{\mu \prime}(x),
\end{equation} 
provided we modify accordingly $\varepsilon(x)$, $p(x)$, $\varrho_0(x)$, 
$q^\mu(x)$, $\tau^{\mu\nu}(x)$ and $j^\mu(x)$ so that the energy--momentum tensor, the conserved current and the entropy current remain unaltered. This is the gauge symmetry, associated with local Lorentz transformations, known as \emph{hydrodynamic-frame invariance}. It translates, that the velocity field has no first-principle definition  in relativistic hydrodynamics.\footnote{In fact, the two extreme options embraced by Eckart and Landau--Lifshitz  may not be achievable in all systems. The former demands a time-like conserved current $J^\mu $ and selects a velocity field aligned with it. The latter requires that the energy--momentum tensor has a time-like eigenvector with positive eigenvalue and alignes the velocity with this eigenvector. In both cases, the defining assumption is reasonable, and the output unique, leaving no residual hydrodynamic-frame invariance. The equivalence amongst these frames relies on a deep interplay between the energy--momentum and the conserved current, when it exists. This interplay reveals in thermodynamics and specifically in the entropy current. However, although equivalent, the two frames are not equally suited for the non-relativistic limit, as we will see in Sec. \ref{galdyn} (see also \cite{RZ}).\label{ELL}}

It should be stressed here that irrespective of the aforementioned local Lorentz freedom,  dealing with out-of-equilibrium phenomena brings its share of indetermination. Once the global equilibrium is abandoned, thermodynamic functions become local and their very existence relies on local thermodynamic equilibrium. They are furthermore supposed to be slow-varying for hydrodynamics to make sense. Nonetheless, even within these assumptions, for non-perfect fluids, neither $p(x)$,  $\varepsilon(x)$ and $\varrho_0(x)$ appearing in the fluid equations, nor  $T(x)$ and $\mu_0(x)$ 
entering the constitutive relations need a priori to be identified with the corresponding local-equilibrium thermodynamic quantities. We have made that choice here, and we will not further discuss this side of hydrodynamic-frame invariance, but rather focus on the kinematical aspects. More information can be found in the already quoted literature, and in particular in \cite{Kovtun:2012rj, Bemfica:2017wps, Grozdanov:2019kge, Grozdanov:2019uhi, Kovtun:2019hdm, Bemfica:2019knx, Hoult:2020, Bemfica:2020zjp, Dore:2021xqq, Hoult:2021gnb}.

Suppose we perform a local Lorentz transformation on the fluid, i.e. a transformation on the velocity vector field
\begin{equation}
\label{loclor}
\text{u}\to \text{u}+\delta \text{u},\quad \delta \text{u}\cdot\text{u}=0.
\end{equation}
One can transform accordingly the various pieces that appear in the decomposition of $ T_{\mu\nu}$ \eqref{T} so that $ \delta T_{\mu\nu}=0$:\footnote{We cannot disentangle at this stage the transverse components $p$ and $\tau_{\mu\nu}$, as their separation relies on thermodynamics ($p$ is the equilibrium pressure, $\tau_{\mu\nu}$ stands for the non-equilibrium stress and its trace is the non-equilibrium pressure). 
%, which requires further discussing the entropy current.
}
\begin{eqnarray}
\label{deleps}
\delta \varepsilon &=&-\frac{2}{c^2}q^\mu \delta u_\mu,\\
\label{delq}
\delta q_\nu&=&\frac{ u_\nu}{c^2} q^\mu \delta u_\mu-{\cal w} \delta u_\nu- \tau_{\nu\mu}\delta u^\mu,\\
\delta \left(p h_{\mu\nu}+\tau_{\mu\nu}\right)&=& \frac{p}{c^2}\left(u_\mu \delta u_\nu+u_\nu \delta u_\mu 
\right)+
 \frac{1}{c^2}\left(u_\mu \tau_{\nu\rho}+u_\nu \tau_{\mu\rho}\right)\delta u^\rho
\nonumber
\\
\label{delptau}
&& -\frac{1}{c^2}\left(\delta u^\mu q^\nu+\delta u^\nu q^\mu\right)
 .
\end{eqnarray}
Similarly, in the presence of a matter current, one requires $\delta J_{\mu}=0$, and 
using  \eqref{curdec} one obtains:
\begin{equation}
\label{delj}
\delta j_{\mu}=-u_{\mu}\delta \varrho_0-  \varrho_0\delta u_{\mu}. 
\end{equation}

The interplay between the energy--momentum tensor and the matter current is ensured by the thermodynamic laws and the entropy current. The latter must also be hydrodynamic-frame-invariant, as are the thermodynamic relations, in particular the Gibbs--Duhem equation and the equation of state.  These requirements provide the tools for computing $\delta \sigma$, $\delta p$, $\delta T$, and $\delta \mu_0$, in terms of the above variations already expressed with the basic data $\delta u_{\mu}$ and $\delta \varrho_0$. We will not delve in analyzing the effects induced on thermodynamic observables  by the local Lorentz transformations (for that see e.g. \cite{Kovtun:2012rj, Bemfica:2017wps, Grozdanov:2019kge, Grozdanov:2019uhi, Kovtun:2019hdm, Bemfica:2019knx, Hoult:2020, Bemfica:2020zjp, Dore:2021xqq, Hoult:2021gnb, Ciambelli:2017wou}), but will conclude this discussion with some remarks. At the first place, as already stressed, there is no microscopic definition of the entropy current. Hydrodynamic-frame invariance serves therefore as a prerequisite for completing expressions like \eqref{entrcur}, order by order in the derivative expansion. Secondly, $q_\mu$,  $\tau_{\mu\nu}$ and $j_\mu$ are given in the form of constitutive relations, which are asymptotic series, terminated at finite order. This blurs the fluid-frame invariance, and is at the heart of the caveats mentioned in the introduction (stability and causality). Privileged hydrodynamic frames unavoidably emerge, depending on the physical situation under consideration. Last but not least, even when the previous items are under control (as in two dimensions \cite{Campoleoni:2018ltl}), the question of global issues remains, in the lines discussed in Refs. \cite{Campoleoni:2018ltl,CMPR, CMPRpos}: like any gauge symmetry, hydrodynamic-frame invariance suffers from possible global breaking.

\subsubsection*{A remark on perfect fluids}

For a perfect fluid, the heat current, the viscous stress tensor and the transverse part of the matter current vanish. From an intrinsic view, given an energy--momentum tensor and a matter current, these requirements are met if the matter current is an eigenvector of the energy--momentum tensor, and if the transverse part of the energy--momentum tensor with respect to this eigenvector is proportional to the projector orthogonal to the current. Given the normalisation of the velocity, here aligned with the current, this provides unambiguously $\varrho_0$, $p$ and $\varepsilon$ (see \eqref{T},  \eqref{curdec}, \eqref{trans} and  \eqref{transcurr}), and guarantees at the same time the absence of  $j^\nu$, $q^\nu$ and $\tau^{\mu\nu}$. The entropy current \eqref{curs} in now given by \eqref{entrcur} and is proportional to the velocity. From \eqref{en2} we infer it has zero divergence.

Formally, one could perform local hydrodynamic-frame transformations. However, following the rules 
\eqref{loclor},  \eqref{deleps},  \eqref{delq}, \eqref{delptau} and \eqref{delj}, no transformation exists, which preserves the  
perfect forms of the matter current and of the energy--momentum tensor. They all generate non-perfect components, which should in this case be considered spurious because they do not reflect any genuine microscopic interaction.\footnote{Although unphysical, this formal freedom is important from a holographic perspective, and was discussed extensively in \cite{Campoleoni:2018ltl, CMPR, CMPRpos}.} In the absence of dissipative phenomena and heat transport, energy is carried by matter and there is no ambiguity in defining a physical fluid velocity, in line with the presentation of Ref.\cite{Landau}.

\subsubsection*{Weyl-invariant fluids}

A physical system such as a fluid can be invariant under Weyl transformations. Those act on the background metric  
and fluid velocity as
\begin{equation}
\text{d}s^2\to\mathcal{B}^{-2} \text{d}s^2, \quad u^\mu\to \mathcal{B} u^\mu,
\end{equation}
and more general tensors are Weyl-covariant if they rescale with some power of  $\mathcal{B}$ (weight $w$)
-- not to be confused with the relativistic enthalpy $\mathcal{w}$).
A Weyl-covariant derivative $\mathscr{D}_\mu$ maintains the canonical Weyl transformation of a Weyl-covariant tensor, and calls for a Weyl connection one-form:\footnote{The explicit form of $\text{A}$ is obtained by demanding $\mathscr{D}_{\mu}u^{\mu}=0$ and $u^{\lambda}\mathscr{D}_{\lambda}u_{\mu}=0
$. The Weyl connection is not unique. Any weight-$w$ conformal vector $\text{V}$ is associated with a bona fide Weyl connection $\text{A}_V=\frac{1}{\| \text{V}\|^2 }\left(\frac{(2-w)\Theta_V }{d+1-w}\text{V}-\text{a}_V\right)$ with $\Theta_V$ and $\text{a}_V$, the expansion and acceleration of $\text{V}$. An example of alternative Weyl connection will be presented in Sec. \ref{conFLUIDS}.
\label{confcongen}}
\begin{equation}
\label{Wconc}
\text{A}=\frac{1}{c^2}\left(\text{a} -\frac{\Theta}{d} \text{u}\right).
\end{equation}
The Weyl covariant derivative is metric-compatible:
\begin{eqnarray}
\mathscr{D}_\rho g_{\mu\nu}&=&0,\\
\left(\mathscr{D}_\mu\mathscr{D}_\nu -\mathscr{D}_\nu\mathscr{D}_\mu\right) f&=& w f F_{\mu\nu},
\end{eqnarray}
where  the action on a weight-$w$ scalar $f$ is
\begin{equation}
\label{WCscalar}
\mathscr{D}_\lambda f=\nabla_\lambda  f +w A_\lambda f,
\end{equation}
and
\begin{equation}
\label{F}
F_{\mu\nu}=\partial_\mu A_\nu-\partial_\nu A_\mu
\end{equation}
is the Weyl curvature (weight-$0$). 
For concreteness, the action of $\mathscr{D}_\lambda$  on a weight-$w$ form $v_\mu$ and a  weight-$w$ tensor $t_{\mu\nu}$ is
\begin{eqnarray}
\label{Wv}
\mathscr{D}_\lambda v_\mu&=&\nabla_\lambda v_\mu+(w+1)A_\lambda v_\mu + A_\mu v_\lambda-g_{\mu\lambda} A^\rho v_\rho,
\\
\mathscr{D}_\lambda  t_{\mu\nu}
&=&\nabla_\lambda  t_{\mu\nu}+(w+2)A_\lambda  t_{\mu\nu} 
+A_\mu  t_{\lambda\nu} +A_\nu  t_{\mu\lambda} 
-g_{\mu\lambda} A^\rho  t_{\rho\nu} 
-g_{\nu\lambda} A^\rho  t_{\mu\rho} 
. 
\end{eqnarray}

Commuting the Weyl-covariant derivatives acting on vectors, as usual one defines the Weyl-covariant Riemann tensor 
\begin{equation}
\label{relweylcurv}
\left(\mathscr{D}_\mu\mathscr{D}_\nu -\mathscr{D}_\nu\mathscr{D}_\mu\right) v^\rho=
\mathscr{R}^\rho_{\hphantom{\rho}\sigma\mu\nu} v^\sigma+ w v^\rho F_{\mu\nu}
\end{equation}
($v^\rho$ are the components of a weight-$w$ vector)
and the usual subsequent quantities. The Weyl-covariant Ricci (weight $0$) and scalar (weight $2$) curvatures read: 
\begin{eqnarray}
\mathscr{R}_{\mu\nu}&=&{R}_{\mu\nu} + (d-1)\left(\nabla_\nu A_\mu + A_\mu A_\nu-g_{\mu\nu}A_\lambda A^\lambda\right) +
g_{\mu\nu}\nabla_\lambda A^\lambda
-F_{\mu\nu},
\label{curlRic}
\\
\mathscr{R}&=&R +2d\nabla_\lambda A^\lambda- d(d-1) A_\lambda A^\lambda . 
\label{curlRc}
\end{eqnarray}

The fluid dynamics captured by \eqref{conT} and \eqref{conJ} is Weyl-invariant under the necessary and sufficient condition that the energy--momentum tensor and the matter current are Weyl-covariant and such that 
\begin{equation}
 \label{conconTJ} 
\nabla_\mu T^{\mu\nu}=\mathscr{D}_\mu T^{\mu\nu},
\quad
\nabla_\mu J^{\mu}=\mathscr{D}_\mu J^{\mu}.
\end{equation}
This demands the conformal weights of $T_{\mu\nu}$ and $J_{\mu}$ be $d-1$, and $T_{\mu\nu}$ be traceless. 
The required weight for the energy--momentum tensor is the translation of  Weyl invariance for the underlying action, as one infers from Eq. \eqref{varrelT}; this Weyl invariance also imposes the absence of trace.\footnote{We recall that
$\delta S=\int \text{d}^{d+1}x \sqrt{-g}\left(
\frac{1}{2}T^{\mu\nu}\delta g_{\mu\nu}
+
J^\mu \delta B_\mu
\right)$.
Hence for an infinitesimal Weyl rescaling (i.e. $\mathcal{B}$ close to the identity), $\delta_{\mathcal{B}} S=-\int \text{d}^{d+1}x \sqrt{-g}\ln \mathcal{B} T_\mu^{\hphantom{\mu}\mu}$.} 
In the decomposition \eqref{T} the latter condition reads $-\varepsilon+dp +\tau^\mu_{\hphantom{\mu}\mu}=0$, usually split into the conformal global-equilibrium equation of state plus a condition on the piece associated with dynamical irreversible phenomena:
\begin{equation}\label{con}
\varepsilon= dp,\quad \tau^\mu_{\hphantom{\mu}\mu}=0. 
\end{equation} 
Furthermore we learn  from Eq. 
\eqref{varrelJ} that the gauge field $B_\mu $ conjugate to $J^\mu $ is weight-zero to comply with the expected weight for $J_\mu $.
We have summarized the weights of the various physical quantities in the Tab. \ref{weights}.
\begin{table}[h!]
  \begin{center}
       \begin{tabular}{c|c} 
      \textbf{weight} & \textbf{observables}\\
      \hline
      $d+1$ & $\varepsilon$, $p$ \\
      $d$ & $q_\mu$, $\varrho_0$ \\
      $d-1$ & $\tau_{\mu\nu}$, $j_\mu$ \\
    \end{tabular}
    \caption{Conformal weights.} \label{weights}
  \end{center}
\end{table}

\subsection{Fluids in Zermelo coordinates}\label{prrfi}

\subsubsection*{Zermelo frame}

In a pseudo-Riemannian manifold $\mathscr{M}$ of $d+1$ dimensions, one can always assume the Arnowitt--Deser--Misner form of the metric
\begin{equation}
\label{galzerm}
\text{d}s^2 =-\Omega^2 c^2 \text{d}t^2+a_{ij} \left(\text{d}x^i -w^i  \text{d}t\right)\left( \text{d}x^j-w^j \text{d}t\right)
\end{equation}
with  $a_{ij}$, $w^i$ and $\Omega$  functions of $x=(ct,\mathbf{x})=\{x^\mu, \mu=0,1,\ldots,d\}$ and $\mathbf{x}$ stands for $\{x^1,\ldots,x^d\}$.
 These coordinates are well-suited for the implementation of the Galilean limit \cite{CMPPS1}. Indeed, 
\emph{Galilean diffeomorphisms }  
\begin{equation}
\label{galdifs} 
t'=t'(t),\quad \mathbf{x}^{\prime}=\mathbf{x}^{\prime}(t, \mathbf{x})
\end{equation}
have Jacobian 
\begin{equation}
 \label{galj}
J(t)=\frac{\partial t'}{\partial t},\quad j^i(t,\mathbf{x}) = \frac{\partial x^{i\prime}}{\partial t},\quad 
J^i_j(t,\mathbf{x}) = \frac{\partial x^{i\prime}}{\partial x^{j}},
\end{equation}
and
the transformation of the metric components
is nicely reduced:
\begin{equation}
\label{galdifawom}
a^{\prime}_{ij} =a_{kl} J^{-1k}_{\hphantom{-1}i} J^{-1l}_{\hphantom{-1}j} ,
\quad
w^{\prime k}=\frac{1}{J}\left(J^k_i w^i+j^k\right),
\quad
\Omega^{\prime }=\frac{\Omega}{J}.
\end{equation}
This fits with the Newton--Cartan structure emerging at infinite $c$. 

We call \emph{Zermelo} metrics  (see \cite{Gibbons:2008zi}) the restricted class of \eqref{galzerm} for which $\Omega$ depends on $t$ only. This class is stable under Galilean diffeomorphisms because $J$ in \eqref{galdifawom} does not depend on spatial coordinates. The corresponding Newton--Cartan geometries reached in the Galilean limit are torsion-free (see \cite{Festu}) and they feature an absolute, invariant Newtonian time $\int \text{d}t\, \Omega(t) =\int \text{d}t^\prime\, \Omega^\prime(t^\prime)$.

The above reduction with respect to the Galilean diffeomorphisms  \eqref{galj}, can be completed as follows. Any tensor component with an upper time index transforms as a Galilean density, and thus is a scalar upon multiplication by $\Omega$. Similarly the components with lower spatial indices transform as Galilean tensors. As an example, the transformation of the $d+1$ vector components $u^\mu$ under a Galilean diffeomorphism leads to\footnote{When the indices are inverted, the transformations are of the connection type:  $ u^{\prime i} = J^i_k u^k+J^i u^0$, $u^{\prime}_{0}=\frac{1}{J}\left( u_0-u_j J^{-1j}_{\hphantom{-1}k} J^k \right)$. For those, the tensorial structure is restored  at the infinite-$c$ limit, where indices are lowered and raised with $a_{ij}$ and its inverse.
 \label{odowniup}} 
\begin{equation}
u^{\prime 0}=J u^0,\quad u^{\prime}_{i} =u_k J^{-1k}_{\hphantom{-1}i}.
\end{equation} 

A relativistic fluid moving in \eqref{galzerm} is described by the components of its velocity $\text{u}$, normalized as $\| \text{u} \|^2=-c^2$:
\begin{equation}
\label{vel}
u^\mu=\frac{ \text{d}x^\mu}{\text{d}\tau}\Rightarrow
u^0=\gamma c,\ u^i = \gamma v^i,
\end{equation}
where the Lorentz factor $\gamma$ is, in the Zermelo frame\footnote{Expressions as $\mathbf{v}^2$ stand for $a_{ij}v^i v^j$, not to be confused with $\| \text{u} \|^2=g_{\mu\nu}u^\mu u^\nu$.}
\begin{equation}
\label{gamzerm}
\gamma=\frac{\text{d}t}{\text{d}\tau}=
\frac{1}{\Omega\sqrt{1-\left(\frac{\mathbf{v}-\mathbf{w}}{c\Omega}\right)^2}}.
\end{equation}
Under a Galilean diffeomorphism \eqref{galj}, the transformation of $u^\mu$ (see footnote \ref{odowniup})  induces the expected transformation on $v^i$:
\begin{equation}
\label{galdifv}
v^{\prime k}=\frac{1}{J}\left(J^k_i v^i+j^k\right).
\end{equation}

Orthogonality conditions \eqref{trans} and \eqref{transcurr} imply that the fundamental data for the non-perfect matter current, the heat current and the stress tensor are $j_i$, $q_j$ and $\tau_{kl}$. Other components are e.g. 
\begin{equation}
 \label{galconhs} 
j^0=\frac{\left(v^i-w^i\right)j_i}{c\Omega^2},\quad q^0=\frac{\left(v^i-w^i\right)q_i}{c\Omega^2}, 
\quad \tau^{00}=\frac{\left(v^k-w^k\right)\left(v^l-w^l\right)\tau_{kl}}{c^2\Omega^4},
\quad
\tau^0_{\hphantom{0}j}= \frac{\left(v^k-w^k\right)\tau_{kj}}{c\Omega^2},
\end{equation}
which transform as tensors under Galilean diffeomorphisms.

\subsubsection*{Hydrodynamic-frame transformations and invariants}

The fluid velocity is parameterized in \eqref{vel} with $d$ components $v^i$. We can thus formulate the relativistic hydrodynamic-frame transformations in terms of arbitrary $\delta v^i(x)$. In the Zermelo  frame, we obtain:
\begin{equation}
\label{delgamZ}
\delta \gamma =  \frac{\gamma^3}{c^2} \delta v^i \left(
v_ i - w_ i \right),
\end{equation}
hence
\begin{equation}
\label{deluZ}
\delta \text{u}= \gamma\delta v^i\left(\partial_i +\frac{\gamma^2}{c^2}\left(v_i-w_i\right)\left(\partial_t+v^k\partial_k\right)\right).
\end{equation}
Using  Eqs. \eqref{deleps}, \eqref{delq} and \eqref{delptau} together with 
\eqref{galconhs} and \eqref{deluZ} we find:\footnote{Notice that $q_\mu \delta u^\mu= \gamma\delta v^i q_i$.}
\begin{eqnarray}
\label{delepsZ}
\delta \varepsilon&=& -2\frac{\gamma}{c^2}\delta v^i q_i,\\
\label{delqZ}
\delta q_i&=& \gamma \delta v^k\left(\frac{\gamma}{c^2}
\left(v_i-w_i\right)q_k
-{\cal w} h_{ki}
-\tau_{ki}
\right),
\\
\nonumber
\delta \left(p h_{ij}+\tau_{ij}\right)&=& \frac{\gamma^2}{c^2} \delta v^k
\left(\left(v_i-w_i\right)
\left(\tau_{jk}+p h_{jk}\right)
+\left(v_j-w_j\right)
\left(\tau_{ik}+p h_{ik}\right)
\right)
\\
\label{deltau2Z}
&&- \frac{\gamma}{c^2} \delta v^k\left(q_i h_{jk}+q_j h_{ik}
\right),
\end{eqnarray}
where (see \eqref{relproj})
\begin{equation}
\label{hkiZ}
h_{ik}=a_{ik} +\frac{\gamma^2}{c^2}
\left(v_i-w_i\right)\left(v_k-w_k\right)
.
\end{equation}
When a matter current is available, the above is completed with \eqref{delj}, which gives 
\begin{equation}
\label{delrho0Z}
\delta \varrho_0= -\frac{\gamma}{c^2}\delta v^i j_i
\end{equation}
and 
\begin{equation}
\label{deljZ}
\delta j_i=\delta v^k\left(\frac{\gamma^2}{c^2}j_{k}\left(v_i-w_i\right)
- \gamma \varrho_0 h_{ki}
\right).
\end{equation}

The transformations at hand translate the invariance of the energy--momentum tensor and current components. The latter define therefore \emph{invariants}, which are simply the energy density, the heat current, the stress tensor, the matter density and the matter non-perfect current in a privileged frame, that we will call ``at rest'' or ``proper,''\footnote{We call this frame ``fiducial'' in Sec. \ref{galdyn} and show it is associated with an observer at velocity $\text{u}_{\text{Z}} = \text{e}_{\hat t}$ given in \eqref{galet}.} borrowing the standard expressions of special relativity:
\begin{equation}
\label{restTinvZer}
T^{00}=\frac{\varepsilon_{\text{r}}}{\Omega^2},
\quad
T^{0}_{\hphantom{0}i}=\frac{1}{c\Omega}q_{\text{r}i},
\quad
T_{ij}=p_{\text{r}} a_{ij}+\tau_{\text{r}{ij}}
\end{equation}
with trace
\begin{equation}
\label{restraceinvZ}
T_\mu^{\hphantom{\mu}\mu}=-\varepsilon_{\text{r}} +d p_{\text{r}} + a^{ij} \tau_{\text{r}ij}
,
\end{equation}and
\begin{equation}
\label{restJinvZer}
J^{0}=\frac{c}{\Omega} \varrho_{0\text{r}},
\quad
J_{i}=j_{\text{r}i}.
\end{equation}
 We find explicitly
\begin{eqnarray}
\label{invenZer}
\varepsilon_{\text{r}}&=&\varepsilon \gamma^2  \Omega^2+\frac{2}{c^2}\gamma q_i
\left(v^i-w^i\right) + \left(ph_{ij}+\tau_{ij}\right)\frac{
\left(v^i-w^i\right)\left(v^j-w^j\right)}{c^2\Omega^2}
,
\\
\label{invqZer}
q_{\text{r}i}&=&
\varepsilon \gamma^2  \Omega\left(v_i-w_i\right)
+\gamma\Omega q_j\left(
\delta^j_i
+ \frac{\left(v^j-w^j\right)\left(v_i-w_i\right)}{c^2\Omega^2}
\right)
\nonumber\\
&&+\left(ph_{ij}+\tau_{ij}\right)\frac{v^j-w^j}{\Omega}
,\\
\label{invtauZer}
p_{\text{r}} a_{ij}+\tau_{\text{r}{ij}}&=&
\frac{\varepsilon \gamma^2 }{c^2}\left(
v_i-w_i\right)\left(v_j-w_j\right)
+\frac{\gamma}{c^2}
\left(
q_i\left(v_j-w_j\right)+q_j 
\left(
v_i-w_i\right)
\right)
\nonumber\\
&&+p h_{ij}+\tau_{ij},
\end{eqnarray}
and
\begin{eqnarray}
\label{invrhoZer}
\varrho_{0\text{r}}&=& \varrho_0 \Omega \gamma + j_i\frac{v^i-w^i}{c^2\Omega}
,\\
\label{invjZer}
j_{\text{r}i}&=&
j_i+ \varrho_0 \gamma 
\left(v^i-w^i\right)
.
\end{eqnarray}
It should be stressed that the above quantities are hydrodynamic-frame invariant but also covariant under Galilean diffeomorphisms. This latter property will be useful when considering the Galilean limit.

\subsubsection*{Killings and conserved currents}

Consider a Killing field on $\mathscr{M}$ satisfying \eqref{relkill}
\begin{equation}   
\label{genkil}
\upxi=\xi^t\partial_t +\xi^i \partial_i=
\xi^{\hat t}\text{e}_{\hat t}+ \xi^{\hat\imath }\text{e}_{\hat\imath}
,
 \end{equation}
where we have introduced a somewhat more convenient frame and coframe
\begin{equation}   
\text{e}_{\hat t}=\frac{1}{\Omega} \left(\partial_t+ w^j \partial_j
\right), \quad \text{e}_{\hat\imath}=\partial_i,
\quad
\uptheta^{\hat t}=\Omega\text{d}t, \quad \uptheta^{\hat\imath}=\text{d}x^i-w^i \text{d}t,
\label{zerfrcofr}
 \end{equation}
so that the metric \eqref{galzerm} reads:
 \begin{equation}   
 \label{orthonmet}
\text{d}s^2= -c^2 \left(\uptheta^{\hat t}\right)^2+ a_{ij}\uptheta^{\hat\imath}\uptheta^{\hat\jmath}.
 \end{equation}
Hence
 \begin{equation}   
\xi^{\hat t} = \xi^t \Omega, \quad \xi^{\hat\imath} = \xi^i - \xi^tw^i, \quad \xi_{\hat t} = -c^2  \xi^{\hat t},
\quad  \xi_{\hat\imath} = a_{ij} \xi^{\hat\jmath} =\xi_i.
 \end{equation}
With these data, the components of the conserved current \eqref{relconcur} are\footnote{We use the standard decomposition  $I^\mu=\iota_0 u^\mu +i^\mu$ with $u^\mu  i_\mu =0$ and  
$\iota_0 = - u^\mu I_\mu$, and introduce $ \iota_{0\text{r}}$ as a proper or fiducial density, following the footsteps of the energy--momentum tensor and the matter current, Eqs. \eqref{restTinvZer} and \eqref{restJinvZer}.}
\begin{equation}
\label{restIinvZer}
I^{0}=\frac{c}{\Omega} \iota_{0\text{r}},
\quad
I_{i}=i_{\text{r}i},
\end{equation}
where
\begin{eqnarray}
\label{inviotaZer}
\iota_{0\text{r}}&=& \frac{1}{c^2}\xi^{\hat\imath} q_{\text{r}i}-\xi^{\hat t}  \varepsilon_{\text{r}}
,\\
\label{inviZer}
i_{\text{r}i}&=&\xi^{\hat\jmath}\left(p_{\text{r}} a_{ij}+\tau_{\text{r}{ij}}\right)-
 \xi^{\hat t} q_{\text{r}i}
.
\end{eqnarray}
The associated conserved charge is obtained using \eqref{relconch}:
\begin{eqnarray}
Q_I&=&\int_{\upSigma_d}\sqrt{a}\iota_{0\text{r}}
\left(\text{d}x^1-w^1 \text{d}t\right)\wedge\ldots\wedge\left(\text{d}x^d-w^d \text{d}t\right)
\nonumber
\\
&&-\int_{\upSigma_d}\sqrt{a}\sum_{i=1}^d \left(\text{d}x^1-w^1 \text{d}t\right)\wedge\ldots\wedge a^{ij}i_{\text{r}j}\Omega \text{d}t\wedge\ldots\wedge\left(\text{d}x^d-w^d \text{d}t\right),
\label{relconch-Z}
\end{eqnarray}
where $a^{ij}i_{\text{r}j}\Omega \text{d}t$ is the $i$th factor in the exterior product of the last term.

\subsection{Fluids in Papapetrou--Randers coordinates}

\subsubsection*{Papapetrou--Randers frame}

An alternative frame for a $d+1$-dimensional pseudo-Riemannian manifold $\mathscr{M}$ is defined as 
\begin{equation}
\label{carrp}
\text{d}s^2 =- c^2\left(\Omega \text{d}t-b_i \text{d}x^i
\right)^2+a_{ij} \text{d}x^i \text{d}x^j,
\end{equation}
where all functions are $x$-dependent -- again $x\equiv(x^0=ct,\mathbf{x})$. It is known as \emph{Papapetrou--Randers}, and this form is stable under  \emph{Carrollian diffeomorphisms} 
 \begin{equation}
\label{cardifs} 
t'=t'(t,\mathbf{x})\quad \text{and} \quad \mathbf{x}^{\prime}=\mathbf{x}^{\prime}(\mathbf{x})
\end{equation}
with Jacobian
\begin{equation}
 \label{carj}
J(t,\mathbf{x})=\frac{\partial t'}{\partial t},\quad j_i(t,\mathbf{x}) = \frac{\partial  t'}{\partial x^{i}},\quad 
J^i_j(\mathbf{x}) = \frac{\partial x^{i\prime}}{\partial x^{j}}.
\end{equation}
Under Carrollian diffeomorphisms, $\Omega$ and the metric transform as in \eqref{galdifawom} i.e.
\begin{equation}
\label{cardifa}
\Omega^{\prime }=\frac{\Omega}{J}, \quad
a^{\prime ij}=J^i_k J_l^ja^{kl}, 
\end{equation}
while $b^i$ obeys a connection transformation
\begin{equation}
\label{cardifb}
b^{\prime}_{k}=\left( b_i+\frac{\Omega}{J} j_i\right)J^{-1i}_{\hphantom{-1}k}.
\end{equation}

The Papapetrou--Randers frame realizes a reduction with respect to the Carrollian diffeomorphisms  \eqref{cardifs}. Any tensor component with a lower time index transforms as a Carrollian density and provides a scalar upon division by $\Omega$; the components with upper spatial indices transform as Carrollian tensors. The $d+1$ vector components $u^\mu$ transform under a Carrollian diffeomorphism as 
\begin{equation}
u^{\prime}_{ 0}= \frac{u_0}{J},\quad u^{\prime i} =u^k J^{i}_{k}.
\end{equation} 

One can again express the components of a velocity field normalized  to $-c^2$ as $u^0=\gamma c$ and $u^i = \gamma v^i$. It is furthermore convenient to parameterize $v^i$ as 
\begin{equation}
\label{vbetacar}
v^{i}=\frac{c^2\Omega\beta^i}{1+c^2\beta^j b_j}
\Leftrightarrow
\beta^{i}=\frac{v^i}{c^2\Omega\left(1-\frac{v^j b_j}{\Omega}\right)},
\end{equation}
because of future use in the Carrollian limit, and due to the simple Carrollian transformation property this definition leads to\footnote{This is easily proven by observing that
$\beta_i+b_i=-\frac{\Omega u_i}{cu_0}$. We define as usual $b^i=a^{ij}b_j$, $\beta_i = a_{ij}\beta^j$, $v_i = a_{ij}v^j$, 
$\pmb{b}^2=b_ib^i$,  $\pmb{\beta}^2=\beta_i \beta^i$ and $\pmb{b}\cdot \pmb{\beta}=b_i \beta^i$.}
\begin{equation}
\label{car-beta-tran}
\beta^{i\prime}=J^i_j \beta^{j}.
\end{equation}
Now the Lorentz factor reads:
\begin{equation}
\label{gammabeta}
\gamma=\frac{1+c^2 \pmb{\beta}\cdot\pmb{b}}{{\Omega}\sqrt{1-c^2\pmb{\beta}^2}}.
\end{equation}

In Papapetrou--Randers frame  \eqref{carrp}, the fundamental hydrodynamic variables are naturally chosen as $j^i$, $q^j$ and $\tau^{kl}$. Using the transversality conditions \eqref{trans} we find:
\begin{equation}
 \label{carmhstr}
 j_0=-c\Omega \beta_i j^i
 ,\quad
q_0=-c\Omega \beta_i q^i
 ,\quad
 \tau_{00}=c^2\Omega^2\beta_k\beta_l\tau^{kl}
 ,\quad
\tau^i_{\hphantom{i}0}=
-c\Omega \beta_k\tau^{ik}.
\end{equation}
These are all Carrollian tensors (or densities).

\subsubsection*{Hydrodynamic-frame transformations and invariants}

Following the same pattern as for the Zermelo frame, we  investigate the hydrodynamic-frame transformations, i.e. local Lorentz transformations  captured here in the $d$ components   $\delta\beta^i(x)$.
We obtain
\begin{equation}
\label{deluRP}
\delta u^0= c\delta \gamma,\quad
\delta u^i= 
c^2\frac{h^{ik}\delta \beta_k}{\sqrt{1-c^2\pmb{\beta}^2}}
\end{equation}
with 
\begin{equation}
\label{delgamRP}
\delta \gamma = c^2 \gamma \delta \beta^i \left(
\frac{b_i}{1+c^2 \pmb{\beta}\cdot \pmb{b}}
+\frac{\beta_i}{1-c^2\pmb{\beta}^2}
\right),
\end{equation}
and
\begin{equation}
\label{hkiRP}
h^{ik}=a^{ik} +\frac{c^2 \beta^i\beta^k}{1-c^2\pmb{\beta}^2}.
\end{equation}

Using the general transformation rules \eqref{deleps}, \eqref{delq} and \eqref{delptau} together with \eqref{carmhstr} and \eqref{deluRP} 
we find (${\cal w}$ is the relativistic enthalpy \eqref{renth}):\footnote{Notice in passing $q_\mu \delta u^\mu= c^2\frac{q^i\delta \beta_i }{\sqrt{1-c^2\pmb{\beta}^2}}$.}
\begin{eqnarray}
\label{delepsRP}
\delta \varepsilon&=& -2 \frac{q^i\delta \beta_i }{\sqrt{1-c^2\pmb{\beta}^2}},
\\
\label{delqRP}
\delta q^i&=& \frac{c^2\delta \beta_k }{\sqrt{1-c^2\pmb{\beta}^2}}\left(\frac{q^k\beta^i}{\sqrt{1-c^2\pmb{\beta}^2}}
-{\cal w} h^{ki} -
\tau^{ki} 
\right),
\\
\nonumber
\delta \left(p h^{ij}+\tau^{ij}\right)&=&  \frac{c^2\delta \beta_k }{1-c^2\pmb{\beta}^2}
\left(\beta^i
\left(p h^{jk}+\tau^{jk}\right)
+\beta^j
\left(p h^{ik}+\tau^{ik}\right)
\right)
\\
\label{deltau2RP}
&&- \frac{\delta \beta_k }{\sqrt{1-c^2\pmb{\beta}^2}}\left(q^i h^{jk}+q^j h^{ik}
\right).
\end{eqnarray}
Similarly, using Eq. \eqref{delj}, we obtain:
\begin{equation}
\label{delrho0RP}
\delta \varrho_0= -\frac{j^i\delta \beta_i }{\sqrt{1-c^2\pmb{\beta}^2}},
\end{equation}
and 
\begin{equation}
\label{deljRP}
\delta j^i= \frac{c^2\delta \beta_k }{\sqrt{1-c^2\pmb{\beta}^2}}\left(\frac{ j^k\beta^i}{\sqrt{1-c^2\pmb{\beta}^2}}
-
\varrho_0 h^{ki} 
\right).
\end{equation}

The energy--momentum tensor is by definition invariant under hydrodynamic-frame transformations. This invariant can be nicely tamed in three canonical objects, which are the energy density $\varepsilon_{\text{r}}$, the heat current $q^i_{\text{r}}$ and the stress tensor $\tau^{ij}_{\text{r}}$, in the fluid proper hydrodynamic frame:\footnote{In Sec. \ref{carfluids} this  frame is referred to as ``fiducial'' and is associated with 
 an observer moving at velocity $\text{u}_{\text{PR}}$ given in \eqref{fidRP}.} 
\begin{equation}
\label{restTinvRP}
T_{00}=\varepsilon_{\text{r}} \Omega^2,
\quad
T_{0}^{\hphantom{0}i}=-\frac{\Omega}{c}q^i_{\text{r}},
\quad
T^{ij}=p_{\text{r}} a^{ij}+\tau^{ij}_{\text{r}}
\end{equation}
with trace
\begin{equation}
\label{restraceinvRP}
T_\mu^{\hphantom{\mu}\mu}=-\varepsilon_{\text{r}} +d p_{\text{r}} + a_{ij} \tau^{ij}_{\text{r}}
,
\end{equation}
and 
\begin{equation}
\label{restJinvRP}
J_{0}=-c\Omega \varrho_{0\text{r}},
\quad
J^{i}=j^{i}_{\text{r}}.
\end{equation}
It is easy to relate these \emph{invariants} to the fluid data in an arbitrary frame encoded in $\beta^i$. We find
\begin{eqnarray}
\label{invenRP}
\varepsilon_{\text{r}}&=&\frac{\varepsilon}{1-c^2\pmb{\beta}^2}+ \frac{2\beta_iq^i}{\sqrt{1-c^2\pmb{\beta}^2}}+c^2 \beta_i\beta_j\left(p h^{ij}+\tau^{ij}\right),\\
\label{invqRP}
q^i_{\text{r}}&=&\frac{c^2\varepsilon \beta^i}{1-c^2\pmb{\beta}^2}
+ \frac{q^j}{\sqrt{1-c^2\pmb{\beta}^2}}\left(\delta_j^i+c^2\beta^i\beta_j\right)+c^2 \beta_j\left(p h^{ij}+\tau^{ij}\right),\\
\label{invtauRP}
p_{\text{r}} a^{ij}+\tau^{ij}_{\text{r}}&=&\frac{c^2\varepsilon \beta^i\beta^j}{1-c^2\pmb{\beta}^2}
+ 
 \frac{\beta^iq^j+\beta^jq^i}{\sqrt{1-c^2\pmb{\beta}^2}}
 +p h^{ij}+\tau^{ij}
,
\end{eqnarray}
and similarly 
\begin{eqnarray}
\label{invrhoRP}
\varrho_{0\text{r}}&=&\frac{\varrho_0}{\sqrt{1-c^2\pmb{\beta}^2}}+ \beta_ij^i,\\
\label{invjRP}
j^{i}_{\text{r}}&=&j^{i}
+ \frac{c^2 \varrho_0\beta^i}{\sqrt{1-c^2\pmb{\beta}^2}}
.
\end{eqnarray}

\subsubsection*{Killings and conserved currents}

Consider a Killing field of the generic form \eqref{genkil}, satisfying \eqref{relkill} on $\mathscr{M}$  in Papapetrou--Randers coordinates,
where the convenient frame and coframe are now\footnote{Later on $ \text{e}_{\hat\imath}$ will be alternatively displayed as $\hat\partial_i$.}
\begin{equation}   
\label{RPframe}
\text{e}_{\hat t}=\frac{1}{\Omega} \partial_t, \quad \text{e}_{\hat\imath}=\partial_i+\frac{b_i}{\Omega}\partial_t,
\quad
\uptheta^{\hat t}=\Omega\text{d}t-b_i \text{d}x^i, \quad \uptheta^{\hat\imath}=\text{d}x^i,
 \end{equation}
so that the metric \eqref{carrp} becomes \eqref{orthonmet}.
The Killing components are 
 \begin{equation}   
  \xi^{\hat t} = \xi^t \Omega-\xi^ib_i, \quad \xi^{\hat\imath} = \xi^i, \quad \xi_{\hat t} = -c^2  \xi^{\hat t}, \quad \xi_{\hat\imath} =a_{ij}\xi^{\hat\jmath} =  \xi_i+ \xi_{\hat t}  b_i,
\label{xicarhat}
 \end{equation}
and those of the conserved current \eqref{relconcur}, following the familiar decomposition procedure in a proper frame:
\begin{equation}
\label{restIinvcar}
I_{0}=-c\Omega \iota_{0\text{r}},
\quad
I^{i}=i_{\text{r}}^i,
\end{equation}
where
\begin{eqnarray}
\label{inviotacar}
\iota_{0\text{r}}&=& \frac{1}{c^2}\xi_{\hat\imath} q_{\text{r}}^i-\xi^{\hat t}  \varepsilon_{\text{r}}
,\\
\label{invicar}
i_{\text{r}}^i&=&\xi_{\hat\jmath}\left(p_{\text{r}} a^{ij}+\tau_{\text{r}}^{ij}\right)-
 \xi^{\hat t} q_{\text{r}}^i
.
\end{eqnarray}
Using \eqref{relconch}, one can  express the conserved charge in the Papapetrou--Randers frame as follows:
\begin{equation}
Q_I=\int_{\upSigma_d}\sqrt{a}\iota_{0\text{r}}
\text{d}x^1\wedge\ldots\wedge\text{d}x^d
-\int_{\upSigma_d}\sqrt{a}\sum_{i=1}^d \text{d}x^1\wedge\ldots\wedge i_{\text{r}}^i\left(\Omega \text{d}t-b_j  \text{d}x^j\right)\wedge\ldots\wedge\text{d}x^d,
\label{relconchPR}
\end{equation}
where  in the exterior product of the second term, $ i_{\text{r}}^i\left(\Omega \text{d}t-b_j  \text{d}x^j\right)$ is the $i$th factor.

\section{Galilean fluid dynamics}

\subsection{Newton--Cartan manifolds  and general Galilean covariance} \label{FLUIDS}

\subsubsection*{Newton--Cartan in a nutshell}

The natural geometric framework for describing non-relativistic fluids is torsionless Newton--Cartan -- see e.g. 
\cite{NC, Duv, Bekaert:2014bwa, Bekaert:2015xua,Hartong:2015xda, Festu, Figueroa-OFarrill:2018ilb,Morand:2018tke} for a comprehensive presentation and further reading suggestions. Newton--Cartan manifolds are members of a wide web including Bargmann spaces or Leibnizian structures (see \cite{Duval:2014uoa, Bekaert:2014bwa, Bekaert:2015xua}).
We will here consider a manifold $\mathscr{M}= \mathbb{R} \times \mathscr{S}$ equipped with coordinates $(t,\mathbf{x})$ and a degenerate cometric\footnote{We systematically omit the tensor product symbol $\otimes$ in the metric and in the cometric.}
\begin{equation}   
\label{dmetinv}
\partial_a^2=a^{ij}\, \partial_i  \partial_j,\quad i,j\ldots \in \{1,\ldots,d\},
\end{equation}
as well as a \emph{clock form} 
\begin{equation}
\label{clock}
\uptheta^{\hat t}=\Omega \text{d}t.
\end{equation}
The dual vector of the latter, referred to as a \emph{field of observers}, is
\begin{equation}   
\label{galet}
\text{e}_{\hat t}=\frac{1}{\Omega} \left(\partial_t +w^j\partial_j\right).
\end{equation}
Here, $a^{ij}$ and $w^i$ are general functions of $(t,\mathbf{x})$ whereas $\Omega = \Omega(t)$. This last feature makes the clock form $\uptheta^{\hat t}$ in \eqref{clock} \emph{exact} and this qualifies for the torsionless nature of the Newton--Cartan manifold. As pointed out in Sec. \ref{prrfi}, this guarantees the existence of an absolute time $\int \text{d}t\, \Omega(t) =\int \text{d}t^\prime\, \Omega^\prime(t^\prime) $, invariant under Galilean diffeomorphisms \eqref{galdifs} -- the only allowed now. For completeness, one should emphasize that even in general, torsionfull Newton--Cartan spacetimes the time interval is invariant but depends on the location $\mathbf{x}$ of the clock: $\int \text{d}t\, \Omega(t, \mathbf{x})$. One still -- abusively -- call it absolute as a way to stress that the differences are due to the location of the clock and not directly to its motion, if any. Motion affects directly the measurements of spatial distances.

The submanifold $\mathscr{S}$ plays the role of $d$-dimensional Newtonian space, endowed with a positive-definite metric, inverse of  $a^{ij}$
\begin{equation}
\label{dmet}
\text{d}\ell^2=a_{ij}(t, \mathbf{x}) \text{d}x^i \text{d}x^j,
\end{equation}
and observed from a frame with respect to which
the locally inertial frame has velocity $\mathbf{w}=w^i \partial_i$ (see footnote \ref{LIF}). A moving particle or a fluid cell will have velocity $\mathbf{v}=v^i \partial_i$ with $v^i=\frac{ \text{d}x^i}{ \text{d}t}$. Under Galilean diffeomorphisms \eqref{galdifs} with Jacobian \eqref{galj}, the transformation rules are as in \eqref{galdifawom}, \eqref{galdifv}, and 
\begin{eqnarray}
\label{galdelt}
\partial^\prime_t&=&\frac{1}{J}\left(\partial_t-j^kJ^{-1i}_{\hphantom{-1}k}\partial_i\right),\\
\label{galdelj}
\partial^\prime_j&=&J^{-1i}_{\hphantom{-1}j}\partial_i.
\end{eqnarray}
The clock form and the field of observers remain invariant:
\begin{equation}
\label{galdelthat}
\uptheta^{\hat t\prime}=\uptheta^{\hat t},
\quad
\text{e}_{\hat t}^\prime=\text{e}_{\hat t}.
\end{equation}

Galilean tensors carry only spatial indices $i, j, \ldots \in \{1, \ldots, d \}$, which are lowered and raised with $a_{ij}$ and $a^{ij}$. They transform covariantly under Galilean diffeomorphisms \eqref{galdifs} with Jacobian $J_i^j$ and $J^{-1i}_{\hphantom{-1}j}$ defined in \eqref{galj}.\footnote{For a vector e.g. the transformation is $V^{\prime k}=J^k_i V^i$.} Tensors depend generically on time $t$ and space $\mathbf{x}$.
Connections can be defined on the geometries at hand, which lead to space and time derivatives, covariant with respect to Galilean diffeomorphisms. These are not uniquely defined (see the literature quoted above) as are torsion-free and metric-compatible connections in Riemannian geometries. We will here make the specific choice, which naturally emerges when the present geometry is reached as an infinite-$c$ limit of a pseudo-Riemannian manifold in the Zermelo frame \eqref{galzerm} (see App. A.1 of Ref. \cite{CMPPS1}). This choice   
makes a sharp separation between space materialized in $\mathscr{S}$ and time. Our spatial connection is 
\begin{equation}
\label{dgamma}
\gamma^i_{jk}=\frac{a^{il}}{2}\left(\partial_j a_{lk}+\partial_k a_{lj}-\partial_l a_{jk}\right).
\end{equation}
The associated covariant derivative is spelled $\hat\nabla_i$, as opposed to  $\nabla_i$, the spatial component of the Levi--Civita covariant derivative $\nabla_\mu$ defined on the ascendent pseudo-Riemannian spacetime.\footnote{In \cite{CMPPS1} the hat was not used in the Galilean covariant derivative, and this might have caused confusion.} This connection is torsionless
\begin{equation}
\label{carcontorsion}
\hat t^k_{\hphantom{k}ij}=2\gamma^k_{[ij]}=0,
\end{equation}
and metric-compatible
\begin{equation}
\label{carconmet}
\hat\nabla_ia_{jk}=0.
\end{equation}
Its Riemann, Ricci and scalar curvature tensors are defined as usual $d$-dimensional Levi--Civita curvature tensors would be on $\mathscr{S}$, except that they are $t$-dependent:
\begin{equation}
\left[\hat\nabla_k,\hat\nabla_l\right]V^i=\left(
\partial_k\gamma^i_{lj}
-\partial_l\gamma^i_{kj}
+\gamma^i_{km}\gamma^m_{lj}
-\gamma^i_{lm}\gamma^m_{kj}
\right)V^j
= \hat r^i_{\hphantom{i}jkl}V^j.
\label{galriemann}
\end{equation}

It is worth stressing that Galilean tensors can be constructed from an object which is not a vector but rather transforming like a connection, 
\begin{equation}
\label{gal-trans1}
A^{\prime k}=\frac{1}{J}\left(J^k_i A^i+j^k\right).
\end{equation}
Indeed 
\begin{equation}
\frac{1}{\Omega}\hat \nabla^{(k}A^{l)}-\frac{1}{2\Omega}\partial_ta^{kl}
=- \frac{1}{2\Omega}\left(\mathscr{L}_{\mathbf{A}}a^{kl}
+
 \partial_t a^{kl}
\right)
\end{equation}
($\mathscr{L}_{\mathbf{A}}$ is the Lie derivative along $\mathbf{A}=A^i\partial_i$) and
\begin{equation}
\frac{1}{\Omega}\hat \nabla_{(k}A_{l)}+\frac{1}{2\Omega}\partial_ta_{kl} 
=\frac{1}{2\Omega}\left(\mathscr{L}_{\mathbf{A}}a_{kl}
+ \partial_t a_{kl}
\right)
\end{equation}
have tensorial transformation rules, and their trace is a scalar.\footnote{Observe that neither $\frac{1}{\Omega}\partial_t $ nor  $\frac{1}{\Omega}\mathscr{L}_{\mathbf{A}}$ acting on Galilean tensors give separately tensors because of  \eqref{galdelt} and $A^i$ transforming as \eqref{gal-trans1}.} We can apply this to $\mathbf{w}$ or $\mathbf{v}$ (see \eqref{galdifawom} and \eqref{galdifv}) and define
\begin{equation}
\hat\gamma^{w}_{\hphantom{w}ij}=\frac{1}{\Omega}\left(\hat \nabla_{(i}w_{j)}+\frac{1}{2}\partial_ta_{ij} \right),
\quad 
\hat\gamma^{v}_{\hphantom{v}ij}=\frac{1}{\Omega}\left(\hat \nabla_{(i}v_{j)}+\frac{1}{2}\partial_ta_{ij} \right),
\end{equation}
where the upper indices refer to the vectors $\mathbf{w}$ and $\mathbf{v}$, corresponding to the geometry and fluid respectively.
The former is purely geometrical (and emerges in the large-$c$ expansion of the relativistic-spacetime Levi--Civita connection in Zermelo frame\footnote{As a general comment, valid both in the present chapter on Galilean dynamics as well as in the next on Carrollian, the $c$-dependence of our relativistic metrics is always explicit and in line with  the Galilean (or Carrollian) reduction. Hence, every term in the power expansions is Galilean-covariant (or Carrollian-covariant).}); the latter is associated with a fluid of velocity $v^i$. They coincide for a fluid at rest in the locally inertial frame, i.e. for $v^i = w^i$. From these tensors, one defines their traceless relatives and the traces: 
the \emph{geometric Galilean shear} 
\begin{equation}
\label{galwshear} 
\xi^{w}_{\hphantom{w}ij}=\frac{1}{\Omega}\left(\hat \nabla_{(i}w_{j)}+\frac{1}{2}\partial_ta_{ij} \right)
-\frac{1}{d} a_{ij}\theta^{w},
\end{equation}
and the \emph{geometric Galilean expansion} 
\begin{equation}
\label{galwexp} 
\theta^{w} = \frac{1}{\Omega}\left(
\partial_t \ln \sqrt{a}+\hat \nabla_i w^i\right),
\end{equation}
as well as the \emph{fluid Galilean shear} 
\begin{equation}
\label{galshear} 
\xi^{v}_{\hphantom{v}ij}=\frac{1}{\Omega}\left(\hat \nabla_{(i}v_{j)}+\frac{1}{2}\partial_ta_{ij} \right)
-\frac{1}{d} a_{ij}\theta^{v},
\end{equation}
and the  \emph{fluid Galilean expansion}
\begin{equation}
\label{galexp} 
\theta^{v} = \frac{1}{\Omega}\left(
\partial_t \ln \sqrt{a}+\hat \nabla_i v^i\right).
\end{equation}

One similarly defines a time, metric-compatible covariant derivative (again emerging in the Galilean expansion of the spacetime Levi--Civita covariant derivative in the time direction of a Zermelo frame). For a scalar function $\Phi$ it is simply
\begin{equation}
\frac{1}{\Omega}\frac{\hat{\text{D}}\Phi}{\text{d}t}=\text{e}_{\hat t}(\Phi)
= \frac{1}{\Omega} \partial_t\Phi+\frac{w^{ j}}{\Omega}\partial_j\Phi 
,
\label{galfcovderfdhatPhi}
\end{equation}
whereas for vectors one finds
\begin{eqnarray}
\frac{1}{\Omega}\frac{\hat{\text{D}}V^i}{\text{d}t}
&=&\frac{1}{\Omega}\partial_t V^i+\frac{w^{j}}{\Omega}\partial_j V^i  -V^{j}\partial_j \frac{w^{i}}{\Omega} 
+\hat\gamma^{wi}_{\hphantom{wi}j}V^{j}
\nonumber\\
&=&\frac{1}{\Omega}\left(\partial_t V^i+ \mathscr{L}_{\mathbf{w}}V^i\right)+\hat\gamma^{wi}_{\hphantom{wi}j}V^{j}.
\label{galfcovdervdhatVup}
\end{eqnarray}
More generally, the Leibniz rule leads to 
\begin{equation}
\frac{1}{\Omega}\frac{\hat{\text{D}}K^{i\ldots}_{\hphantom{i\ldots}j\ldots}}{\text{d}t}=
\frac{1}{\Omega}\left(\partial_t K^{i\ldots}_{\hphantom{i\ldots}j\ldots}+ \mathscr{L}_{\mathbf{w}}K^{i\ldots}_{\hphantom{i\ldots}j\ldots}\right)
+\hat\gamma^{wi}_{\hphantom{wi}k}K^{k\ldots}_{\hphantom{k\ldots}j\ldots}+\cdots
-\hat\gamma^{wk}_{\hphantom{wk}j}K^{i\ldots}_{\hphantom{i\ldots}k\ldots}-\cdots,
\label{galfcovderdhat}
\end{equation}
and as a consequence
\begin{equation}
\frac{1}{\Omega}\frac{\hat{\text{D}}a^{ij}}{\text{d}t}=
\frac{1}{\Omega}\frac{\hat{\text{D}}a_{ij}}{\text{d}t}=0.
\end{equation}

In the presence of a fluid one can also introduce the more physical  \emph{material derivative}
\begin{equation}
\label{galfder} 
\frac{\text{d}}{\text{d}t}= \partial_t +v^i\hat \nabla_i,
\end{equation}
which produces a scalar density (or a scalar upon division by $\Omega$) when acting on a scalar function. When acting on arbitrary tensors, it should be supplemented with appropriate $\mathbf{w}$ and/or $\mathbf{v}$ terms in order to maintain the tensorial transformation properties. Several options exist and we here quote the most physical (see \cite{CMPPS1}):\footnote{For a detailed and general presentation of Galilean affine connections see
\cite{Bekaert:2014bwa, Bekaert:2015xua}.}
\begin{equation}
\label{galfcovder-f} 
\frac{1}{\Omega}\frac{\text{D}V^i}{\text{d}t}= 
\frac{1}{\Omega}\frac{\text{d}V^i}{\text{d}t}-\frac{1}{\Omega}V^j \hat \nabla_j w^i
,\quad \frac{1}{\Omega}\frac{\text{D}V_i}{\text{d}t}=\frac{1}{\Omega}\frac{\text{d}V_i}{\text{d}t}+\frac{1}{\Omega}V_j\hat \nabla_i w^j,
\end{equation}
resulting in genuine tensors under Galilean diffeomorphisms. As opposed to \eqref{galfcovderdhat}, this time-covariant derivative is not metric compatible:
\begin{equation}
\label{galfcovdermet} 
\frac{1}{\Omega}\frac{\text{D}a_{ij}}{\text{d}t}
=2\hat\gamma^{w}_{\hphantom{w}ij}.
\end{equation}

Space and time Galilean covariant derivatives do not commute. They define a Galilean tensor, rooted in the Riemann tensor of the ascendent relativistic spacetime at finite velocity of light. We find  
\begin{eqnarray}
\left[\frac{1}{\Omega}\frac{\hat{\text{D}}}{\text{d}t},\hat\nabla_i\right]\Phi&=& -\hat\gamma^{wk}_{\hphantom{wk}i}\partial_k \Phi,
\\
\left[\frac{1}{\Omega}\frac{\hat{\text{D}}}{\text{d}t},\hat\nabla_i\right]V^j&=& -\hat\gamma^{wk}_{\hphantom{wk}i}
\hat\nabla_kV^j +\hat r ^j_{\hphantom{j}ik} V^k,
\end{eqnarray}
and similarly for higher-rank Galilean tensors, where
\begin{equation}
\hat r ^j_{\hphantom{j}ik} =\frac{1}{\Omega}
\left( \partial_t \gamma^j_{ik}+\hat\nabla_i \hat\nabla_k w^j
-\hat\nabla_i \hat\gamma^{wj}_{\hphantom{wj}k}
+w^l \hat r ^j_{\hphantom{j}kli}
\right).
\end{equation}

\subsubsection*{Galilean diffeomorphisms and conservation equations}  

Without referring specifically to a fluid, one may consider an effective action describing the dynamics of a system defined on the geometry $\mathscr{M}= \mathbb{R}\times \mathscr{S}$ discussed previously. This effective action is thus a functional of $a^{ij}$, $\Omega$ and $w^i$: $S=\int \text{d}t\,  \text{d}^{d}x \sqrt{a}\Omega\mathcal{L}$. The standard relativistic energy--momentum tensor \eqref{varrelT} is now traded for the following \emph{Galilean momenta}, namely the \emph{energy--stress tensor}, the \emph{momentum} and the \emph{energy density}:
\begin{eqnarray}
\label{galmompiij}
\Pi_{ij}&=&-\frac{2}{\sqrt{a} \Omega}\frac{\delta S}{\delta a^{ij}},\\
\label{galmompii}
P_{i}&=&-\frac{1}{\sqrt{a} \Omega}\frac{\delta S}{\delta \frac{w^i}{\Omega}},\\
\label{galmompi}
\Pi&=&-\frac{1}{\sqrt{a}\Omega}\left(\Omega\frac{\delta S}{\delta \Omega}-\frac{w^i}{\Omega}\frac{\delta S}{\delta \frac{w^i}{\Omega}}\right),
 \end{eqnarray}
which can likewise be combined as $\frac{\delta S}{\delta \Omega}=-\sqrt{a}\left(\Pi+\frac{w^i}{\Omega}P_i\right)$. These momenta are summarized in the following variation -- at least for the gravitational sector:
\begin{equation}
\label{delSgravzer}
\delta S=-\int\text{d}t \, \Omega\int \text{d}^dx \sqrt{a}  \left(\frac{1}{2}\Pi_{ij}\delta a^{ij} + P_i \delta\frac{w^i}{\Omega}
+\left(\Pi+ \frac{w^i}{\Omega}P_i\right)\delta \ln \Omega\right).
\end{equation}

The above momenta obey conservation equations as a consequence of the assumed invariance of the action under Galilean diffeomorphisms, which simultaneously guarantees their Galilean-covariant transformation rules. Galilean diffeomorphisms \eqref{galdifs} are generated by vector fields on $\mathscr{M}$ whose time component depends only on $t$: 
\begin{equation}   
\label{galkill}
\upxi=\xi^t\partial_t +\xi^i \partial_i=
\xi^{\hat t}\text{e}_{\hat t}+ \xi^{\hat\imath }\partial_i 
 \end{equation}
(this is the same expression as \eqref{genkil}), where  $\xi^{\hat t}(t) = \xi^{t}(t)\Omega(t)$ is a Galilean scalar, and $ \xi^{\hat\imath }(t,\mathbf{x})=\xi^i - \xi^t w^i$ are the components of a Galilean vector. The variation under diffeomorphisms is implemented through the Lie derivative (the minus sign is conventional):
 \begin{equation}   
-\delta_\upxi a^{ij} =\mathscr{L}_\upxi a^{ij}=-2\left(\hat  \nabla^{(i}\xi^{\hat \jmath)}+\hat\gamma^{wij}\xi^{\hat t} +\frac{1}{\Omega} w^{(i} a^{j)k}\partial_k \xi^{\hat t} \right),
\label{Liedaijgal}
 \end{equation}
where the last term drops for Galilean diffeomorphisms.
 Furthermore 
\begin{equation}   
\label{Liedfieldgal}
\mathscr{L}_\upxi \text{e}_{\hat t}= -\frac{1}{\Omega} \left(\partial_t\xi^{\hat t}+ \mathscr{L}_{\mathbf{w}}\xi^{\hat t}\right)
\text{e}_{\hat t} -\frac{1}{\Omega} \left(\partial_t \xi^{\hat \imath}+ \mathscr{L}_{\mathbf{w}} \xi^{\hat \imath}\right) \partial_i,
 \end{equation}
 from which, using \eqref{galet}, we infer
 \begin{equation}   
-\delta_\upxi \Omega=\mathscr{L}_\upxi \Omega =\partial_t\xi^{\hat t}+ \mathscr{L}_{\mathbf{w}}\xi^{\hat t}, \quad
\delta_\upxi w^i=-\mathscr{L}_\upxi w^i
= \partial_t \xi^{\hat \imath}+ \mathscr{L}_{\mathbf{w}} \xi^{\hat \imath}. 
\label{Liedwiomgal}
\end{equation}
Notice also the action on the clock form:
\begin{equation}   
\label{Liedclockgal}
\mathscr{L}_\upxi \uptheta^{\hat t}= \frac{1}{\Omega} \left(\partial_t\xi^{\hat t}+ \mathscr{L}_{\mathbf{w}}\xi^{\hat t}\right)
\uptheta^{\hat t}=\frac{1}{\Omega}\frac{\hat{\text{D}}\xi^{\hat t}}{\text{d}t}\uptheta^{\hat t} =\mu \uptheta^{\hat t},
 \end{equation}
where we introduced
\begin{equation}
\label{galmu}
\mu(t,\mathbf{x})= \frac{1}{\Omega}\frac{\hat{\text{D}}\xi^{\hat t}}{\text{d}t}
 \end{equation}
not to be confused with the chemical potential introduced in thermodynamics.
% $\delta_\upxi S$ \eqref{delSgal} computed on-shell 

We can now determine the variation of the action under Galilean diffeomorphisms:
 \begin{eqnarray}
 \delta_\upxi S
  &=&\int \text{d}t \text{d}^dx  \sqrt{a}\Omega 
\left\{-\xi^{\hat t}\left[\frac{1}{\Omega}\frac{\hat{\text{D}}\Pi}{\text{d}t}+
\theta^w \Pi+
 \Pi_{ij} \hat\gamma^{wij} \right]
\right.
\nonumber
\\
&&+\left.\xi^{\hat \imath}\left[\frac{1}{\Omega}\frac{\hat{\text{D}}P_i}{\text{d}t}
+\theta^w P_i
+P_j
 \hat\gamma^{wj}_{\hphantom{wj}i}
+
\hat \nabla^j \Pi_{ij} 
\right]\right\} 
\nonumber\\ 
&&
+\int \text{d}t \text{d}^dx  
 \left\{\partial_t\left(\sqrt{a} \left(\Pi \xi^{\hat t} -P_j \xi^{\hat \jmath}  \right)\right) \right.
\nonumber\\
&&+\partial_i \left(\sqrt{a} w^i\left(\Pi \xi^{\hat t} -P_j \xi^{\hat \jmath} 
\right)\left.-\sqrt{a} \Omega\Pi^{i}_{\hphantom{i}j} \xi^{\hat \jmath} \right)
 \right\}.
 \label{delSgal}
 \end{eqnarray}
Requiring that $ \delta_\upxi S$ vanishes and ignoring the boundary terms (last two lines in Eq. \eqref{delSgal}), we reach two equations. The momentum equation is the simplest because $\xi^{\hat \jmath} $ being functions of both $t$ and $\mathbf{x}$, their factor must vanish:
 \begin{equation}
\boxed{
\left(\frac{1}{\Omega}\frac{\hat{\text{D}}}{\text{d}t}
+\theta^w \right)P_i
+ P_j
\hat\gamma^{wj}_{\hphantom{wj}i}
+
\hat \nabla^j \Pi_{ij} 
=0.}
 \label{delSgalmomeq} 
 \end{equation}
The energy equation is more subtle because $\xi^{\hat t}$ depends on $t$ only. As a consequence it is enough to require that its factor be the Galilean divergence of a vector:
 \begin{equation}
\boxed{
\left(\frac{1}{\Omega}\frac{\hat{\text{D}}}{\text{d}t}+
\theta^w\right) \Pi+
 \Pi_{ij} \hat\gamma^{wij}
=-\hat \nabla_i \Pi^i ,
} \label{delSgaleneq}
 \end{equation}
where $\Pi^i $ is undetermined a priori. Indeed,  $\sqrt{a}\Omega \xi^{\hat t}\hat\nabla_i \Pi^i = 
\partial_i\left(\sqrt{a}\Omega \xi^{\hat t}\Pi^i \right)$, which leads to a boundary term and vanishes inside the integral. One can interpret $\Pi^i $ as the \emph{energy current} (also \emph{energy flux}). 
  
 \subsubsection*{Gauge invariance and matter conservation}  

Besides Galilean covariance, the action might also be invariant under a local $U(1)$
symmetry,
parameterized by $\Lambda (t, \mathbf{x})$ and acting on the components of a gauge field  $\text{B}=B (t, \mathbf{x}) \text{d}t +B_i (t, \mathbf{x}) \text{d}x^i$ as 
 \begin{equation}
 \label{delB}
 \delta_\Lambda B_i = -\partial_i \Lambda, \quad
\delta_\Lambda B = -\partial_t \Lambda.
 \end{equation}
The conjugate momenta are now  the \emph{matter density} and the \emph{matter current}:
\begin{eqnarray}
\label{galmomchi}
\varrho&=&-\frac{1}{\sqrt{a}}\frac{\delta S}{\delta B},\\
\label{galmomphii}
N^{i}&=&\frac{1}{\Omega\sqrt{a}}
\left(w^i \frac{\delta S}{\delta B}-\frac{\delta S}{\delta B_i}\right)
\end{eqnarray}
with $\frac{\delta S}{\delta B_i}=-\sqrt{a}\left(\Omega N^i +\varrho w^i \right)$, and
 \begin{equation}
 \label{delSgauzer}
 \delta S=-
 \int \text{d}t \text{d}^{d}x \sqrt{a} \left(
\varrho  \delta B+
\left(\Omega
N^i+\varrho w^i 
\right) \delta B_i \right) 
 \end{equation}
 for the matter sector. The gauge variation of the action reads:
 \begin{eqnarray}
 \delta_\Lambda S&=&
 \int \text{d}t \text{d}^{d}x \sqrt{a} \left( 
\varrho  \partial_t \Lambda+
\left(\Omega
N^i+\varrho w^i 
\right) \partial_i \Lambda\right) 
\nonumber 
 \\ &=&-\int \text{d}t \text{d}^dx  \sqrt{a}\Omega 
\Lambda\left(\frac{1}{\Omega}\frac{\hat{\text{D}}\varrho}{\text{d}t}+
\theta^w \varrho+\hat \nabla_i N^i
\right)\nonumber\\
&&+\int \text{d}t \text{d}^dx  
 \left\{\partial_t\left(\sqrt{a} \Lambda \varrho \right) +\partial_i \left(\sqrt{a}\Lambda\left( \Omega N^i+\varrho w^i \right)\right)
 \right\}.
 \label{U1gal}
 \end{eqnarray}
Invariance of $S$ leads to the Galilean continuity equation:
 \begin{equation}
\boxed{\left(
\frac{1}{\Omega}\frac{\hat{\text{D}}}{\text{d}t}+
\theta^w \right)\varrho
+\hat \nabla_i N^i=0 .}
  \label{delSgalcon}
 \end{equation}

The continuity equation can be alternatively presented in an integral form, using Stokes and Gauss theorems:\footnote{Stokes theorem is valid irrespective of the metric. Gauss' requires a dual exterior derivative $\text{d}^\dagger$, which can be introduced consistently despite the cometric being degenerate. We will not elaborate on this matter here. For the Carrollian case, this was discussed for $d=1$ in Ref. \cite{CMPR}. \label{ddagger}}
\begin{eqnarray}
\int_{\mathscr{W}} \text{d}t\text{d}^dx \Omega\sqrt{a}\left(\left(
\frac{1}{\Omega}\frac{\hat{\text{D}}}{\text{d}t}+
\theta^w \right)\varrho
+\hat \nabla_i N^i\right)= \oint_{\partial\mathscr{W}}
\sqrt{a}\varrho
\left(\text{d}x^1-w^1 \text{d}t\right)\wedge\ldots\wedge\left(\text{d}x^d-w^d \text{d}t\right)
\nonumber
\\
- \oint_{\partial\mathscr{W}} \sqrt{a}\sum_{i=1}^d \left(\text{d}x^1-w^1 \text{d}t\right)\wedge\ldots\wedge N^i\uptheta^{\hat t}\wedge\ldots\wedge\left(\text{d}x^d-w^d \text{d}t\right),
\label{gaustok-NC}
\end{eqnarray}
where $\mathscr{W}\subset \mathscr{M}= \mathbb{R} \times \mathscr{S}$ and $ N^i\uptheta^{\hat t}$ is the $i$th factor in the exterior product of the last term ($\uptheta^{\hat t}$ is the clock form given in \eqref{clock}).  From this we obtain a conserved charge -- under the usual assumptions for the behaviour of the fields -- expressed as an integral over an arbitrary hypersurface $\upSigma_d$ of $\mathscr{M}= \mathbb{R} \times \mathscr{S}$. It coincides with the relativistic Zermelo result captured e.g. in \eqref{relconch-Z}.

Although not compulsory, it is convenient to chose $\upSigma_d\equiv  \mathscr{S}$ i.e. a constant-$t$ hypersurface. We then find
 \begin{equation}
Q_N=\int_{\mathscr{S}}\text{d}^{d}x \sqrt{a}\varrho,
  \label{galconch}
 \end{equation}
which fits the usual definition of charge in Galilean physics. In this case, the conservation of $Q_N$ is often phrased as independence with respect to ordinary time $t$, although it is actually a stronger statement, even in Newton--Cartan spacetimes, where there is a privileged time direction (for the relativistic case, see the comment in footnote \ref{comconservation}). Time-independence appears explicitly if one trades $\mathscr{S}$ in the integral 
\eqref{galconch} with
$\mathscr{V}\subset \mathscr{S}$. Assuming for simplicity that the boundary $\partial\mathscr{V}$ of that domain does not depend on $t$ and using \eqref{delSgalcon}, the time evolution of the matter/charge content of  $\mathscr{V}$ is 
  \begin{equation}
\frac{1}{\Omega}\frac{\text{d}}{\text{d}t}\int_{\mathscr{V}}\text{d}^{d}x \sqrt{a}\varrho=-\int_{\mathscr{V}}\text{d}^{d}x \, 
\partial_i\left(\sqrt{a} \left( N^i+\varrho \frac{w^i}{\Omega} \right)\right)
=-\int_{\partial\mathscr{V}}\star\left( \mathbf{N}+\varrho \frac{\mathbf{w}}{\Omega} \right), 
  \label{galcontime}
 \end{equation}
where $\star$ stands for the $d$-dimensional $\mathscr{S}$-Hodge dual based on $\sqrt{a}$ and on the antisymmetric symbol
 $\epsilon_{i_1 \ldots i_d}$ with $\epsilon_{1\ldots d}=1$. If the integral is performed over the entire $\mathscr{S}$ it vanishes (assuming a reasonable asymptotic behaviour), and 
$Q_N$ in
  \eqref{galconch}
is conserved.  
 
Equation \eqref{delSgalcon}   and its variables are a priori independent of the energy--momentum equations \eqref{delSgalmomeq},  \eqref{delSgaleneq} and their variables. As we will see, thermodynamics sets a relationship among the momentum $P_i$ and the current $N_i$.

 \subsubsection*{Isometries, conservation and non-conservation laws}

In (pseudo-)Riemannian geometry, isometries are diffeomorphisms generated by vectors leaving the metric invariant, i.e. requiring \eqref{relkill}.\footnote{For historical reasons, some authors use the name ``Killing fields'' for generators of isometries in a (pseudo-)Riemannian manifold exclusively. We take the freedom here to call every isometry generator a Killing field, be it for weak or strong, Newton--Cartan or Carrollian manifolds.} Newton--Cartan spacetimes may also have isometries. Due to the degenerate cometric, however, their definition and the determination of the Killing fields are more subtle. This subject has been abundantly discussed in the literature. We will summarize the features important for our purpose regarding the Galilean fluid dynamics.

The Killing fields are of the Galilean type \eqref{galkill} and are required to obey  
\begin{equation}
\label{NCkill}
\mathscr{L}_\upxi a^{ij}=0, \quad\mathscr{L}_\upxi \uptheta^{\hat t}=0,
\end{equation}
since the fundamental geometric data are the cometric \eqref{dmetinv} and the clock form \eqref{clock}.
Using expressions \eqref{Liedaijgal} and  \eqref{Liedclockgal} for Galilean diffeomorphisms ($\xi^{\hat t}$ is only $t$-dependent), we obtain the Galilean Killing equations: 
\begin{equation}   
\label{galkilleq}
\boxed{\hat  \nabla^{(i}\xi^{\hat \jmath)}
+\hat\gamma^{wij}\xi^{\hat t}  =0
,\quad
\frac{1}{\Omega}\frac{\hat{\text{D}}\xi^{\hat t}}{\text{d}t}=0.}
 \end{equation}
These equations generally admit an infinite number of solutions. The reason is that they refer to the \emph{weak definition} of Newton--Cartan spacetimes \cite{Duval:2014lpa} in terms of the cometric and the clock form, and express exclusively the invariance of these data. A \emph{strong definition} exists and requires additionally a symmetric affine connection, which is metric-compatible and parallel-transports the clock form.\footnote{ It turns out that the connection we have introduced in this chapter obeys these properties.} Isometries are thus restricted to comply with the strong definition, and leave the affine connection invariant. This reduces the set of generators to a finite number \cite{Duval:1993pe}. 

Notice also that the fundamental geometric piece of data for a Newton--Cartan spacetime is genuinely the clock form $\uptheta^{\hat t}$ rather than the field of observers $\text{e}_{\hat t}$. The latter is not required to have a vanishing Lie derivative along Killing vectors, and using \eqref{Liedfieldgal} and \eqref{galkilleq} we indeed find
\begin{equation}   
\label{Liedkilfieldgal}
\mathscr{L}_\upxi \text{e}_{\hat t}= -\frac{1}{\Omega} \left(\partial_t \xi^{\hat \imath}+ \mathscr{L}_{\mathbf{w}} \xi^{\hat \imath}\right) \partial_i,
 \end{equation}
for a generic Killing field $\upxi$.

Consider as an example the Newton--Cartan manifold with $a^{ij}=\delta^{ij}$, $\Omega = 1$ and $w^i$ constant. This is our familiar $\mathbb{R}\times \mathbb{E}_3$ spacetime, which is flat for the connection introduced  earlier. Equations \eqref{galkilleq} possess an infinite number of solutions:
\begin{equation}   
\label{galkillflatsol}
\upxi= \left(\Omega_i^{\hphantom{i}j}(t) x^i +Z^j(t)\right)\partial_j + T \partial_t
 \end{equation}
with $T$ a constant and $\Omega_{ij}= \Omega_i^{\hphantom{i}k}\delta_{kj}$ antisymmetric. Imposing the invariance of the affine connection, one recovers \cite{Duval:1993pe} the $\nicefrac{(d+2)(d+1)}{2}$-dimensional Galilean algebra
$\mathfrak{gal}(d+1)$ with contant $\Omega_{ij}$ generating the $\mathfrak{so}(d)$ rotations,  $Z^j(t)=V^j t+X^j$ for the Galilean boosts and spatial translations, and $T$ for the time translations. We find in particular that 
\begin{equation}   
\label{galkillflatsol-lie}
\mathscr{L}_\upxi \text{e}_{\hat t}= - \left(V^i+ w^k \Omega_k^{\hphantom{k}i}\right) \partial_i \neq 0, 
 \end{equation}
showing among others that the boosts produce a displacement in the field of observers. This is expected because $w^i$ describes the constant  velocity of the original inertial frame, which is shifted during a Galilean boost, and nicely illustrates why it would have been unnatural to impose the invariance of $\text{e}_{\hat t}$ under the action of an isometry. 

Assuming the existence of an isometry, we can now address the conservation law that would take the Galilean form \eqref{delSgalcon} with a Galilean scalar $\kappa$ and a Galilean vector $K^i$ determined from the Killing components 
$\xi^{\hat t}$ and $\xi^{\hat\imath}$, and from the Galilean momenta, i.e. the energy density $\Pi$, the momentum $P_i$ and the energy--stress tensor $\Pi_{ij}$ defined in Eqs. \eqref{galmompiij}, \eqref{galmompii}, \eqref{galmompi}, as well as the energy flux $\Pi_i$, and satisfying the conservation equations \eqref{delSgalmomeq} and \eqref{delSgaleneq}. The  Galilean scalar
 \begin{equation}
\mathcal{K}=\left(\frac{1}{\Omega}\frac{\hat{\text{D}}}{\text{d}t}+
\theta^w \right)\kappa
+\hat \nabla_i K^i
  \label{kilgalcon}
 \end{equation}
would then vanish on-shell. The components of the Galilean current  $\kappa$ and $K^i$ are read off in the boundary terms of  $\delta_\upxi S$ given in \eqref{delSgal} and set on-shell:\footnote{They are in fact  inherited from the relativistic-current components  i.e. as a large-$c$ expansion of \eqref{inviotaZer} and \eqref{inviZer}, and the precise computation is performed at the end of Sec. \ref{comgal}, Eq. \eqref{galkappaK}.}
\begin{eqnarray}
\label{inviotagal}
\kappa&=& \xi^{\hat\imath} P_i-\xi^{\hat t}  \Pi,\\
\label{invigal}
K_i&=&\xi^{\hat\jmath}\Pi_{ij}-
 \xi^{\hat t} \Pi_i
.
\end{eqnarray}
Using the conservation equations \eqref{delSgalmomeq} and \eqref{delSgaleneq} we obtain the following result:
 \begin{eqnarray}
\label{galcon}
\mathcal{K}&=& -\frac{\Pi}{\Omega}\frac{\hat{\text{D}}\xi^{\hat t}}{\text{d}t}+\frac{P_i}{\Omega} \left(\partial_t \xi^{\hat \imath}+ \mathscr{L}_{\mathbf{w}} \xi^{\hat \imath}\right) +\Pi_{ij}\left(\hat  \nabla^{i}\xi^{\hat \jmath}
+\hat\gamma^{wij}\xi^{\hat t}  \right)
\\
\label{galnoncon}
&=&\frac{P_i}{\Omega} \left(\partial_t \xi^{\hat \imath}+ \mathscr{L}_{\mathbf{w}} \xi^{\hat \imath}\right).
\end{eqnarray}
We have used the Killing equations \eqref{galkilleq} to reach \eqref{galnoncon} from \eqref{galcon}, which shows that \emph{in Newton--Cartan spacetimes, a Killing field fails to systematically support an on-shell conservation law for Galilean dynamics}. This could have been anticipated, actually. Indeed, $P_i$ is conjugate to $\nicefrac{w^i}{\Omega}$ (see \eqref{galmompii}) and $w^i$ transforms under diffeomorphisms according to \eqref{Liedwiomgal}, \emph{even when this diffeomorphism is generated by a Killing vector field}. This is precisely what Eq. \eqref{galnoncon} conveys. Still, a conservation law exists for those Killing fields, which happen to satisfy $\mathscr{L}_\upxi \text{e}_{\hat t}\equiv\left[\upxi ,\text{e}_{\hat t}\right]= 0$ -- using Jacobi identity,  one checks that the commutator of two such Killings leaves also $\text{e}_{\hat t}$ invariant. As we mentioned earlier, ordinary Galilean boosts in $\mathbb{R}\times \mathbb{E}_3$ do not, Eq. \eqref{galkillflatsol-lie}. In other instances with multiple degrees of freedom arising from a Laurent  expansion in powers of $c^2$ (see Sec. \ref{comgal}), several currents coexist in the presence of a Killing, and some can be conserved due to the accidental -- as opposed to a priori demanded -- absence of a $P_i$-like component. 

The above result sounds iconoclastic, in view of the robustness  and generality of N\oe ther's theorems. In Ref. \cite{Festu}, for instance, the existence of an isometry-related conserved current in torsional Newton--Cartan spacetimes has not been demonstrated, not even questioned -- it was just assumed to be true.  However, \emph{it is not ipso facto}, for the simple reason that isometries are less restraining in Newton--Cartan spacetimes, so that  only a restricted class of Galilean Killing fields might indeed be eligible for conservation laws -- boosts in flat spacetime aren't. 

One may naturally contemplate that $\upxi $ being a Killing field, the right-hand side  of Eq. \eqref{galnoncon} is  associated with a \textsl{boundary term}. If this happens, the boundary term contributes the current components $\kappa$ and $K^i$, leading to an effective $\kappa'$ and $K^{i\prime}$, truly conserved. However, this does not seem to be the rule, not even in flat space, where $\mathcal{K}=P_i\left(V^i+ w^k \Omega_k^{\hphantom{k}i}\right)$, obtained using  Eqs. \eqref{galkillflatsol} and \eqref{galkillflatsol-lie}.\footnote{For the conservation to occur in flat space, $P_i$ should be a \emph{potential flow} (also called \emph{irrotational}, see \cite{Landau}  \textsection 9) i.e. obey $P_i = \left(\partial_t +w^j\partial_j\right) \phi_i + \partial_i \phi$ for some set of functions $\phi(t,\mathbf{x})$ and $\phi_i(t,\mathbf{x})$ -- scalar and vector potentials. Then   $\mathcal{K}=\left(\partial_t +w^j\partial_j\right) \phi_iW^i + \partial_i \phi W^i$ with $W^i = V^i+ w^k \Omega_k^{\hphantom{k}i}$ and the conservation works out (\eqref{kilgalcon} vanishes) with current components $\kappa -\phi_iW^i $ and $K^i - \phi W^i $. Notice that in contrast to the present framework,  the conservation is generally valid for free-particle motion on Newton--Cartan flat spacetimes, as in that case the momentum is a total derivative (typically $\dot x^i$). Appendix \ref{frmotNC} summarizes this instance and provides the necessary details for making the statement sound. \label{33}}

The same conclusion will be reached for Carrollian spacetimes in Sec. \ref{cardifcons}, and further discussed from the large-$c$ perspective in the upcoming paragraphs, as well as in App. \ref{garbage}.

 \subsubsection*{Weyl invariance, conformal isometries, conservation and non-conservation laws}

Fluids involving massless excitations have observables with  remarkable scaling properties. 
We can introduce Weyl transformations acting as follows on the fundamental geometric data of a Newton--Cartan geometry:
\begin{equation}
\label{weyl-geometry-NC}
a^{ij}\to \mathcal{B}^2 a^{ij},\quad \Omega\to \frac{1}{\mathcal{B}}\Omega,\quad w^{i}\to w^{i},\quad w_{i}\to \frac{1}{\mathcal{B}^2}w_{i}.
\end{equation} 
Since $\Omega$ is a function of $t$ only, the second of \eqref{weyl-geometry-NC} imposes $\mathcal{B}=\mathcal{B}(t)$. 
Weyl-invariance requirement of an effective action $S$ leads to following weights for the Galilean momenta in \eqref{galmompiij}, \eqref{galmompii}, \eqref{galmompi}: the energy--stress tensor $\Pi_{ij}$ has weight $d-1$, the momentum $P_i$, $d$, and the energy density $\Pi$, $d+1$. The energy flux $\Pi_i$ introduced in \eqref{delSgaleneq} has also weight $d$. Furthermore, using \eqref{delSgravzer} with  $\delta_{\mathcal{B}}S=0$ implies that 
\begin{equation}
\label{gal-conf-cond}
\Pi_i^{\hphantom{i}i}=\Pi.
\end{equation} 
On the matter sector, the gauge fields $B$ and $B_i$ are weight-zero, whereas $\varrho$ is weight-$d$ and $N_i$, $d-1$. 

Weyl covariance can be implemented with the appropriate Galilean--Weyl covariant derivatives for both time and space. These will be introduced latter in Sec. \eqref{conFLUIDS}, our aim being here to foster on the case of conformal Killing fields and their possible role in supporting conservation laws within Weyl-invariant Galilean dynamics. 

Following \cite{Duval:2014uoa,  Duval:2014lpa} a conformal isometry is generated by a vector field $\upxi$ satisfying
\begin{equation}
\label{NCkillconf}
\mathscr{L}_\upxi a^{ij}=\lambda a^{ij},
\end{equation}
where
\begin{equation}
\label{galkilleqconf}
\lambda(t,\mathbf{x}) =\frac{2}{d}
\left(\hat  \nabla_{i}\xi^{\hat \imath}
+\theta^{w}\xi^{\hat t}  
\right).
 \end{equation}
This set of partial differential equations is not sufficient for defining conformal Killing vectors -- again a consequence of the  cometric degeneration. Besides, imposing a requirement similar to \eqref{NCkillconf} on the clock form  $\uptheta^{\hat t}$ does not help since according to \eqref{Liedclockgal}, the Lie derivative of $\uptheta^{\hat t}$ is already proportional to $\uptheta^{\hat t}$ with factor \eqref{galmu}. What is rather natural is to tune $\mu$ versus $\lambda$ so that the scaling of the metric be twice that of the clock form:\footnote{More generally, one considers $2\mu + z\lambda=0$, where $z$ is the \emph{dynamical exponent} i.e. minus the conformal weight of $\Omega$. Here, due to the close relationship of our Newton--Cartan spacetimes with relativistic ascendents,  the weight of  $\Omega$ is inherited from the latter and $z=1$. One also defines the level  $N= \nicefrac{2}{z}$, which appears in the conformal algebras emerging in flat Newton--Cartan spacetimes. \label{dynexp}} 
\begin{equation}
\label{extra-conf-cond}
2\mu + \lambda=0.
 \end{equation}
This is a consistent Weyl-covariant condition, leading to a reasonable set of conformal Killing fields.  Using the strong versus the weak definition of Newton--Cartan structures further reduces the freedom for conformal isometries, and opens the Pandora box for investigating conformal Galilean algebras in flat Newton--Cartan spacetimes:  $\mathfrak{cga}(d+1)$ and their multiple variations. Discussing these in detail is beyond our scope and ample information can be found in the quoted references. 

Assuming Weyl invariance i.e. \eqref{gal-conf-cond}, and the presence of a conformal Killing field,  the conservation equations \eqref{delSgalmomeq} and \eqref{delSgaleneq} can be exploited for computing $\mathcal{K}$ defined in \eqref{kilgalcon}, \eqref{inviotagal} and \eqref{invigal}:
 \begin{equation}
\label{galconconf}
\mathcal{K}= -\Pi\left(\frac{\lambda}{2}+\mu\right)
+\frac{P_i}{\Omega} \left(\partial_t \xi^{\hat \imath}+ \mathscr{L}_{\mathbf{w}} \xi^{\hat \imath}\right).
\end{equation}
The extra condition \eqref{extra-conf-cond} defining the conformal Killing vectors of Newton--Cartan spacetimes emerges naturally in the quest of conserved Galilean currents and gives $\mathcal{K}=\frac{P_i}{\Omega} \left(\partial_t \xi^{\hat \imath}+ \mathscr{L}_{\mathbf{w}} \xi^{\hat \imath}\right)\neq 0$.
As for the ordinary Killings, \emph{a conformal Killing vector does not guarantee a conservation law for Weyl-invariant Galilean dynamics}.

\subsection{Galilean hydrodynamics as a non-relativistic limit}\label{galdyn}

\boldmath
\subsubsection*{Philosophy and large-$c$ behaviour}
\unboldmath
Equations \eqref{delSgalmomeq}, \eqref{delSgaleneq} and \eqref{delSgalcon} summarize the conservation properties of Galilean dynamics on a general, curved and time-dependent space $\mathscr{S}$, spatial section of a torsionless Newton--Cartan spacetime $\mathbb{R}\times \mathscr{S}$. They are general-Galilean-covariant and are a consequence of this invariance, as much as the relativistic equations \eqref{conT} and 
\eqref{conJ} reflect general Riemannian covariance. 

Fluid dynamics is more. Its description requires expressing the momenta in terms of the velocity field $\mathbf{v}$, the heat current $\mathbf{Q}$, the stress tensor $\pmb{\Sigma}$, together with local-equilibrium thermodynamic variables such as $e$, $p$, $h$, $\varrho$, $\mu$, $T$ and $s$, obeying further thermodynamic laws (see App. \ref{thermo}), and ultimately entering the constitutive relations. A systematic approach to this programme is based on the large-$c$ expansion of relativistic hydrodynamics in Zermelo frame, which is the natural pseudo-Riemannian ascendent of the present Galilean framework. This method was applied successfully in \cite{CMPPS1}, where the Eckart frame was implicitly assumed on the relativistic side, subsequently ignoring the role of the current and the chemical potential. We will generalize it here in order to address the issue of the non-relativistic hydrodynamic-frame invariance, which relies intimately on the behaviour of the various observables.

The dependence with respect to the velocity of light is explicit in the relativistic geometric data (see the metric \eqref{galzerm}). This is our choice for making the bridge with Newton--Cartan straightforward. As a consequence, the behaviour of kinematical observables is also known -- see \eqref{def21}, \eqref{def23}, \eqref{def24}, \eqref{vel} and \eqref{gamzerm}. We find in particular
\begin{equation}
\label{galvel}
u^0=\frac{c}{\Omega}+\text{O}\left(\nicefrac{1}{c}\right),
\quad
u_i=\frac{v_i-w_i}{\Omega}+\text{O}\left(\nicefrac{1}{c^2}\right),
\end{equation}
and
\begin{eqnarray}
\label{galsiglim} 
\sigma_{ij}&=&\xi^{v}_{\hphantom{v}ij}+
\text{O}\left(\nicefrac{1}{c^2}\right),
\\
\label{galexplim}
\Theta&=&\theta^{v} +
\text{O}\left(\nicefrac{1}{c^2}\right),
\\
\label{galomlim}
\omega_{ij}&=&\frac{1}{\Omega}\left(\partial_{[i}(v-w)_{j]}\right)+
\text{O}\left(\nicefrac{1}{c^2}\right).
\end{eqnarray}

The non-relativistic limit of thermodynamic variables is standard and is recalled in App. \ref{thermo}.
The precise relation of the Galilean density $\varrho$ to the proper density $\varrho_0$ deserves however a comment. Indeed, the density measured by an observer is the projection of the current \eqref{curdec} 
onto the observer's velocity. For instance, an observer at rest with respect to the fluid, i.e. running with velocity $\text{u}$, finds 
\begin{equation}
\varrho_0 = -\frac{1}{c^2} J_\mu u^\mu.
\end{equation}
The density $\varrho$ is the one measured by a fiducial observer of Zermelo frame. Since $\varrho$ is the ``non-relativistic density,'' this fiducial observer should have some ``absolute'' status. In Minkowski spacetime we would have simply taken the inertial observer at rest in the inertial frame at hand. In Zermelo it is natural to consider an observer with velocity $\text{u}_{\text{Z}} \equiv \text{e}_{\hat t}$ as given in \eqref{galet} because 
\begin{equation}
\label{covdivuzer}
u_{\text{Z}}^\mu \nabla_\mu u^\nu_{\text{Z}} =0,
\end{equation}
and thus this observer indeed defines a locally inertial frame.\footnote{This is precisely why it was stated earlier that a frame with velocity $\mathbf{w}=w^i \partial_i$ was inertial in the Newton--Cartan geometry \eqref{galet},  \eqref{dmet}. Notice that the property \eqref{covdivuzer} holds because $\Omega$ is a function of time only, which is the emanation of the torsionless nature of the limiting Newton--Cartan structure, and enables to define in turn an absolute Newtonian time in this Galilean limit. This framework, where no isometry is assumed, is the closest we can go to the standard classical Newtonian physics on $\mathbb{R}\times \mathbf{E}_d$ with clock form $\text{d}t$. \label{LIF}} Hence we obtain
\begin{equation}
\label{newvarrho}
\varrho = -\frac{1}{c^2} J_\mu u^\mu_{\text{Z}}=\frac{ \Omega}{c}J^0= \Omega \gamma \varrho_0  + j_i\frac{v^i-w^i}{c^2\Omega},
\end{equation}
which naturally coincides with the hydrodynamic-frame invariant $\varrho_{0\text{r}}$ introduced in \eqref{invrhoZer}. This expression agrees with Ref. \cite{CMPPS1} only in the \emph{Eckart frame}, i.e. when $ j_i=0$.

The behaviour of $ \varrho_0$ in terms of $ \varrho$ at large $c$ depends on the behaviour of $ j_i$ and this brings us to the heart of the discussion of the non-relativistic limit in hydrodynamics: \emph{how do  
the stress tensor $ \tau_{ij}$, 
the heat current $ q_i$ 
and the
non-perfect current $ j_i$
behave at large $c$?}  There is no absolute answer to these questions, it depends on the microscopic properties of the system (the interactions in particular). These properties are encapsulated in the transport coefficients appearing in the constitutive relations for  $ \tau_{ij}$, $ q_i$  and $ j_i$ as derivative expansions. For instance, at first-derivative order $\tau_{(1)ij}=-2\eta \sigma_{ij}-\zeta h_{ij}\Theta$, $q_{(1)i}=-\kappa h_i^{\hphantom{i}\nu}\left(\partial_\nu T+\frac{T}{c^2}\, a_\nu \right)$ and $j_{(1)i}=-h_i^{\hphantom{i}\nu}
\sigma_T T \partial_\nu\frac{\mu_0}{T}$ with a typical relationship among the transport coefficients: $\kappa =\frac{\sigma_T {\cal w}^2}{T\varrho_0^2}$ (see e.g. Ref. \cite{Kovtun:2012rj}, where $\sigma_T$ is referred to as the charge conductivity). Under the reasonable assumption that $\eta$, $\zeta$ and $\kappa$ (the shear and bulk viscosities, and the heat conductivity) are of dominant order $1$, 
$\sigma_T $ is of order $\nicefrac{1}{c^4}$ (due to ${\cal w}^2$). Since $\mu_0$ is of order  $c^{2}$ (rest-mass contribution in \eqref{mumu0}) we conclude that $\tau_{ij}$ and $q_i$ are of order $1$, whereas $j_i$ is of order $\nicefrac{1}{c^2}$. For reasons that will become clear in the discussion on hydrodynamic-frame invariance, we would like to be slightly more general at this stage, and assume the following behaviour: 
\begin{eqnarray}
\label{sigexpG}
\tau_{ij}&=&-\Sigma_{ij} +\text{O}\left(\nicefrac{1}{c^2}\right),
\\
\label{qgalun}
q_l&=&c^2r_l+k_l+\text{O}\left(\nicefrac{1}{c^2}\right),
\\
\label{jgalun}
j_l&=&n_l+\frac{m_l}{c^2}+\text{O}\left(\nicefrac{1}{c^4}\right).
\end{eqnarray}
Following the above discussion, $r_l$ and $n_l$ are expected to vanish for ordinary non-relativistic fluids. Their presence will disclose some interesting properties though. 

Before we proceed with the Galilean fluid equations a comment should be made. Considering several distinct orders in the expansion of the relativistic data amounts to allow for a multiplication of the degrees of freedom in the Galilean limit (see the end of Sec. \ref{comgal}). This is accompanied with further dynamical (or possibly constraint) equations. Sometimes, these degrees of freedom and their dynamics can emerge separately, by performing an appropriate $c^2$ rescaling in the relativistic data before taking the limit. This is how the electric versus magnetic options occur, as e.g. in Refs. \cite{Duval:2014uoa,henneaux2021carroll}. 

Inserting the expression \eqref{jgalun} in \eqref{newvarrho}, we find:\footnote{Although $\lim\limits_{c\to \infty} \Omega \gamma=1$, we must keep terms of order $\nicefrac{1}{c^2}$ because of the rest mass contributions. }
\begin{equation}
\label{newvarrhoexp}
\varrho_0= \varrho-\frac{1}{c^2}
\left(
\frac{\varrho}{2} \left(\frac{\mathbf{v}-\mathbf{w}}{\Omega}\right)^2+\frac{\mathbf{n}\cdot\left(\mathbf{v}-\mathbf{w}\right)}{\Omega}
\right)+\text{O}\left(\nicefrac{1}{c^4}\right).
\end{equation}
The latter can in turn be used inside \eqref{epsrho} leading to
\begin{equation}
\label{newepsexp}
\varepsilon= c^2 \varrho +   \varrho \left(e-
\frac{1}{2} \left(\frac{\mathbf{v}-\mathbf{w}}{\Omega}\right)^2\right)-\frac{\mathbf{n}\cdot\left(\mathbf{v}-\mathbf{w}\right)}{\Omega}+\text{O}\left(\nicefrac{1}{c^2}\right),
\end{equation}
where the first term is the rest energy, the second is the internal energy corrected by the kinetic energy with respect to the local inertial frame, and the third is a contribution originating from the leading term in the matter current \eqref{jgalun}. We already foresee that this amounts to the presence of a spring or a sink that create or consume matter, and this will be confirmed when reaching the conservation equation. This situation is usually not considered except when discussing diffusion or similar phenomena (see for instance \cite{LandauF2}, chapter \RomanNumeral6). 

A last comment concerns the role of \eqref{epsrho}, or its large-$c$ emanation \eqref{newepsexp}. This equation of thermodynamic nature, makes the bridge between energy--momentum and matter conservations. As we will see soon, it establishes a relationship amongst $P_i$ and $N_i$ respectively defined in \eqref{galmompi} and \eqref{galmomphii} as the fluid momentum and the matter current. 

Notice that if several charges and associated conserved currents are present, only one will have a privileged relationship with the energy--momentum, being therefore affiliated with mass conservation -- and entering the thermodynamic relation \eqref{epsrho}, as mentioned in App. \ref{thermo}.

\subsubsection*{Galilean momenta}

The Galilean momenta defined earlier in \eqref{galmompiij}, \eqref{galmompii}, \eqref{galmompi},  \eqref{galmomchi} and \eqref{galmomphii} will now appear in the large-$c$ expansion of the relativistic energy--momentum and current components \eqref{restTinvZer} and \eqref{restJinvZer} in Zermelo frame, as explicit expressions in terms of the various hydrodynamic and thermodynamic observables. This includes the energy current $\Pi^i$ introduced in Eq. \eqref{delSgaleneq} -- and not a priori defined as a variation of the effective action with respect to some conjugate variable. As we have already emphasized, the thermodynamic laws set relationships amongst the energy--momentum and the matter.

Using Eqs. \eqref{invenZer}, \eqref{invqZer}, \eqref{invtauZer}, 
\eqref{invrhoZer}, \eqref{invjZer} and \eqref{newvarrho}, 
we obtain:
\begin{eqnarray}
\label{nlimgalrho0}
\varrho_{0\text{r}}&=&
\varrho,
\\
\label{nlimgaljri}
j_{\text{r}i}&=&
N_i +\frac{1}{c^2}p_i
+
\text{O}\left(\nicefrac{1}{c^4}\right),
\end{eqnarray}
where we introduced the leading and subleading matter currents 
\begin{eqnarray}
N^i&=&\varrho \frac{v^i-w^i}{\Omega} +n^i,
\\
\label{galsubcurr}
p_i&=&m_{i}
-\frac{\mathbf{n}\cdot\left(\mathbf{v}-\mathbf{w}\right)\left(v_i-w_i\right)}{\Omega^2}.
\end{eqnarray}
The subleading terms must be kept because they are multiplied in the expansions by the rest-mass term and  contribute the equations. Anticipating the next steps, we set  
\begin{equation}
\label{galmomP}
P^i= \varrho \frac{v^i-w^i}{\Omega} +r^i.
\end{equation}
We recognize in $P_i$ (defined generically in \eqref{galmompii} -- indices raised with $a^{ij}$) a slight extension of the usual fluid momentum, while the matter current $N^i$ (introduced in \eqref{galmomphii}) is related to the former as 
\begin{equation}
\label{PhiP}
N^i=P^i+n^i -r^i.
\end{equation}
The more standard equality $N^i=P^i$ occurs when $n^i =r^i$.  

Similarly we find for the energy--momentum
\begin{eqnarray}
\label{nlimgalvrep}
\varepsilon_{\text{r}}&=&  c^2 \varrho+ \Pi 
+ \text{O}\left(\nicefrac{1}{c^2}\right)
,\\
\label{nlimgalqri}
q_{\text{r}i}&=&
c^2 P_i+\Pi_{i}+p_{i} +
\text{O}\left(\nicefrac{1}{c^2}\right),
\\
p_{\text{r}} a_{ij}+\tau_{\text{r}{ij}}&=&
\Pi_{ij}
+ \text{O}\left(\nicefrac{1}{c^2}\right)
\label{nlimgalij}
\end{eqnarray}
with\footnote{The energy current $\Pi_i$ defined in \eqref{encurgalgeni} differs slightly from the expression (3.44) of \cite{CMPPS1}, even at vanishing $\mathbf{r}$. In that reference, neither were the momenta defined as variations of an effective action,  nor was the hydrodynamic-frame invariance a guide.}
\begin{eqnarray}
\Pi_{ij}&=&\varrho  \frac{\left(v_i-w_i\right)\left(v_j-w_j\right)}{\Omega^2}
+p a_{ij}-\Sigma_{ij}+2\frac{(v_{(i}-w_{(i})r_{j)}}{\Omega},
\label{stressgalgeni}
\\
\Pi&=&\varrho  \left( e+\frac{1}{2}\left(\frac{\mathbf{v}-\mathbf{w}}{\Omega} \right)^2\right)
+\frac{\left(2 \mathbf{r}-  \mathbf{n} \right) \cdot\left(\mathbf{v}-\mathbf{w}\right)}{\Omega} 
,
\label{endengalgeni}
\\
\Pi_{i}&=&\varrho \frac{v_i-w_i}{\Omega}  \left( h+\frac{1}{2}\left(\frac{\mathbf{v}-\mathbf{w}}{\Omega} \right)^2\right)-\frac{v^j-w^j}{\Omega}
\Sigma^{\hphantom{j}i}_{j}\nonumber
\\
&&
+\frac{r_i}{2}\left(\frac{\mathbf{v}-\mathbf{w}}{\Omega}\right)^2+
\frac{\mathbf{r}\cdot\left(\mathbf{v}-\mathbf{w}\right)\left(v_i-w_i\right)}{\Omega^2}+k_i-m_i ,
\label{encurgalgeni}
\end{eqnarray}
the explicit expressions for  \eqref{galmompiij} and \eqref{galmompi}. Observe that we have split the order-$1$ contribution in \eqref{nlimgalqri} as $\Pi_{i}+p_{i} $. This is not arbitrary because $p_{i} $,  defined in 
\eqref{galsubcurr}, occurs also as subleading term in the matter current \eqref{nlimgaljri}, and these will annihilate in the final equations, which will turn exactly as \eqref{delSgalmomeq}, \eqref{delSgaleneq} and \eqref{delSgalcon}.
We recover in the above formulas the fluid energy--stress tensor, energy  density and energy current,  
as defined in Ref. \cite{Landau} \S\S\ 15 and 49, generalized though in a covariant fashion for arbitrary torsionless Newton--Cartan geometries. They all receive exotic contributions from the $n_i$ and $r_i$, absent in standard Galilean fluids. The combination 
\begin{equation}
\label{QexpG}
Q_{l}=k_l-m_{l},
\end{equation}
inside the energy current, appears as  the Galilean heat current. It receives contributions from both the relativistic heat current $q_i$ and the relativistic non-perfect matter current $j_i$ (see \eqref{qgalun} and \eqref{jgalun}). This is exactly how it should: in Landau--Lifshitz frame  $q_i=0$ and the Galilean heat current originates exclusively from the relativistic non-perfect matter current, whereas in Eckart frame where $j_i=0$ it is the other way around.  A more complete discussion on hydrodynamic frames will be brought off in a short while. As anticipated in footnote \ref{ELL}, we can  however notice that for fluids without conserved current, although sensible, the Landau--Lifshitz hydrodynamic frame is not suited for the Galilean limit -- it leads to $Q_i=0$ and Eckart frame  is always preferred \cite{RZ}.

\subsubsection*{Hydrodynamic equations}

The fluid equations \eqref{conT} and \eqref{conJ} translate the vanishing of 
 \begin{eqnarray} 
\label{conTgalexp0} 
c\Omega\nabla_\mu T^{\mu0}&=& c^2\left(\mathcal{C}+ \hat\nabla_i\left(r^i-n^i\right)\right)+ \hat\nabla_ip^i+\mathcal{E}+ \text{O}\left(\frac{1}{c^2}\right),
\\
\label{conTgalexpi} 
\nabla_\mu T^{\mu}_{\hphantom{\mu}i}&=&\mathcal{M}_i+ \text{O}\left(\frac{1}{c^2}\right),
\\
\label{conJgalexp} 
\nabla_\mu J^{\mu}&=&\mathcal{C}+ \frac{1}{c^2} \hat\nabla_ip^i+ \text{O}\left(\frac{1}{c^4}\right)\,,
\end{eqnarray}
with
\begin{eqnarray}
\label{Eeq} 
\mathcal{E}&=& \frac{1}{\Omega}\frac{\hat{\text{D}}\Pi}{\text{d}t}+
\theta^w \Pi+
 \Pi_{ij} \hat\gamma^{wij}
+\hat \nabla_i \Pi^i,
\\
 \label{Meq}
 \mathcal{M}_i&=&\frac{1}{\Omega}\frac{\hat{\text{D}}P_i}{\text{d}t}
+\theta^w P_i
+P_j
 \hat\gamma^{wj}_{\hphantom{wj}i}
+
\hat \nabla^j \Pi_{ij}, 
\\ 
\label{Ceq} 
\mathcal{C}&=&\frac{1}{\Omega}\frac{\hat{\text{D}}\varrho}{\text{d}t}+
\theta^w \varrho
+\hat \nabla_i N^i.
\end{eqnarray}
Two fluid equations emerge from  \eqref{conTgalexp0} because of the presence of order-$1$ and order-$c^2$ terms that should both be zero in the infinite-$c$ limit. The latter should be taken after we combine  \eqref{conJgalexp} with 
 \eqref{conTgalexpi} as 
\begin{equation}
 c\Omega\nabla_\mu T^{\mu0}= c^2\hat\nabla_i\left(r^i-n^i\right)+ \mathcal{E}+ \text{O}\left(\nicefrac{1}{c^2}\right),
 \end{equation}
leading to
\begin{equation}
\label{nSteq}
\hat \nabla_i \left(r^i-n^i\right)=0.
\end{equation}
and 
\begin{equation}
\label{Eeq0}
\boxed{
\mathcal{E}=0,}
\end{equation}
which is the fluid energy equation. Using \eqref{PhiP}, Eq. \eqref{nSteq} is recast as
\begin{equation}
\label{nablaPhiP}
\boxed{\hat \nabla_i \left(N^i-P^i\right)=0,}
\end{equation}
which is a constraint equation for the divergences of the matter current and the fluid momentum. Finally, 
\eqref{conTgalexpi} provides the fluid momentum equation 
\begin{equation}
\label{Meq0}
\boxed{
\mathcal{M}_i=0,}
\end{equation}
whereas \eqref{conJgalexp} exhibits the continuity equation, which thanks to \eqref{nablaPhiP} also reads:\footnote{This is the typical equation describing phenomena, where several fluid components are present but are not separately conserved. Examples are diffusion or superfluid dynamics (e.g. \cite{LandauF2}, chapters \RomanNumeral{6} and \RomanNumeral{16}, Eqs. (58,3) or (139,3)).}
\begin{equation}
\label{Ceq0}
\boxed{
\mathcal{C}=0,}
\end{equation}
possibly recast in several forms
\begin{equation}
\label{Ceq0exp}
\mathcal{C}=\frac{1}{\Omega}\frac{\hat{\text{D}}\varrho}{\text{d}t}+
\theta^w \varrho
+\hat \nabla_i P^i=\frac{1}{\Omega}\frac{\text{d}\varrho}{\text{d}t}+
\theta^v \varrho
+\hat \nabla_i r^i=\frac{1}{\Omega}\frac{\text{d}\varrho}{\text{d}t}+
\theta^v \varrho
+\hat \nabla_i n^i=0.
\end{equation}

We can now summarize our findings. The Galilean fluid equations on a general background  are by essence  fully covariant under Galilean diffeomorphisms (alternative views about this statement are available in Ref. \cite{CMPPS1}) and extend the standard hydrodynamic equations on flat Euclidean space with absolute time. The momentum equation \eqref{Meq0} coincides with \eqref{delSgalmomeq}, whereas the energy equation \eqref{Eeq0} is  \eqref{delSgaleneq}. Similarly, the continuity equation  \eqref{Ceq0exp} is identified with  \eqref{delSgalcon}, provided 
the matter current and the fluid momentum have equal divergence \eqref{nablaPhiP}. This requirement is subsequent to the relationship \eqref{newepsexp}, which finally relates the energy--momentum equations  with the matter equation. 

Strictly speaking, Eq. \eqref{Ceq0exp}  \emph{is not} a conservation law. Integrated inside a static domain $\mathscr{V}$, the density $\varrho$ varies in time, not only because of the expansion or the contraction of $\mathscr{V}$ (term $\varrho\theta^{v}$), but also due to the flux of $\mathbf{n}$ through $\partial\mathscr{V}$. 
Using \eqref{galcontime} we find:
  \begin{equation}
\frac{1}{\Omega}\frac{\text{d}}{\text{d}t}\int_{\mathscr{V}}\text{d}^{d}x \sqrt{a}\varrho+\int_{\partial\mathscr{V}}\varrho  \frac{\star\mathbf{v}}{\Omega} 
=-\int_{\partial\mathscr{V}}\star\mathbf{n}.
  \label{galcontimerealfluids}
 \end{equation}
Allegedly, $\mathbf{n}$ appears as the flux of matter brought about by a sink or a spring. Furthermore, Eq. \eqref{nSteq} 
transcribes
that matter loss or gain goes along with heat loss or gain. In usual Galilean hydrodynamics  $r^i$ is required to vanish, which then forbids such spring or  of matter. At the same time, in those cases $n^i=0$, and the fluid momentum and matter current are identical: $P_i=N_i$. The systems under investigation here are more general and possess a remarkable property, which is broken in ordinary non-relativistic fluids: Galilean hydrodynamic frame invariance.

\subsubsection*{The fate of hydrodynamic-frame invariance}

The relativistic fluid equations are invariant under arbitrary unimodular transformations of the velocity field $\text{u}$,  captured in\footnote{The infinitesimal local Lorentz transformations are parameterized with Lorentz boost and rotation generators, $V^i(t,\mathbf{x})$ and  $\Omega^{ij}(t,\mathbf{x})$ -- antisymmetric,  as follows:
$ \delta v^i= V^i -V^j \frac{\left(v_j-w_j\right)\left(v^i-w^i\right)}{c^2\Omega^2}+ \Omega^{ij}\left(v_j-w_j\right)$. In the Galilean limit the general local velocity transformation is thus 
$ \delta v^i= V^i + \Omega^{ij}\left(v_j-w_j\right)$ -- Galilean boosts and rotations.} $v^k\to v^k + \delta v^k(t,\mathbf{x})$, provided they are accompanied with the transformations of all other dynamical quantities, as described in Eqs. \eqref{delepsZ}, \eqref{delqZ}, \eqref{deltau2Z}, \eqref{delrho0Z} and  \eqref{deljZ}. Does this survive in the Galilean limit? 

The intuitive answer to this question is no. The velocity field is a physical and observable quantity and only variations by constant values in directions associated with isometries of the underlying Galilean spacetime, if any, might leave the equations invariant. The fluid density $\varrho$ is also physical and has furthermore a microscopic definition in terms of an observable expectation value. It is hard to imagine how one could maintain the continuity equation invariant without altering the density. Although this might still be considered as an abstract field redefinition,\footnote{This point of view, slightly different from ours, is adopted to some extent in \cite{Jensen:2014ama, Karch}. We  acknowledge a rich exchange with P.~Kovtun on that matter.} it would be at the expense of giving up the physical meaning of the various quantities at hand.

The above intuitive answer seems to contradict the mathematical structure of the equations describing the dynamics. Indeed, on the one hand the operators entering equations \eqref{Eeq}, \eqref{Meq} and \eqref{Ceq} are velocity-independent; on the other hand, the momenta  $\varrho$, $N_i$, $P_i$, $\Pi$, $\Pi_i$ and $\Pi_{ij}$, appear as coefficients in the expansion of the hydrodynamic-frame-invariant relativistic momenta \eqref{nlimgalrho0}, \eqref{nlimgaljri}, \eqref{nlimgalvrep}, \eqref{nlimgalqri} and \eqref{nlimgalij}. It is however too naive to infer that the Galilean momenta are automatically invariant. They are obtained assuming a behaviour of the stress, and of the heat and matter currents with respect to $c^2$ (Eqs. \eqref{sigexpG}, \eqref{qgalun}, \eqref{jgalun}, \eqref{newvarrhoexp}, \eqref{newepsexp}), and this behaviour may or may not be stable under velocity transformations. 

The precise answer calls for a thorough examination of the transformations \eqref{delepsZ}, \eqref{delqZ}, \eqref{deltau2Z}, \eqref{delrho0Z} and  \eqref{deljZ} in the infinite-$c$ limit, and of their effect on the non-relativistic quantities introduced through \eqref{sigexpG}, \eqref{qgalun}, \eqref{jgalun}, \eqref{newvarrhoexp}  and \eqref{newepsexp}. We find the following transformations under the action of general local Galilean boosts and rotations:\footnote{Notice the following useful formulas: 
$\delta\theta^{v}=\frac{1}{\Omega}\hat\nabla_i \delta v^i$ and $\delta \xi^{v}_{ij}=\frac{1}{\Omega}\left(\hat\nabla_{(i}\delta v_{j)}-\frac{1}{d} a_{ij}\hat\nabla_k \delta v^k\right)$, whereas $\delta\theta^{w}=\delta \xi^{w}_{ij}=0$.
}
\begin{eqnarray}
\label{delgalpaik}
a_{ij}\delta p-\delta \Sigma_{ij}&=&-\frac{2}{\Omega}r_{(i} \delta v_{j)},
\\
\label{delgalrk}
\delta  r_i&=&-\frac{\varrho}{\Omega} \delta v_i,
\\
\label{ndelgalki}
\delta k_i&=&\frac{ \delta v_i}{\Omega}\frac{\left(\mathbf{v}-\mathbf{w}\right)\cdot\mathbf{n}}{\Omega}+\frac{ \delta v^j}{\Omega}\left(\frac{v_i-w_i}{\Omega}r_j-\varrho\frac{(v_i-w_i)(v_j-w_j)}{\Omega^2}-\varrho ha_{ij}+\Sigma_{ij}\right),\qquad
\\
\label{ndelgalQi}
\delta Q_i&=&\frac{ \delta v^j}{\Omega}\left(\left(r_j-n_j\right)\frac{v_i-w_i}{\Omega}-\varrho ha_{ij}+\Sigma_{ij}\right),
\\
\label{ndelgale}
\varrho\delta e&=&\left(n_i-2r_i\right)\frac{ \delta v^i}{\Omega},
\\
\delta \varrho&=&0,
\label{delgalvrho}
\\
\label{delgalnk}
\delta  n_i&=&-\frac{\varrho}{\Omega} \delta v_i,
\\
\label{ndelgalik}
\delta m_i&=&\frac{ \delta v^j}{\Omega}\left(n_j-\varrho\frac{v_j-w_j}{\Omega}\right)
\frac{v_i-w_i}{\Omega}+\frac{ \delta v_i}{\Omega}\frac{(\mathbf{v}-\mathbf{w})\cdot\mathbf{n}}{\Omega}
.
\end{eqnarray}
and thus
\begin{equation}
\label{delgalrn}
\delta  \left(n^i-r^i\right)=0\,.
\end{equation}
In turn, this action translates into the invariance of the fundamental momenta, i.e. the fluid energy density, the fluid energy current, the fluid energy--stress tensor, the fluid momentum, the matter density and the matter current: 
\begin{equation}
\label{delgalPis}
\delta  \Pi= 0, \quad \delta \Pi^{i}= 0,\quad \delta \Pi^{ij}=0, \quad \delta P^i=0,\quad \delta  \varrho=0, \quad \delta N^i=0.
\end{equation}
These imply that\emph{ the Galilean fluid equations are invariant under an arbitrary local redefinition of the fluid velocity field $v^i(t,\mathbf{x})$}. 

This result is important but should be reckoned with great care. On the one hand, the non-relativistic density $\varrho$ is not sensitive to the velocity field $\mathbf{v}$ (in contrast with the relativistic result \eqref{delrho0Z}). On the other hand, the standard momentum $ \varrho \frac{v^i-w^i}{\Omega} $ does depend on the velocity. The actual momentum emerging here, $P^i$ displayed in Eq. \eqref{galmomP}, is therefore invariant thanks to $n^i$. Hence, the non-conservation of matter discussed previously saves hydrodynamic-frame invariance. Put differently, this invariance is disabled upon demanding a genuine conservation i.e. $n^i=0$:  \emph{truly conserved, ordinary non-relativistic fluids are not hydrodynamic-frame-invariant}.
For these fluids the behaviours of the relativistic heat and matter currents are $q_l=k_l+\text{O}\left(\nicefrac{1}{c^2}\right)$ and $j_l=\frac{m_{l}}{c^2}+\text{O}\left(\nicefrac{1}{c^4}\right)$. Following our earlier discussion these behaviours are physical, but are at the same time unstable under velocity transformations.\footnote{The requirements $r^i=0$ or $n^i=0$ are not compatible with the transformations \eqref{delgalrk} or  \eqref{delgalnk}. 
Observe however that a choice, stable under hydrodynamic-frame transformations, 
is $n_i=r_i$, thanks to \eqref{delgalrn}. With this, $P^i=N^i$ and we are the closest possible to ordinary non-relativistic fluids, without genuine conservation though.}

Although negative, our last conclusion makes clear the origin of the physically expected breaking of velocity invariance in non-relativistic fluids obeying an authentic conservation equation. In this respect, it is worth quoting a substantial  literature on trials to explore the symmetries of Navier--Stokes equations (i.e. first-derivative truncated Galilean fluid equations) in their compressible or incompressible form. A nice overview is given in \cite{Horvathy:2009kz}, from which it comes out that most of these extra symmetries are probably accidental, and bound to the specific truncation or system (compressible/incompressible), rather than emanating from the original relativistic hydrodynamic-frame invariance.

\subsubsection*{Heat and entropy equations}

As we have emphasized in Sec. \ref{rreleq}, the investigation of hydrodynamic-frame transformations should be 
completed with the analysis of the entropy current, which is also invariant in the relativistic theory. We will not pursue this endeavour any further, which provides ultimately the transformations $\delta s$, $\delta p$, $\delta T$, and $\delta \mu$. Instead, we will combine the above results in order to reach the non-relativistic entropy equation. 

As a first step we can perform the usual combination $\mathcal{E} - \frac{v^i-w^i}{\Omega}\mathcal{M}_i$, which leads to the heat equation. It takes the following form: 
\begin{eqnarray}
\frac{1}{\Omega}\frac{\text{d}}{\text{d}t}\left(e \varrho+\frac{(\mathbf{v}-\mathbf{w})\cdot(\mathbf{r}-\mathbf{n})}{\Omega}\right) +\left(h \varrho+\frac{(\mathbf{v}-\mathbf{w})\cdot(\mathbf{r}-\mathbf{n})}{\Omega}  -\frac{\Sigma}{d}  \right)\theta^{v}
-\xi^{v}_{\hphantom{v}ij} \Sigma^{ij}&&
\nonumber \\
+\hat \nabla_i \left(Q^{i}
-\frac{v^i - w^i}{\Omega} \frac{(\mathbf{v}-\mathbf{w})\cdot(\mathbf{r}-\mathbf{n})}{\Omega}
\right)
+\frac{r^i}{\Omega}\frac{\text{D}}{\text{d}t}\frac{v_i - w_i}{\Omega} 
-\frac12\left(\frac{\mathbf{v}-\mathbf{w}}{\Omega}\right)^2 \hat \nabla_jr^j
&=&0.
\label{galEEuler2} 
\end{eqnarray}
This equation can be alternatively established within the relativistic framework, by considering 
$-u_\nu\nabla_\mu T^{\mu\nu}$ as in \eqref{en1},  and its subsequent Galilean limit.
Notice that due to the explicit appearance of the velocity field ($u_\nu$ or $v^i$), the equation at hand is hydrodynamic-frame invariant only on-shell. 

Assuming the thermodynamic properties\footnote{This might turn naive due to the extra underlying degrees of freedom carried by the \emph{effective} creation/destruction currents $\mathbf{n}$ and $\mathbf{r}$. Investigating this issue in not in our agenda here.} of Sec. \ref{thermo} and using in particular Eq. \eqref{derhoh}
together with the continuity equation \eqref{Ceq0exp}, Eq. \eqref{galEEuler2} is recast as follows: 
\begin{eqnarray}
\frac{\varrho T}{\Omega}\frac{\text{d}s}{\text{d}t} +\frac{1}{\Omega}\frac{\text{d}}{\text{d}t}\left(\frac{(\mathbf{v}-\mathbf{w})\cdot(\mathbf{r}-\mathbf{n})}{\Omega}\right) &=&
\left(\frac{\Sigma}{d}-\frac{(\mathbf{v}-\mathbf{w})\cdot(\mathbf{r}-\mathbf{n})}{\Omega} \right)\theta^v
+\xi^{v}_{\hphantom{v}ij} \Sigma^{ij}
\nonumber\\
&&-\hat \nabla_i \left(Q^{i}-\frac{v^i - w^i}{\Omega} \frac{(\mathbf{v}-\mathbf{w})\cdot(\mathbf{r}-\mathbf{n})}{\Omega}-r^i
\left(h
+\frac12\left(\frac{\mathbf{v}-\mathbf{w}}{\Omega}\right)^2\right)\right)
\nonumber\\
&&-r^i\left(\frac{1}{\Omega}\frac{\text{D}}{\text{d}t}\frac{v_i - w_i}{\Omega} +\hat \nabla_i \left(h
+\frac12\left(\frac{\mathbf{v}-\mathbf{w}}{\Omega}\right)^2\right)\right),
\label{galentropy30} 
\end{eqnarray}
which can also be reached from the relativistic equation \eqref{en2} in the infinite-$c$ limit. This equation is intricate and can be somehow simplified by setting $\mathbf{n}=\mathbf{r}$, which does not spoil the Galilean hydrodynamic-frame invariance (again realized on-shell):  
\begin{eqnarray}
\frac{\varrho T}{\Omega}\frac{\text{d}s}{\text{d}t} &=&
\frac{\Sigma}{d}\theta^{v}
+\xi^{v}_{\hphantom{v}ij} \Sigma^{ij}
-\hat \nabla_i \left(Q^{i}-r^i
\left(h
+\frac12\left(\frac{\mathbf{v}-\mathbf{w}}{\Omega}\right)^2\right)\right)
\nonumber\\
&&-r^i\left(\frac{1}{\Omega}\frac{\text{D}}{\text{d}t}\frac{v_i - w_i}{\Omega} +\hat \nabla_i \left(h
+\frac12\left(\frac{\mathbf{v}-\mathbf{w}}{\Omega}\right)^2\right)\right)
.
\label{galentropy3} 
\end{eqnarray}
The latter resembles the entropy equations found when studying diffusion phenomena or dissipative processes in superfluids (see Eqs. (58,6) and (140,4) in \cite{LandauF2}). The combination $h
+\frac12\left(\frac{\mathbf{v}-\mathbf{w}}{\Omega}\right)^2$ materializes  a sort of effective chemical potential for the current $\mathbf{r}$.  For vanishing $\mathbf{r}$, \eqref{galentropy3} is the standard non-relativistic entropy equation 
in arbitrary Galilean backgrounds. We will not pursue this discussion any longer.

\subsection{A comment on Galilean conservation versus non-relativistic limit}
\label{comgal}

\boldmath
\subsubsection*{The hidden local $U(1)$}
\unboldmath

We have so far pursued two distinct approaches. The first (Sec. \ref{FLUIDS}) relies on the requirement of Galilean general covariance for a system defined on a (torsionless) Newton--Cartan spacetime -- possibly but not necessarily -- obtained as an infinite-$c$ limit of a pseudo-Riemannian geometry in Zermelo frame. The second (Sec. \ref{galdyn}) amounts to taking the $c\to \infty$ limit after the general-covariance conservation has been imposed on the relativistic system. 

It is legitimate to wonder whether the two approaches are equivalent, in other words, whether the conservation requirement and the $c\to \infty$ limit commute. 

In order to present a clean answer to this question, we must consider the simplest possible situation. Aiming at this, we focus on the energy and momentum only (no matter) and make no reference to their expressions in terms of fluid variables such as density, thermodynamic quantities, velocity etc. Starting with a Newton--Cartan set up, we define the energy--stress $\Pi_{ij}$, the momentum $P_i$ and the energy density $\Pi$ as in \eqref{galmompiij},  \eqref{galmompii} and   \eqref{galmompi}. General Galilean covariance translates into two equations, 
\eqref{delSgalmomeq} and  \eqref{delSgaleneq}, which reveal a novel, a priori  undetermined vector $\Pi_i$, interpreted as the energy current.   

Alternatively, one may work in a relativistic spacetime equipped with Zermelo coordinates, and a full-fledged and conserved energy--momentum tensor $T_{\mu\nu}$ with the following large-$c$ expansion:
\begin{equation}
\label{gal-no-curem}
 \begin{cases}
\Omega^2T^{00}=\varepsilon_{\text{r}}=  \Pi 
+ \text{O}\left(\nicefrac{1}{c^2}\right)
\\
c \Omega T^0_{\hphantom{0}i}=q_{\text{r}i}=
c^2 P_i+\Pi_{i}+
\text{O}\left(\nicefrac{1}{c^2}\right)
\\
T_{ij}=p_{\text{r}} a_{ij}+\tau_{\text{r}{ij}}=
\Pi_{ij}
+ \text{O}\left(\nicefrac{1}{c^2}\right).
\end{cases}
\end{equation}
This \emph{determines the energy current $\Pi_{i}$ as a subleading term with respect to the momentum $P_i$ }in the expansion of the relativistic heat current. Furthermore, the expansion of the relativistic equations is now
\begin{equation}
\label{gal-no-cureq}
 \begin{cases}
c\Omega\nabla_\mu T^{\mu0}= c^2 \hat \nabla_jP^j+ \mathcal{E}+ \text{O}\left(\nicefrac{1}{c^2}\right)=0
\\
\nabla_\mu T^{\mu}_{\hphantom{\mu}i}=\mathcal{M}_i+ \text{O}\left(\nicefrac{1}{c^2}\right)=0.
\end{cases}
\end{equation}
with $ \mathcal{E}$ and $\mathcal{M}_i$ given in Eqs. \eqref{Eeq} and  \eqref{Meq}. We recover, as in the first way, the Galilean conservation equations \eqref{delSgalmomeq} and  \eqref{delSgaleneq}, \emph{supplemented now by an extra constraint on the current $P_i$}:
\begin{equation}
\label{extra-co-gal}
\hat \nabla_jP^j
=0.
\end{equation}

The punch line of the current discussion is that taking the $c \to \infty$ limit followed by the general-covariance requirement is less restrictive than following the pattern in the reverse order. With this latter order, not only the energy current is provided explicitly but the fluid current obeys an extra  constraint equation. The reason for this is simple. When the infinite-$c$ limit is the  last step, the system secretly remembers the full diffeomorphism invariance present at the first step, which contracts during the limit into the Galilean covariance accompanied with a central extension \cite{Bergshoeff:2017btm}. This extra hidden local $U(1)$ invariance accounts for the supplementary equation \eqref{extra-co-gal}.  When the Galilean diffeomorphism invariance is the second step, the $\mathbf{x}$-independence of $\xi^t$ leaves $\Pi_i$ undetermined, and no further equation is found. 

All this shows how the full -- as opposed to \eqref{extra-co-gal} -- continuity equation \eqref{Ceq0exp} 
\begin{equation}
\label{Ceq0exprep}
\frac{1}{\Omega}\frac{\hat{\text{D}}\varrho}{\text{d}t}+
\theta^w \varrho
+\hat \nabla_i P^i=0,
\end{equation}
can emerge provided $\varepsilon_{\text{r}}$ contains an extra $c^2\varrho$ term, \emph{without} introducing a relativistic conserved current with a local $U(1)$ symmetry, as in \cite{CMPPS1}.\footnote{With this method, however, the relativistic hydrodynamic frame is locked to Eckart's since the Galilean heat current $Q^i$ in \eqref{QexpG} receives only the contribution $k^i$.} When such an explicit $U(1)$ current is present, as in Secs. \ref{FLUIDS} and \ref{galdyn}, it is promptly  identified with the hidden one through \eqref{nablaPhiP}, originating in the deep relationship between energy and mass, Eq. \eqref{epsrho}.

\subsubsection*{More general abstract equations}

The large-$c$ behaviours \eqref{nlimgalrho0},  \eqref{nlimgaljri},  \eqref{nlimgalvrep},  \eqref{nlimgalqri} and \eqref{nlimgalij} (or a simpler version of the latter without $\varrho$ \eqref{gal-no-curem}) are motivated by the physics standing behind these momenta,  captured in the behaviours of transport coefficients and materialized in \eqref{sigexpG},  \eqref{qgalun} and \eqref{jgalun}. As an echo to the comments made when setting the latter, one could be more abstract and consider order-$c^2$ terms in the stress tensor $\tau_{ij}$ as we have already introduced in the heat and matter currents, and possibly further powers in those ($r_l$ and $n_l$ were already beyond normalcy, but introduced as a mean of restoring hydrodynamic-frame invariance -- or demonstrating its breaking in genuine non-relativistic fluids). The net effect of this sort of  options is to bring the energy--momentum tensor in the form
\begin{equation}
\label{gal-no-curem-extra}
 \begin{cases}
\Omega^2T^{00}=\varepsilon_{\text{r}}=  c^2 \varrho +\Pi 
+ \text{O}\left(\nicefrac{1}{c^2}\right)
\\
c \Omega T^0_{\hphantom{0}i}=q_{\text{r}i}=
c^4 \tilde P_i+c^2 P_i+\Pi_{i}+
\text{O}\left(\nicefrac{1}{c^2}\right)
\\
T_{ij}=p_{\text{r}} a_{ij}+\tau_{\text{r}{ij}}=
c^2 \tilde \Pi_{ij}+
\Pi_{ij}
+ \text{O}\left(\nicefrac{1}{c^2}\right),
\end{cases}
\end{equation}
and produce at infinite $c$ a genuine hierarchy of equations, which are replicas of those we have already met: 
\begin{equation}
\label{gal-no-eq-extra}
 \begin{cases}
 \left(\frac{1}{\Omega}\frac{\hat{\text{D}}}{\text{d}t}+
\theta^w\right) \Pi+
 \Pi_{ij} \hat\gamma^{wij}
+\hat \nabla_i \Pi^i =0 \\
\left(\frac{1}{\Omega}\frac{\hat{\text{D}}}{\text{d}t}+
\theta^w\right) \varrho+
 \tilde \Pi_{ij} \hat\gamma^{wij}
+\hat \nabla_i P^i =0 \\
\hat \nabla_j\tilde P^j
=0\\
\left(\frac{1}{\Omega}\frac{\hat{\text{D}}}{\text{d}t}
+\theta^w \right)P_i
+ P_j
\hat\gamma^{wj}_{\hphantom{wj}i}
+
\hat \nabla^j \Pi_{ij} 
=0
\\
\left(\frac{1}{\Omega}\frac{\hat{\text{D}}}{\text{d}t}
+\theta^w \right)\tilde P_i
+ \tilde P_j
\hat\gamma^{wj}_{\hphantom{wj}i}
+
\hat \nabla^j \tilde \Pi_{ij} 
=0.
\end{cases}
\end{equation}

This sort of situation is the archetype of multiplication of degrees of freedom, mentioned earlier. It comes naturally thanks to the existence of a parameter $c$, which makes it possible to organize a Laurent expansion. It is more artificial to interpret this system as conservation equations portraying local symmetries, because this would require introducing further variables, conjugate to the new momenta, such as $\tilde a^{ij}$, $\tilde w^i$, $\tilde \Omega$ etc. We will not elaborate on that, but keep the structure in mind for comparison with the
 forthcoming analysis of Sec. \eqref{carfluids2} about Carrollian fluids.  For the latter, no physical intuition can possibly serve a as guide -- basic thermodynamics is even missing. Only a blind $\nicefrac{1}{c^2}$ expansion applies, as suggested by the only known application field of Carrollian fluids, which is flat holography \cite{Campoleoni:2018ltl, CMPR,CMPRpos, CMPPS2}. Then the hierarchy obtained is dual to \eqref{gal-no-eq-extra}, and this plainly justifies our present excursion from standard, physical non-relativistic fluids. 

Multiplication of degrees of freedom occurs also in the matter sector. On could indeed abstractly assume  
 that some matter current behaves like
\begin{equation}
\label{jexpGextinv}
\frac{\Omega}{c}I^0=\iota_{0\text{r}}=c^2 \tilde \kappa + \kappa + \text{O}\left(\nicefrac{1}{c^2}\right),\quad
I_k=i_{\text{r}k}=c^4 \tilde{\tilde{K}}_k+c^2 \tilde K_k+K_k+ \text{O}\left(\nicefrac{1}{c^2}\right).
\end{equation}
Using these expansions in the relativistic divergence of the matter current $J^\mu$ in Zermelo background
we find:
\begin{equation}
 \label{conJgalrexpn} 
\nabla_\mu I^{\mu}= c^4\tilde{\tilde{\mathcal{K}}}+ c^2\tilde{\mathcal{K}}+ \mathcal{K} + \text{O}\left(\nicefrac{1}{c^2}\right)
\end{equation}
with
\begin{equation}
\label{galN}
 \begin{cases}
\tilde{\tilde{\mathcal{K}}}=
\hat\nabla_j \tilde{\tilde{ K}}^j
 \\\tilde{\mathcal{K}}=
 \left(\frac{1}{\Omega}\frac{\hat{\text{D}}}{\text{d}t} +\theta^w\right)\tilde\kappa+\hat\nabla_j \tilde K^j
 \\ \mathcal{K}=
 \left(\frac{1}{\Omega}\frac{\hat{\text{D}}}{\text{d}t} +\theta^w\right)\kappa+\hat\nabla_j K^j,
\end{cases}
\end{equation}
which must vanish if $\nabla_\mu I^{\mu}=0$.

As an aside application of the latter results, we can insert the behaviour \eqref{gal-no-curem-extra} inside the components  \eqref{inviotaZer} and \eqref{inviZer} of a relativistic conserved current resulting from the combination of the energy--momentum tensor with a Killing field. We find thus
\begin{equation}
\label{galkappaK}
 \begin{cases}
\kappa= \xi^{\hat\imath} P_i-\xi^{\hat t}  \Pi
\\
\tilde\kappa=\xi^{\hat\imath} \tilde P_i-\xi^{\hat t}  \varrho
\\
K_i=\xi^{\hat\jmath}\Pi_{ij}-
 \xi^{\hat t} \Pi_i
\\
\tilde K_i =\xi^{\hat\jmath}\tilde \Pi_{ij}-
 \xi^{\hat t} P_i
\\
\tilde{\tilde{ K}}_i= -
 \xi^{\hat t} \tilde P_i,
\end{cases}
\end{equation}
where $\kappa$ and $K_i$ are precisely as anticipated in \eqref{inviotagal} and \eqref{invigal}. On-shell i.e. assuming \eqref{gal-no-eq-extra}, and using \eqref{galkilleq} 
we find for \eqref{galN}
\begin{equation}
\label{galNkill}
 \begin{cases}
\tilde{\tilde{\mathcal{K}}}=
0
 \\\tilde{\mathcal{K}}=
\frac{\tilde P_i}{\Omega} \left(\partial_t \xi^{\hat \imath}+ \mathscr{L}_{\mathbf{w}} \xi^{\hat \imath}\right)
 \\ \mathcal{K}
=\frac{P_i}{\Omega} \left(\partial_t \xi^{\hat \imath}+ \mathscr{L}_{\mathbf{w}} \xi^{\hat \imath}\right),
\end{cases}
\end{equation}
in agreement with the result \eqref{galnoncon} for the last two. The first vanishes, and this shows that even though the existence of a Killing field does not guarantee the conservation of a Galilean current, such a conservation can occur if the appropriate vector vanishes. Ordinary Galilean fluids as those studied in Sec. \ref{galdyn} have $\tilde P_i=\tilde\Pi_{ij}=0$ in \eqref{gal-no-curem-extra} so that $\tilde{\tilde{ K}}^j= 0$. Hence two currents survive, one with  $\mathcal{K}\neq 0$ (non-conserved) and another, which is conserved ($\tilde{\mathcal{K}}= 0$) but already known. Indeed, it is the very same that appears inside the ordinary continuity equation, and no extra conservation arises as a consequence of isometries. 

One might be legitimately skeptical about the validity of the above conclusion on non-conservation: how can the bona fide
 law $\nabla_\mu I^{\mu}= 0$ 
of a relativistic current $I^\mu = \xi_\nu T^{\mu\nu}$ based on a Killing field $\upxi$ of a pseudo-Riemannian spacetime,  break down suddenly in the infinite-$c$ limit? The answer is captured by the very definition of a Galilean Killing, which ultimately leaves non-vanishing terms in the divergence. The precise way this comes about is  exposed in App. \eqref{galgarbage}.

\subsection{Massless carriers and Weyl properties} \label{conFLUIDS}

\subsubsection*{Generic hydrodynamic equations}

Although massless particles are ultra-relativistic, a macroscopic collection of them forming a fluid can be compatible with Galilean symmetries. The latter appear as a phenomenological emanation and are agnostic on their microscopic origin. Such a system can also have a conserved current and a chemical potential, which are not related to mass, but to some charge. Some examples of this sort are mentioned in App. \ref{thermo} (see Ref. \cite{GM}) together with their basic thermodynamic properties. 

The difference with respect to the previous chapter (Sec. \ref{galdyn}) is the Galilean limit. Here, instead of \eqref{newvarrhoexp}, we can simply consider 
\begin{equation}
\label{newvarrhoeepsexpmassless}
\varrho_0= \varrho+\text{O}\left(\nicefrac{1}{c^2}\right), \quad \varepsilon=  \varrho e+\text{O}\left(\nicefrac{1}{c^2}\right),
\end{equation}
where $e$ is the energy per charge unit and $\varrho$ is the charge volume density -- as opposed to proper volume 
-- and $\upvarepsilon = \varrho e $ the non-relativistic energy density (see also App. \ref{thermo}). Again, our goal is to find the fundamental variables as well as the dynamical equations, and probe the behaviour of the latter under Galilean hydrodynamic-frame transformations. 

Why do we expect this sort of system be potentially Galilean hydrodynamic-frame invariant? A gas of photons, perhaps under some isotropy and homogeneity requirement regarding interactions, gives no handle for measuring a global velocity. In other words, we do not expect $v^i$  to enter the hydrodynamic equations. This reasoning is not a proof, but a guideline to pursue here. For that, we will adopt again the behaviours \eqref{sigexpG}, \eqref{qgalun} and \eqref{jgalun} for the stress tensor, the heat and the charge currents, even though we are aware that on physics grounds $r_i$ and $n_i$ are bound to vanish. For the charge current, the subleading term $m_l$ turns out to be irrelevant here because of the absence of rest mass. 

Equations \eqref{invenZer}, \eqref{invqZer}, \eqref{invtauZer}, 
\eqref{invrhoZer} and \eqref{invjZer}, 
now give:
\begin{equation}
\varrho_{0\text{r}}=
\varrho,
\quad
j_{\text{r}i}=
N_i 
+
\text{O}\left(\nicefrac{1}{c^2}\right)
\label{nlimgalrhojrimasl}
\end{equation}
with Galilean charge current
\begin{equation}
\label{Phivarrhon}
N^i=\varrho \frac{v^i-w^i}{\Omega} +n^i.
\end{equation}
From the energy--momentum tensor one obtains
\begin{eqnarray}
\label{nlimgalvrepmasl}
\varepsilon_{\text{r}}&=&  \Pi 
+ \text{O}\left(\nicefrac{1}{c^2}\right)
,\\
\label{nlimgalqrimasl}
q_{\text{r}i}&=&
c^2 P_i+\Pi_{i}+
\text{O}\left(\nicefrac{1}{c^2}\right),
\\
p_{\text{r}} a_{ij}+\tau_{\text{r}{ij}}&=&
\Pi_{ij}
+ \text{O}\left(\nicefrac{1}{c^2}\right),
\label{nlimgalijmasl}
\end{eqnarray}
which coincides with \eqref{gal-no-curem}, where
\begin{eqnarray}
\Pi_{ij}&=&p a_{ij}-\Sigma_{ij}+2\frac{(v_{(i}-w_{(i})r_{j)}}{\Omega},
\label{galstressmasl}
\\
\label{galmomPmasl}
P^i&=& r^i,
\\
\label{galenPmasl}
\Pi&=&\upvarepsilon+\frac{2 \mathbf{r} \cdot\left(\mathbf{v}-\mathbf{w}\right)}{\Omega} 
,
\\
\Pi_{i}&=&\left((\upvarepsilon+p) a_{ij}
-\Sigma_{ij} \right)\frac{v^j-w^j}{\Omega}
+  \frac{r_i}{2}\left(\frac{\mathbf{v}-\mathbf{w}}{\Omega}\right)^2+\frac{\mathbf{r}\cdot\left(\mathbf{v}-\mathbf{w}\right)\left(v_i-w_i\right)}{\Omega^2}+k_i ,
\label{galenfluxmasl}
\end{eqnarray}
are the explicit expressions for  \eqref{galmompiij}, \eqref{galmompii}  and \eqref{galmompi}, as well as for the energy current 
$\Pi_{i}$, which will appear in the energy equation \eqref{delSgaleneq}. The Galilean heat current receives now a single contribution as 
\begin{equation}
\label{QexpGml}
Q_{l}=k_l.
\end{equation}

For the fluid under consideration, the structure of  the conservation equations is as follows (actually as in \eqref{gal-no-cureq} because the energy--momentum \eqref{nlimgalvrepmasl}, \eqref{nlimgalqrimasl}, \eqref{nlimgalijmasl}  is as in \eqref{gal-no-curem}):
 \begin{eqnarray}
 \label{conTgalexp0ml} 
c\Omega\nabla_\mu T^{\mu0}&=& c^2 \hat \nabla_jr^j+ \mathcal{E}+ \text{O}\left(\nicefrac{1}{c^2}\right),
\\
\label{conTgalexpiml} 
\nabla_\mu T^{\mu}_{\hphantom{\mu}i}&=&\mathcal{M}_i+ \text{O}\left(\nicefrac{1}{c^2}\right),
\\
\label{conJgalexpmasl} 
\nabla_\mu J^{\mu}&=&\mathcal{C}+ \text{O}\left(\nicefrac{1}{c^2}\right),
\end{eqnarray}
with  $\mathcal{E}$, $\mathcal{M}_i$ and $\mathcal{C}$ as in \eqref{Eeq}, \eqref{Meq}, \eqref{Ceq}. At infinite $c$ the hydrodynamic equations are again  \eqref{Eeq0}, \eqref{Meq0}, \eqref{Ceq0}, and we recover Eqs.  \eqref{delSgalmomeq}, \eqref{delSgaleneq}  and \eqref{delSgalcon}, as expected, plus the extra equation (same as \eqref{extra-co-gal})
\begin{equation}
\label{extramaslr}
 \hat \nabla_jr^j=0,
\end{equation}
which is absent when the unphysical vector $r^j$ originating from the $c^2$ term of the relativistic heat current vanishes. The difference with respect to the massive case studied in the previous section dwells in the expression of the momenta (energy--stress tensor, fluid current, fluid energy density and fluid energy current -- the charge current is the same as the matter current before). 

We can now combine the above results in order to reach the heat and next the entropy equations. Equivalently these  are obtained as infinite-$c$ limits of Eqs. \eqref{en1} and \eqref{en2}. We find for the former 
\begin{eqnarray}
\mathcal{E} -\frac{v_i-w_i}{\Omega}\mathcal{M}_i =\frac{1}{\Omega}\frac{\text{d}}{\text{d}t}\left(\upvarepsilon+\frac{(\mathbf{v}-\mathbf{w})\cdot\mathbf{r}}{\Omega}\right) +\left(\upvarepsilon+p+\frac{(\mathbf{v}-\mathbf{w})\cdot\mathbf{r}}{\Omega}  -\frac{\Sigma}{d}  \right)\theta^{v}
-\xi^{v}_{\hphantom{v}ij} \Sigma^{ij}
\nonumber \\
+\hat \nabla_i \left(Q^{i}
-\frac{v^i - w^i}{\Omega} \frac{(\mathbf{v}-\mathbf{w})\cdot\mathbf{r}}{\Omega}
\right)
+\frac{r^i}{\Omega}\frac{\text{D}}{\text{d}t}\frac{v_i - w_i}{\Omega}=0.
\label{galEEuler2ml} 
\end{eqnarray}
For the entropy equations there are two options. If no conserved charge current exists, the equation \eqref{Ceq0} is immaterial, the chemical potential vanishes and \eqref{2stlmassless} gives $ \text{d}\upvarepsilon=T\text{d}\upsigma $,
which can be substituted in \eqref{galEEuler2ml}. This happens e.g. for a gas of photons. If a conserved charge current is available then  $\upvarepsilon$ can be traded for $\varrho e$, $\upvarepsilon + p $ for  $\varrho h$,  $\upsigma$ for $\varrho s$, and 
using \eqref{GDGM},   \eqref{2stlmassless}  and \eqref{Ceq0} one obtains
\begin{equation}
\frac{1}{\Omega}\frac{\text{d}\varrho e}{\text{d}t}=\frac{\varrho T}{\Omega}\frac{\text{d}s}{\text{d}t} -\varrho h \theta^v-h\hat\nabla_i n^i, 
\end{equation}
which can be inserted back in \eqref{galEEuler2ml}:
\begin{eqnarray}
\frac{\varrho T}{\Omega}\frac{\text{d}s}{\text{d}t}
+
\frac{1}{\Omega}\frac{\text{d}}{\text{d}t}\left(\frac{(\mathbf{v}-\mathbf{w})\cdot\mathbf{r}}{\Omega}\right) +\left(\frac{(\mathbf{v}-\mathbf{w})\cdot\mathbf{r}}{\Omega}  -\frac{\Sigma}{d}  \right)\theta^{v}
-\xi^{v}_{\hphantom{v}ij} \Sigma^{ij}-h\hat\nabla_i n^i
\nonumber \\
+\hat \nabla_i \left(Q^{i}
-\frac{v^i - w^i}{\Omega} \frac{(\mathbf{v}-\mathbf{w})\cdot\mathbf{r}}{\Omega}
\right)
+\frac{r^i}{\Omega}\frac{\text{D}}{\text{d}t}\frac{v_i - w_i}{\Omega}=0.
\label{galentrml} 
\end{eqnarray}

Given the above Galilean hydrodynamical equations, one may reconsider their behaviour under velocity local transformations. The absence of rest mass for the carriers modifies the scalings with respect to the speed of light, and possibly the invariance properties. Bringing together the transformations \eqref{delepsZ}, \eqref{delqZ}, \eqref{deltau2Z}, \eqref{delrho0Z} and  \eqref{deljZ}, and the scalings  \eqref{sigexpG}, \eqref{qgalun},  \eqref{jgalun} and \eqref{newvarrhoeepsexpmassless},  we find in the infinite-$c$ limit that 
$\delta \varrho$ and $\delta  n_i$ are still as  in \eqref{delgalvrho} and \eqref{delgalnk},
while  
\begin{eqnarray}
\label{delgalpaikmasl}
a_{ij}\delta p-\delta \Sigma_{ij}&=&-\frac{2}{\Omega}r_{(i} \delta v_{j)},
\\
\label{delgalrkmasl}
\delta  r_i&=&0,
\\
\delta k_i&=& \delta Q_i
\nonumber
\\
&=&\frac{ \delta v^j}{\Omega}\left(\frac{v_i-w_i}{\Omega}r_j-\varrho ha_{ij}+\Sigma_{ij}\right),
\label{ndelgalkimasl}
\\
\delta \upepsilon
  &=&\varrho\delta e
\nonumber
\\&=&-2r_i\frac{ \delta v^i}{\Omega}.
\label{ndelgalemasl}
\end{eqnarray}
These transformations ensure the invariance of the Galilean momenta as in Eqs. \eqref{delgalPis}, which thus implies that \emph{for fluids consisting of massless particles, the Galilean fluid equations established above are invariant under arbitrary hydrodynamic frame transformations}.\footnote{Equations \eqref{galEEuler2ml} and \eqref{galentrml} are hydrodynamic-frame-invariant only on-shell.}

One should stress that, thanks to \eqref{delgalrkmasl}, the hydrodynamic-frame invariance holds even when $r_i=0$, which is the physically interesting situation, following our previous discussion on the behaviour of the relativistic heat current. The momentum equation obtained from \eqref{Meq} greatly simplifies in this case: 
\begin{equation}
\mathcal{M}_i= \partial_i p -\nabla_j  \Sigma^{\hphantom{i}j}_{i}=0.
\label{galMEulermassless} 
\end{equation}
Nonetheless, due to  \eqref{delgalnk}, hydrodynamic-frame invariance  does not resist when $n_i$ is required to vanish in the charge current, which is necessary for the continuity equation \eqref{Ceq}, \eqref{Ceq0}  to be a genuine conservation. This caveat, which opposes again hydrodynamic-frame invariance to conservation, is evaded precisely for fluids without conserved charge, as are photon gases, which are therefore truly hydrodynamic-frame-invariant in the Galilean regime with entropy equation (for the physical situation where $r_i=0$) 
\begin{equation}
\frac{T}{\Omega}\frac{\text{d}\upsigma}{\text{d}t} +\left(\upvarepsilon+p -\frac{\Sigma}{d}  \right)\theta^{v}
-\xi^{v}_{\hphantom{v}ij} \Sigma^{ij}+\hat \nabla_i Q^{i}=0.
\label{galEntr2ml} 
\end{equation}

\subsubsection*{Weyl invariance}

Galilean groups can accommodate conformal extensions. The subject has generated an abundant literature, part of which is already quoted here \cite{Duval:1990hj, Duval:2014uoa,Duv, Bagchi2010e, Bagchi2} -- more can be found in those references. The analysis of conformal symmetry in non-relativistic fluid dynamics has been in the agenda of many groups.  No real guiding principle has been followed though, the search has been usually blind and the output often looks accidental.\footnote{Reference \cite{Horvathy:2009kz} makes better contact with the relativistic fluid equations, and provides a nice and critical overview of the field.} From the fluid perspective on non-isometric and non-conformal-isometric backgrounds, conformal symmetry is rather meant to be Weyl symmetry, which is expected to portray the dynamics when no massive excitations are present. This  instance was touched upon in Sec. \ref{FLUIDS}, when presenting the basic features of Newton--Cartan geometry, and will be considered in the present chapter from the perspective of the large-$c$ limit in Zermelo backgrounds.

The fundamental quantities of the Zermelo geometry \eqref{galzerm}
behave as follows under a Weyl transformation:
\begin{equation}
\label{weyl-geometry-Z}
a_{ij}\to \frac{1}{\mathcal{B}^2}a_{ij},\quad w^{i}\to w^{i},\quad w_{i}\to \frac{1}{\mathcal{B}^2}w_{i},\quad \Omega\to \frac{1}{\mathcal{B}}\Omega\,,
\end{equation} 
and since $\Omega$ depends on time only, the last of \eqref{weyl-geometry-Z} imposes $\mathcal{B}=\mathcal{B}(t)$. The velocity components $u^\mu$ have weight $1$. This gathers the following for the ordinary spatial fluid velocity:
\begin{equation}
\label{W-velocity-Z}
v^{i}\to v^{i},\quad v_{i}\to \frac{1}{\mathcal{B}^2}v_{i}.
\end{equation} 

With this at hand, one can wonder what is the Galilean Weyl-covariant derivative, acting on Galilean Weyl-covariant tensors. From a purely mathematical perspective taming the connections on Newton--Cartan (or Carrollian) geometries is a thriving subject (see e.g. \cite{Duv, Bekaert:2014bwa, Bekaert:2015xua}). We will here answer modestly this question 
by examining the infinite-$c$ limit of the connection \eqref{Wconc} and the corresponding Weyl-covariant derivative used in the relativistic case. 
As we will also witness later in the Carrollian side, this splits into time and space Weyl derivatives, associated with time and space Weyl connections, inherited from the limit of $\text{A}$ given in \eqref{Wconc}. Owing to the fact that 
\begin{equation}
\label{GlimA}
\lim_{c\to \infty}\Omega c A^0 = -\frac{\theta^{v}}{d}, \quad \lim_{c\to \infty} A_i = 0, 
\end{equation} 
there is no spatial Weyl connection in the Galilean limit. The ordinary Galilean spatial covariant derivative $\hat \nabla_i$ used here as the usual $d$-dimensional metric-compatible and torsionless covariant derivative with connection coefficients \eqref{dgamma} (possibly time-dependent since generally $a_{ij}=a_{ij}(t,\mathbf{x})$) is thus Weyl-covariant on its own right. This is not a surprise since a Weyl rescaling with $\mathcal{B}(t)$ leaves the Christoffel symbols  \eqref{dgamma} unaltered.

The Galilean time covariant derivative $\frac{\text{D}}{\text{d}t}$ given in  \eqref{galfcovder-f} is not Weyl-covariant, though. It can be promoted to a Weyl-covariant Galilean time derivative  $\mathscr{D}_t$ thanks to $\theta^{v}$, which transforms indeed as a connection:
\begin{equation}
\label{thetaGWeyl}
\theta^{v}\to \
\mathcal{B} \theta^{v}-\frac{d}{\Omega} \partial_t\mathcal{B}.
\end{equation} 
Consequently, if $S^{ij\ldots}_{\hphantom{ij\ldots}kl\ldots}$ are the components of a weight-$w$ Galilean tensor, then  
\begin{equation}
\label{GWeyltd}
\frac{1}{\Omega}\mathscr{D}_tS^{ij\ldots}_{\hphantom{ij\ldots}kl\ldots}=
\left(\frac{1}{\Omega}\frac{\text{D}}{\text{d}t}+\frac{w}{d} \theta^{v}\right)S^{ij\ldots}_{\hphantom{ij\ldots}kl\ldots}
\end{equation} 
are the components of Galilean tensor with weight $w+1$ . Observe that the components of the Galilean shear given in \eqref{galshear} is of weight $-1$:  
 \begin{equation}
 \label{weyl-geometry-2-Z}
\xi^v_{\hphantom{v}ij}\to \frac{1}{\mathcal{B}}\xi^v_{\hphantom{v}ij}.
\end{equation}

The Weyl transformation \eqref{thetaGWeyl} holds equally for $\theta^{w}$ defined in \eqref{galwexp},  and so does in fact \eqref{weyl-geometry-2-Z} for $\xi^w_{\hphantom{v}ij}$ defined in \eqref{galwshear}. One can therefore introduce an alternative Galilean Weyl-covariant  time derivative, defined in purely geometrical terms:\footnote{This sort of Weyl-covariant derivative is insensitive to the fluid velocity and is thus better suited for discussing hydrodynamic-frame invariance. Its relativistic ascendent is a Weyl connection $\text{A}^{\text{Z}}$ constructed, as explained generally in footnote \ref{confcongen}, with the vector field $\text{u}_{\text{Z}}=\text{e}_{\hat t}$ defined in \eqref{galet} (and used in Sec. \ref{galdyn}), which has norm $-c^2$ in the Zermelo background \eqref{galzerm}. This connection exists irrespective of the fluid velocity: $\text{A}^{\text{Z}}=\frac{\theta^w}{d}\Omega \text{d}t$.}  
\begin{eqnarray}
\frac{1}{\Omega}\hat{\mathscr{D}}_t\Phi&=&
\left(\frac{1}{\Omega}\frac{\hat{\text{D}}}{\text{d}t}+\frac{w}{d} \theta^{w}\right)\Phi,\\
\frac{1}{\Omega}\hat{\mathscr{D}}_t V_i&=&
\left(\frac{1}{\Omega}\frac{\hat{\text{D}}}{\text{d}t}+\frac{w+1}{d} \theta^{w}\right)V_i,\end{eqnarray} 
for weight-$w$ Galilean scalars or forms, extendable by the Leibniz rule. For convenience, both $\mathscr{D}_t $ and $\hat{\mathscr{D}}_t $ will be used in the following and should not be confused.

Equipped with the above tools and imposing Weyl invariance \eqref{gal-conf-cond}, the fundamental equations 
\eqref{delSgalmomeq}, \eqref{delSgaleneq}  and \eqref{delSgalcon} are recast as follows:\footnote{We saw in Sec. \ref{FLUIDS} that
$\Pi_{ij}$ has weight $d-1$, $P_i$ and $\Pi_i$ weight $d$, and $\Pi$ weight $d+1$; similarly $\varrho$ is weight-$d$ and $N_i$ weight  $d-1$.} 
\begin{eqnarray}
\label{delSgalmomeqcon}
 \frac{1}{\Omega}\hat{\mathscr{D}}_tP_i
 +P_j
 \xi^{wj}_{\hphantom{wj}i}
+
\hat \nabla^j \Pi_{ij} 
&=&0,
\\
\label{delSgaleneqcon}
 \frac{1}{\Omega}\hat{\mathscr{D}}_t\Pi
+ \Pi_{ij} \xi^{wij}
+\hat \nabla_i \Pi^i &=&0,
\\
\label{delSgalconconf}
 \frac{1}{\Omega}\hat{\mathscr{D}}_t\varrho
+\hat \nabla_i N^i&=&0.
\end{eqnarray}
They are Weyl-covariant of weights $d+1$, $d+2$ and $d+1$. 

When dealing  with Galilean fluids, the Galilean momenta $\varrho$, $N_i$, $ P_i$, $\Pi$, $\Pi_i$ and $\Pi_{ij}$ 
emerge in the large-$c$ expansion
of $\varrho_{\text{r}0}$, $j_{\text{r}i}$, $\varepsilon_{\text{r}}$,  $q_{\text{r}i}$ and $p_{\text{r}} a_{ij}+\tau_{\text{r}{ij}}$ (see Eqs. \eqref{nlimgalrhojrimasl},   \eqref{nlimgalvrepmasl},  \eqref{nlimgalqrimasl},  \eqref{nlimgalijmasl}). The weights inherited in this limiting procedure (the relativistic weights are available in Tab. \ref{weights}) are in agreement with those previously defined through the effective-action definition of the momenta. These momenta are  expressed in terms of the Galilean velocity $v^i$ together with the usual list of variables emanating from the relativistic stress, heat current and charge/matter current.

From the expressions \eqref{galstressmasl}, \eqref{galmomPmasl}, \eqref{galenPmasl}, \eqref{galenfluxmasl},  we infer that the forms $r_i$, $k_i$ (and thus $Q_i$) have weight $d$, while $n_i$ and the Galilean stress $\Sigma_{ij}$ have weight $d-1$. The Weyl condition\footnote{Notice in passing that the Weyl-invariance requirement \eqref{gal-conf-cond} determined from the effective action, is also the large-$c$ expression of the relativistic condition $T_\mu^{\hphantom{\mu}\mu}=0$ discussed at the end of Sec. \ref{rreleq}, obtained using \eqref{restraceinvZ} with
\eqref{nlimgalvrepmasl}, \eqref{nlimgalqrimasl} and \eqref{nlimgalijmasl}.} \eqref{gal-conf-cond} now reads $\upvarepsilon= dp -\Sigma$. The stress being considered as a correction to perfect fluids, absent at global thermodynamic equilibrium, this condition splits into the  conformal equation of state
\begin{equation}\label{mlGal}
\upvarepsilon= dp,
\end{equation} 
accompanied with the Weyl-invariance  requirement
\begin{equation}\label{conG}
\Sigma\equiv\Sigma_{ij}a^{ij}=0. 
\end{equation} 
Other thermodynamic observables like $e$, $T$, $\mu$ or $h$ have all weight $1$, and $s$ is weight zero. 

It is important to stress that the above analysis is consistent because we have been systematically referring to fluids with microscopic massless degrees of freedom, and have thus used Eqs. \eqref{nlimgalrhojrimasl},   \eqref{Phivarrhon}, \eqref{nlimgalvrepmasl},  \eqref{nlimgalqrimasl},  \eqref{nlimgalijmasl},  \eqref{galstressmasl}, \eqref{galmomPmasl}, \eqref{galenPmasl}, \eqref{galenfluxmasl}. Had one considered fluids with massive carriers, conflicts would have appeared in the conformal weights, as e.g. in Eq. \eqref{nablaPhiP} setting a relationship among $N^i$ and $P^i$, which in a Weyl-covariant system are expected to have different weights ($d+1$ and $d+2$).  

Hydrodynamic equations \eqref{galEEuler2ml} and \eqref{galEntr2ml} are recast  as
\begin{eqnarray}
\frac{1}{\Omega}{\mathscr{D}}_t\left(\upvarepsilon+\frac{(\mathbf{v}-\mathbf{w})\cdot\mathbf{r}}{\Omega}\right)+\frac{r^i}{\Omega}{\mathscr{D}}_t\frac{v_i - w_i}{\Omega} -\xi^{v}_{\hphantom{v}ij} \Sigma^{ij}&&
\nonumber \\
+\hat \nabla_i \left(Q^{i}
-\frac{v^i - w^i}{\Omega} \frac{(\mathbf{v}-\mathbf{w})\cdot\mathbf{r}}{\Omega}
\right)
&=&0,
\label{galEEuler2mlcon} 
\\
\frac{\varrho T}{\Omega}{\mathscr{D}}_t\ s
+
\frac{1}{\Omega}{\mathscr{D}}_t\left(\frac{(\mathbf{v}-\mathbf{w})\cdot\mathbf{r}}{\Omega}\right) 
+\frac{r^i}{\Omega}{\mathscr{D}}_t\frac{v_i - w_i}{\Omega}-\xi^{v}_{\hphantom{v}ij} \Sigma^{ij}&&
\nonumber \\
-h\hat\nabla_i n^i
+\hat \nabla_i \left(Q^{i}
-\frac{v^i - w^i}{\Omega} \frac{(\mathbf{v}-\mathbf{w})\cdot\mathbf{r}}{\Omega}
\right)
&=&0.
\label{galentrmlcon}
\end{eqnarray}
For more conventional conformal fluids with $r^i = 0$ and no conserved charge we find
\begin{equation}
\boxed{\frac{1}{\Omega}\mathscr{D}_t\upvarepsilon 
-\xi^v_{\hphantom{v}ij} \Sigma^{ij}+\hat \nabla_i Q^{i}=\frac{T}{\Omega}\mathscr{D}_t \upsigma 
-\xi^v_{\hphantom{v}ij} \Sigma^{ij}+\hat \nabla_i Q^{i}=0,}
\label{galEEuler2mlcon-nc} 
\end{equation}
which are Weyl-covariant of weight $d+2$.   
The Euler (transverse) equation \eqref{galMEulermassless} remains unchanged and can be expressed in terms of the energy thanks to \eqref{mlGal}, or further using \eqref{GDGM}:
\begin{equation}
\boxed{\frac{1}{d}\partial_i \upvarepsilon  -\hat\nabla^j \Sigma_{ij}=\frac{1}{(d+1)} \partial_i (T \sigma)  -\hat\nabla^j \Sigma_{ij}=0.}
\label{galMEulercon} 
\end{equation}
It is Weyl-covariant of weight $d+1$.

Besides expressing hydrodynamic equations for fluids based on massless microscopic constituents in a Galilean general-covariant fashion, our present analysis exhibits one class of physical fluids, where hydrodynamic-frame invariance survives the Galilean limit: the conformal fluids without any conserved charge. This is in contrast to the more general Galilean hydrodynamics studied in Sec.  \ref{galdyn}, where hydrodynamic-frame invariance was only emerging in exotic fluids, where matter conservation was not fulfilled. Furthermore, under the physical assumption that the heat current remains of order 1 in the infinite-$c$ limit ($r_i = 0$ in \eqref{qgalun}), the energy density is insensitive to the choice of fluid velocity (see \eqref{ndelgalemasl}).

\subsubsection*{Conformal isometries and conservation laws}

There is not much we can add on conservation laws that has not yet been processed. Summarizing, the large-$c$ expansion of the energy--momentum tensor \eqref{nlimgalvrepmasl}, 
\eqref{nlimgalqrimasl}, 
\eqref{nlimgalijmasl}, combined with 
the components 
\eqref{inviotaZer}, \eqref{inviZer} in the presence of a conformal Killing field $\upxi$, produces 
\begin{equation}
\label{jexpGextinvcon}
\iota_{0\text{r}}=\kappa + \text{O}\left(\nicefrac{1}{c^2}\right),\quad
i_{\text{r}k}=K_k+ \text{O}\left(\nicefrac{1}{c^2}\right)
\end{equation}
with $\kappa$ and $K_k$ as in \eqref{inviotagal} and \eqref{invigal}, defining a Galilean current. Using the Weyl-invariance condition \eqref{gal-conf-cond}, the equations \eqref{delSgalmomeqcon}, \eqref{delSgaleneqcon}, and the Galilean Killing equations \eqref{NCkillconf}, \eqref{extra-conf-cond}, the Galilean divergence \eqref{kilgalcon} of this current turns out to satisfy \eqref{galconconf}. 
As usual, conservation demands a symmetry wider than a conformal isometry, for which $\partial_t \xi^{\hat \imath}+ \mathscr{L}_{\mathbf{w}} \xi^{\hat \imath}=0$.

\section{Carrollian fluid dynamics}\label{FLUID2}

\subsection{Carroll structures and general Carrollian covariance}\label{cardifcons}

\subsubsection*{Carrollian manifolds}

Carroll structures are alternatives to Newton--Cartan spacetimes, introduced in  \cite{Duval:2014uoa, Duval:2014uva, Duval:2014lpa}. They consist of a $d+1$-dimensional manifold $\mathscr{M}= \mathbb{R} \times \mathscr{S}$  endowed with a degenerate metric and a vector field, which is the kernel of the metric. These manifolds
are described in terms of fibre bundles\footnote{Carrollian structures were defined as ``ambient structures'' in Refs. \cite{Bekaert:2014bwa,Bekaert:2015xua}. Notice that we use equally the wording ``manifolds,'' ``spacetimes'' and ``structures.'' Mathematically the latter is more precise for it embraces the various attributes. Depending on these attributes, it can be even refined into \emph{weak} or \emph{strong} structure as we will see in the following.} with one-dimensional fibre and a $d$-dimensional base $\mathscr{S}$ thought of as space, the fibre being time. The Carroll group \cite{Levy, SenGupta} emerges as the isometry group of flat Carrollian structures, but our framework is here more general with no assumption about isometries, but Carrollian diffeomorphisms instead; the Carrollian transformations are realized locally, in the tangent space. These diffeomorphisms have the virtue of preserving the time/space separation, as opposed to general diffeomorphisms. 

For concreteness $\mathscr{M}$ will be equipped with coordinates $( t, \mathbf{x})$ and we will restrict to  degenerate metrics of the form
\begin{equation}   
\label{cardegmet}
\text{d}\ell^2=a_{ij}( t, \mathbf{x}) \text{d}x^i \text{d}x^j,\quad i,j\ldots \in \{1,\ldots,d\}
\end{equation}
with kernel generated by 
\begin{equation}   
\label{kert}
\text{e}_{\hat t}
= \frac{1}{\Omega}\partial_t,
\end{equation}
which defines a \emph{field of observers}. This coordinate system is adapted to the fiber/base splitting, which is in turn respected by Carrollian diffeomorphisms \eqref{cardifs}. It is also naturally reached in the Carrollian limit of a pseudo-Riemannian spacetime in Papapetrou--Randers gauge \eqref{carrp}. The Carrollian structure naturally incorporates an \emph{Ehresmann connection}, which is a background gauge field  
$\pmb{b}=b_i \text{d}x^i$, appearing in the dual form of the kernel generator \eqref{kert}:
\begin{equation}   
\label{kertdual}
\uptheta^{\hat t}=\Omega \text{d}t -b_i \text{d}x^i,
\end{equation}
the \emph{clock form}. The scale factor $\Omega$ and the gauge components $b_i$ depend on $t$ and $\mathbf{x}$. 

Under Carrollian diffeomorphisms \eqref{cardifs} (the Jacobian is defined in \eqref{carj}), the transformation rules of the various geometric objects are as in \eqref{cardifa}, \eqref{cardifb} as well as
\begin{eqnarray}
\label{cart}
\partial^\prime_t&=&\frac{1}{J}\partial_t,\\
\label{carjj}
\partial^\prime_j&=&J^{-1i}_{\hphantom{-1}j}\left(\partial_i-
\frac{j_i}{J}\partial_t\right),\\
\uptheta^{\hat t\prime}
&=&
\uptheta^{\hat t},\\
\label{carb2}
\hat\partial_i^\prime &=&
J^{-1j}_{\hphantom{-1}i} \hat\partial_j.
\end{eqnarray}
where 
\begin{equation}
\label{dhat}
\hat\partial_i=\partial_i+\frac{b_i}{\Omega}\partial_t,
\end{equation}
are the vector fields dual to the forms $\text{d}x^i$, also spelled $\text{e}_{\hat\imath}$ in \eqref{RPframe}.

For Carrollian manifolds it is customary to say that space is absolute, whereas time isn't, as opposed to their dual relatives, the Newton--Cartan spacetimes. This is again somehow abusive, except when $a_{ij}$ depends on space only, and it mostly refers to the form of the Jacobian \eqref{carj}. It is rooted to the properties of the primitive Carrollian manifold obtained as the $c\to 0$ limit of Minkowski spacetime.

Carrollian tensors depend generically on time $t$ and space $\mathbf{x}$. They
carry indices $i, j, \ldots \in \{1, \ldots, d \}$, which are lowered and raised with $a_{ij}$ and its inverse spatial cometric $a^{ij}$, and transform covariantly under Carrollian diffeomorphisms \eqref{cardifs} with Jacobian $J_i^j$ and $J^{-1i}_{\hphantom{-1}j}$ defined in \eqref{carj}. Following \cite{CMPPS1}, we introduce a Levi--Civita--Carroll connection with coefficients 
\begin{equation}
\label{dgammaCar}
\hat\gamma^i_{jk}=\frac{a^{il}}{2}\left(
\hat\partial_j
a_{lk}+\hat\partial_k  a_{lj}-
\hat\partial_l a_{jk}\right). 
\end{equation}
This connection is not unique (see the already quoted literature  \cite{Duval:2014lpa,Bekaert:2015xua, Hartong:2015xda}), but emerges naturally in the vanishing-$c$ limit of a Levi--Civita connection in the Papapetrou--Randers coordinates \eqref{carrp}. It defines a spatial Carrollian covariant derivative $\hat \nabla_i$ with tensorial transformation properties under Carrollian diffeomorphisms (details on the transformation properties can be found in the appendix A.2 of Ref. \cite{CMPPS1}).\footnote{\underline{Important remark:} many symbols are common to the Galilean and Carrollian sides investigated in the present paper. The context should leave no room for confusion.}

The Levi--Civita--Carroll connection is torsionless and metric-compatible:
\begin{equation}
\label{carcontormetcomp}
\hat t^k_{\hphantom{k}ij}=2\hat\gamma^k_{[ij]}=0, \quad \hat\nabla_ia_{jk}=0.
\end{equation}
The vectors $\hat\partial_i$ do not commute and define the Carrollian vorticity:
\begin{equation}
\label{carconcomderf}
\left[\hat\partial_i,\hat\partial_j\right]=
\frac{2}{\Omega}\varpi_{ij}\partial_t,\quad\varpi_{ij}=\partial_{[i}b_{j]}+b_{[i}\varphi_{j]}
\end{equation}
with 
\begin{equation}
\label{caromacc}
\varphi_i=\frac{1}{\Omega}\left(\partial_t b_i+\partial_i \Omega\right).
\end{equation}
the Carrollian acceleration. Notice that 
\begin{equation}
\label{dthetat}
\text{d}\uptheta^{\hat t}=
\varphi_i \text{d}x^i\wedge \uptheta^{\hat t}-\varpi_{ij} \text{d}x^i\wedge  \text{d}x^j,
\end{equation}
so that vanishing Carrollian acceleration and vorticity  are necessary and sufficient conditions for $\uptheta^{\hat t}$ be closed and define a family of hypersurfaces inside  $\mathscr{M}= \mathbb{R} \times \mathscr{S}$  as $\tau(t,\mathbf{x})=\text{const.} $, where locally $\uptheta^{\hat t}=\text{d}\tau$.

The ordinary time derivative operator $\frac{1}{\Omega}\partial_t$ acts covariantly on Carrollian tensors. However, it is not metric-compatible because $a_{ij}$ depends on time. Hence one defines a new Carrollian temporal covariant derivative by requiring covariance, i.e. $\frac{1}{\Omega^\prime}{\hat D}^\prime_t=\frac{1}{\Omega}\hat D_t$, and 
\begin{equation}
\label{cartimemet}
\hat D_ta_{jk}=0.
\end{equation}
This is achieved by introducing a temporal Carrollian connection
\begin{equation}
\label{dgammaCartime}
\hat\gamma_{ij}=\frac{1}{2\Omega}\partial_t a_{ij}
=\xi_{ij} + \frac{1}{d}a_{ij}\theta,
\end{equation}
which is a genuine symmetric Carrollian tensor split into a traceless part, the Carrollian shear, and the trace, which is the Carrollian expansion:
\begin{equation}
\label{carexp-tempcon}
\theta=
\frac{1}{\Omega}              
\partial_t \ln\sqrt{a},
\end{equation}
The action of  $\hat D_t$ on scalars is $\partial_t$
\begin{equation}
\label{Cartimecovdersc}
\hat D_t \Phi=\partial_t  \Phi,
\end{equation}
whereas on vectors or forms it
is defined as
\begin{equation}
\label{Cartimecovdervecform}
\frac{1}{\Omega}\hat D_tV^i=\frac{1}{\Omega} \partial_tV^i+\hat\gamma^i_{\hphantom{i}j} V^j,\quad
\frac{1}{\Omega}\hat D_tV_i=\frac{1}{\Omega} \partial_tV_i-\hat\gamma_i^{\hphantom{i}j} V_j.
\end{equation}
Leibniz rule generalizes the latter to any tensor and allows to demonstrate the property \eqref{cartimemet}. 

The commutators of Carrollian covariant spatial derivatives define further Carrollian tensors ($\Phi$ and $V^i$ are a Carrollian scalar and a Carrollian vector):\footnote{In \cite{CMPPS1} an alternative tensor was defined as 
$
\hat r^i_{\hphantom{i}jkl}=\hat R^i_{\hphantom{i}jkl}
+2 \hat\gamma^i_{\hphantom{i}j}\varpi_{kl}
$ with $\left[\hat\nabla_k,\hat\nabla_l\right]V^i= \hat r^i_{\hphantom{i}jkl}V^j+
\varpi_{kl}\frac{2}{\Omega} \partial_{t}V^i$.
}
\begin{eqnarray}
\label{carcontor}
\left[\hat\nabla_i,\hat\nabla_j\right]\Phi&=&
\varpi_{ij}\frac{2}{\Omega}\partial_t\Phi,
\\
\left[\hat\nabla_k,\hat\nabla_l\right]V^i&=&\left(
\hat\partial_k\hat\gamma^i_{lj}
-\hat\partial_l\hat\gamma^i_{kj}
+\hat\gamma^i_{km}\hat\gamma^m_{lj}
-\hat\gamma^i_{lm}\hat\gamma^m_{kj}
\right)V^j+\left[\hat\partial_k,\hat\partial_l\right]V^i
\\
&=& \hat R^i_{\hphantom{i}jkl}V^j+
\varpi_{kl}\frac{2}{\Omega}\hat D_tV^i.
\label{carriemann}
\end{eqnarray}
Similarly, time and space derivatives do not commute:
\begin{equation}
\left[\frac{1}{\Omega}\hat D_t,\hat\nabla_i\right]V^j= 
\varphi_{i}\left(\left(\frac{1}{\Omega}\hat D_t
+
\theta
\right)V^j - \hat\gamma^{j}_{\hphantom{j}k}V^k\right)
-\hat\gamma_{i}^{\hphantom{i}k}\hat\nabla_k V^j
-d \hat r^j_{\hphantom{j}ik}V^k
\label{3carriemanntimetilde}
\end{equation}
with\footnote{Notice that $\frac{1}{\Omega} \partial_t \hat \gamma^j_{ik}
=\left(\hat \nabla_i + \varphi_i\right) \hat\gamma^{j}_{\hphantom{j}k}
+\left(\hat \nabla_k + \varphi_k\right) \hat\gamma^{j}_{\hphantom{j}i}
-
\left(\hat \nabla^j + \varphi^j \right) \hat\gamma_{ik}
$
is a Carrollian tensor, even though $ \hat \gamma^j_{ik}$ is not.}
\begin{equation}
\hat r^j_{\hphantom{j}ik}=\frac{1}{d}
\left(\theta \varphi_i \delta^j_k 
+\hat \nabla_i \hat\gamma^{j}_{\hphantom{j}k}
-\frac{1}{\Omega} \partial_t \hat \gamma^j_{ik}
\right)
\label{3carriemanntime}
\end{equation}
and 
\begin{equation}
\hat r^j_{\hphantom{j}jk}=\hat r_k=\frac{1}{d}\left(\hat\nabla_j\hat\gamma^{j}_{\hphantom{j}k}
 -\hat\partial_k\theta
\right),
\label{carriemanntime}
\end{equation}
further Carrollian curvature tensors. 

\subsubsection*{Carrollian dynamics and Carrollian diffeomorphisms}
 
Consider now a dynamical system on a Carrollian manifold $\mathscr{M}=\mathbb{R}\times \mathscr{S}$ described with an (effective) action $S=\int \text{d}t\,  \text{d}^{d}x \sqrt{a}\Omega\mathcal{L}$, functional of $a_{ij}$, $\Omega$ and $b_i$. The associated \emph{Carrollian momenta}, which replace the corresponding relativistic energy--momentum tensor \eqref{varrelT}
are now (see \cite{CM1,Chandrasekaran:2021hxc})\footnote{The fluid energy density $\Pi$ was spelled $e_{\text{e}}$ in \cite{CMPPS1}.}
\begin{eqnarray}
\label{carvarstren} 
\Pi^{ij}&=&\frac{2}{\sqrt{a} \Omega}\frac{\delta S}{\delta a_{ij}},\\
\label{carvarencur}
\Pi^{i}&=&\frac{1}{\sqrt{a} \Omega}\frac{\delta S}{\delta b_i},\\
\label{carvarenden}
\Pi&=&-\frac{1}{\sqrt{a}}\left(\frac{\delta S}{\delta \Omega}+\frac{b_i}{\Omega}\frac{\delta S}{\delta b_i}\right),
 \end{eqnarray}
with $\frac{\delta S}{\delta \Omega}=-\sqrt{a}\left(\Pi+b_i \Pi^i\right)$. These are the \emph{energy--stress tensor}, the \emph{energy current} and the \emph{energy density}.

Diffeomorphisms are generated by vector fields as in \eqref{genkil}
\begin{equation}   
\label{carkil}
\upxi=\xi^t \partial_t +\xi^i \partial_i= \left(\xi^t -\xi^i\frac{b_i}{\Omega}\right) \partial_t
+ \xi^i \left(\partial_i+\frac{b_i}{\Omega}\partial_t\right)=
\xi^{\hat t}\frac{1}{\Omega} \partial_t+\xi^i \hat \partial_i .
 \end{equation}
Carrollian diffeomorphisms \eqref{cardifs} are restricted to $\xi^i=\xi^i(\mathbf{x})$.
As usual, the variation under diffeomorphisms is implemented through the Lie derivative and we find the following:
\begin{equation}   
-\delta_\upxi a_{ij}=\mathscr{L}_\upxi a_{ij}= 2\hat\nabla_{(i}\xi^ka_{j)k}+2\xi^{\hat t}  \hat \gamma_{ij}
-2b_{(i} a_{j)k} \frac{1}{\Omega}  \partial_t \xi^{k} ,
\label{Liedaijcar}
 \end{equation}
where the last term drops for Carrollian diffeomorphisms. Furthermore 
\begin{equation}   
\label{Liedcfoocar}
\mathscr{L}_\upxi \text{e}_{\hat t}=-\left(\frac{1}{\Omega} \partial_t
\xi^{\hat t }+
\varphi_i \xi^i \right)   \text{e}_{\hat t}
=\mu
\text{e}_{\hat t},
 \end{equation}
and (the form $\uptheta^{\hat t}$ is defined in \eqref{kertdual})
\begin{equation}   
\label{Liedcehrecar}
\mathscr{L}_\upxi \uptheta^{\hat t}=\left(\frac{1}{\Omega} \partial_t
\xi^{\hat t }+
\varphi_i \xi^i \right)  \uptheta^{\hat t} +\left(\left(\hat\partial_i-\varphi_i\right)\xi^{\hat t }-
2\xi^j \varpi_{ji}\right)\text{d}x^i.
 \end{equation}
 From the latter we infer
 \begin{equation}   
-\delta_\upxi \ln \Omega=\frac{1}{\Omega}\mathscr{L}_\upxi \Omega = \frac{1}{\Omega} \partial_t
\xi^{\hat t }+
\varphi_i \xi^i  , \quad
-\delta_\upxi b_i=
\mathscr{L}_\upxi b_i
=b_i\left(\frac{1}{\Omega} \partial_t
\xi^{\hat t }+
\varphi_j \xi^j \right)-\left(\left(\hat\partial_i-\varphi_i\right)\xi^{\hat t }-
2\xi^j \varpi_{ji}\right) . 
\label{Liedwiomcar}
\end{equation}

We can now move to the variation of the action under Carrollian diffeomorphisms:
\begin{equation}   
\label{delSgravRP}
\delta_\upxi S= \int \text{d}t \text{d}^{d}x \sqrt{a}   \Omega \left(\frac{1}{2}\Pi^{ij} \delta_\upxi a_{ij} + 
\Pi^i  \delta_\upxi  b_i -\frac{1}{\Omega}
\left(
\Pi+b_i \Pi^i
\right) \delta_\upxi \Omega
\right).
\end{equation}
Using \eqref{Liedaijcar} and  \eqref{Liedwiomcar} with $\xi^i=\xi^i(\mathbf{x})$, we obtain (indices are here lowered with $a_{ij}$)
 \begin{eqnarray}
 \delta_\upxi S&=&\int \text{d}t \text{d}^dx  \sqrt{a}\Omega 
\left\{-\xi^{\hat t}\left[
\left(\frac{1}{\Omega}\partial_t +\theta\right)\Pi
+\left(\hat \nabla_i+2\varphi_i\right)\Pi^i
+\Pi^{ij} \hat \gamma_{ij}
\right]\right.
 \nonumber\\
&&+\left.\xi^i\left[\left(\hat \nabla_j+\varphi_j\right)\Pi^{j}_{\hphantom{j}i}
+2\Pi^j\varpi_{ji}
+\Pi \varphi_i
\right]\right\} \nonumber\\
&&+\int \text{d}t \text{d}^dx  
 \left\{\partial_t\left[\sqrt{a} \left(\xi^{\hat t}\left(\Pi+b_i \Pi^i\right) -\xi^j b_i \Pi^{i}_{\hphantom{i}j} \right)\right] \right.
\nonumber\\
&&+\partial_i \left[\left.\sqrt{a} \Omega\left(\xi^{\hat t}\Pi^i -\xi^j \Pi^{i}_{\hphantom{i}j} \right) \right]
 \right\}.
\label{carvarS}
 \end{eqnarray}
Ignoring the boundary terms (last two lines of \eqref{carvarS}),  $\delta_\upxi S=0$ implies that the Carrollian momenta defined previously in \eqref{carvarstren}, \eqref{carvarencur}, \eqref{carvarenden}  are Carrollian-covariant, and leads to two equations. The energy equation is simple because $\xi^{\hat t}$ depends on $t$ and $\mathbf{x}$:
 \begin{equation}   
 \label{delScareneq}
\boxed{\left(\frac{1}{\Omega}\partial_t +\theta\right)\Pi
+\left(\hat \nabla_i+2\varphi_i\right)\Pi^i
+\Pi^{ij} \hat \gamma_{ij}
=0.}
\end{equation}
The momentum equation calls for a careful treatment. Indeed, $\xi^i$ is a function of $\mathbf{x}$ only, hence the factor in brackets in the second line of \eqref{carvarS} needs not vanish, but rather
  \begin{equation}   
\label{delScarmomeq}
\boxed{\left(\hat \nabla_j+\varphi_j\right)\Pi^{j}_{\hphantom{j}i}
+2\Pi^j\varpi_{ji}
+\Pi \varphi_i = -\left(\frac{1}{\Omega}\partial_t +\theta\right) P_i }
 \end{equation}
because $ \sqrt{a}\Omega \xi^{i }\left(\frac{1}{\Omega}\partial_t +\theta\right) P_i = \partial_t\left(\sqrt{a}\xi^i  P_i \right)$, 
which is a boundary term and vanishes inside the integral. 

The new vector $ P_i $, which we will refer to as \emph{momentum}, is not defined directly through a variation of the action with respect to some conjugate variable. It is  however inescapable, and this can be verified whenever a microscopic action is available in terms of fundamental fields leading to full-fledged equations of motion. In this instance, Eqs. \eqref{delScareneq} and \eqref{delScarmomeq} must be obeyed on-shell, and this procedure determines the momentum $ P_i$ (for a Carrollian scalar field see \cite{RBV22}). 

A plethora of comments and comparison to the existing literature is appropriate at this point.  The Carrollian equations at hand, which are ultimately the Carrollian fluid equations, have been the source of confusion or  misinterpretation, and unfortunately this is not fading. 

As a first and minor remark, the term in the right-hand side of \eqref{delScarmomeq} was missing in\cite{CM1,Chandrasekaran:2021hxc}.{\footnote{More precisely, Eq. \eqref{delScarmomeq} here is in disagreement with Eq. (2.30) in \cite{CM1} resulting from the invariance of the action under Carrollian diffeomorphisms. It agrees however with Eq. (4.4) in the same reference, reached via a vanishing-$c$ limiting procedure (see also our Eq. \eqref{carG}).}
More importantly, it has been claimed that both vectors, the energy current $\Pi_i $ and the momentum $ P_i$, should vanish \cite{dutch, newdutch}. Systems where this happens are not excluded, but there is no principle that demands a priori such a property -- as no reason exist for the Galilean energy current and momentum to vanish.\footnote{As an aside comment, in flat as in AdS holography, the momentum $ P_i$ of the boundary fluid is mapped onto the bulk angular-momentum aspect \cite{CMPPS2}, and this is not expected to vanish. From a different perspective, as already mentioned, microscopic systems such as the Carrollian scalar field exhibit non-vanishing  $\Pi_i $ and $ P_i$
 \cite{CM1, RBV22}.}

Another feature of the above equations \eqref{delScareneq} and \eqref{delScarmomeq} is their superficial resemblance to Eqs. \eqref{delSgaleneq} and \eqref{delSgalmomeq}. This is superficial because when coming to the genuine fluid equations, the various momenta are expressed in terms of kinematical and physical parameters, which are different in the two instances (e.g. there is no ``velocity'' in Carrollian dynamics -- see next section for details on the Carrollian fluids). But even at the superficial level, the resemblance is alleviated by the symmetries, which are undoubtedly distinct: Carrollian versus Galilean. Nonetheless, confusion has settled for good in the literature around the membrane paradigm, which was originally carefully stated \cite{Damour:1979}, but has drifted in time giving an  overwhelming importance to Navier--Stokes equations, which are Galilean par excellence, for the description of phenomena occurring in the vicinity of black-hole horizons, which call for Carrollian physics (Ref. \cite{Bredberg:2011jq} is a good example of abuse). Recently, efforts were made to clarify this issue  \cite{Penna1, Donnay:2019jiz}.
 
 \boldmath
 \subsubsection*{A $U(1)$ local invariance and conservation law}  
\unboldmath

If the action on the Carrollian manifold is further invariant under a local $U(1)$ associated with a gauge field  $\text{B}=B (t, \mathbf{x}) \text{d}t +B_i (t, \mathbf{x}) \text{d}x^i$ as in \eqref{delB}, then a further conservation is available. This sort of conservation is not as useful as it was in the Galilean framework. Indeed, the thermodynamic law \eqref{epsrho} that sets the relationship between a conserved charge and the energy (see also the discussion in Sec. \ref{comgal}) is invalidated here by the vanishing-$c$ limit, and plays no subsequent role in the fluid dynamics.

The conjugate momenta are again the \emph{charge density} and the \emph{charge current}:
\begin{eqnarray}
\label{carmomchi}
\varrho&=&\frac{1}{\sqrt{a}}\left(\frac{\delta S}{\delta B}-\frac{b_i}{\Omega}\frac{\delta S}{\delta B_i}\right),\\
\label{carmomphii}
N^{i}&=&\frac{1}{\Omega\sqrt{a}}
\frac{\delta S}{\delta B_i}
\end{eqnarray}
with $\frac{\delta S}{\delta B}=\sqrt{a}\left(\varrho+b_iN^i \right)$. 
 The gauge variation of the action is here:
 \begin{eqnarray}
  \label{delScauzer}
  \delta_\Lambda S&=&
 \int \text{d}t \text{d}^{d}x \sqrt{a} \left(
\left(\varrho+b_iN^i \right)\delta_\Lambda  B+
\Omega
N^i \delta_\Lambda  B_i \right) 
\\ 
&=&-
 \int \text{d}t \text{d}^{d}x \sqrt{a} \left( 
\left(\varrho+b_iN^i \right)  \partial_t \Lambda+
\Omega
N^i \partial_i \Lambda\right) 
\nonumber 
 \\
 &=&
 \int \text{d}t \text{d}^dx  \sqrt{a}\Omega 
\Lambda\left(\frac{1}{\Omega}\partial_t\varrho+
\theta \varrho+\left(\hat \nabla_i +\varphi_i\right)N^i
\right)\nonumber\\
&&-\int \text{d}t \text{d}^dx  
 \left\{\partial_t\left(\sqrt{a} \Lambda \left(\varrho+b_iN^i \right)  \right) +\partial_i \left(\sqrt{a}\Lambda \Omega N^i\right)
 \right\}.
 \label{U1car}
 \end{eqnarray}
Invariance of $S$ leads to a Carrollian continuity equation:
 \begin{equation}
\boxed{\left(
\frac{1}{\Omega}\partial_t+
\theta \right)\varrho
+\left(\hat \nabla_i +\varphi_i\right)N^i=0 .
}
  \label{delScarcon}
 \end{equation}
 
Using Stokes and Gauss theorems (see also footnote \ref{ddagger}) and 
the Carrollian continuity equation \eqref{delScarcon} we find
\begin{eqnarray}
\int_{\mathscr{W}} \text{d}t\text{d}^dx\, \Omega\sqrt{a}\left(\left(
\frac{1}{\Omega}\partial_t+
\theta \right)\varrho
+\left(\hat \nabla_i +\varphi_i\right)N^i\right)= \oint_{\partial\mathscr{W}}
\sqrt{a}\varrho
\text{d}x^1\wedge\ldots\wedge\text{d}x^d
\nonumber
\\
- \oint_{\partial\mathscr{W}} \sqrt{a}\sum_{i=1}^d \text{d}x^1\wedge\ldots\wedge N^i\uptheta^{\hat t}\wedge\ldots\wedge\text{d}x^d,
\label{gaustok-PR}
\end{eqnarray}
where $\mathscr{W}\subset \mathscr{M}= \mathbb{R} \times \mathscr{S}$ and $ N^i\uptheta^{\hat t}$ ($\uptheta^{\hat t}$ given in \eqref{kertdual}) is the $i$th factor in the exterior product of the last term.  Assuming a good behaviour for the fields, a conserved charge exists and 
can be expressed as an integral over an arbitrary space-like hypersurface $\upSigma_d$ of $\mathscr{M}= \mathbb{R} \times \mathscr{S}$. This conserved charge is identical to the relativistic Papapetrou--Randers result obtained e.g. in \eqref{relconchPR}. As for the Galilean instance, it is suitable to chose $\upSigma_d\equiv  \mathscr{S}$ i.e. a constant-$t$ hypersurface, and the charge then reads:\footnote{It should be noticed  that the presence of $b_i$ apparently breaks the manifest covariance, since according to \eqref{cardifb} the form of the integrand is respected only by coordinate transformations such that $t'=t'(t)$ i.e. a subset of Carrollian diffeomorphisms. This is actually innocuous, and merely translates  a feature of the hypersurface chosen for computing the charge, which is otherwise an absolute constant, as emphasized in footnote \ref{comconservation} for the relativistic case, and further discussed in Sec. \ref{FLUIDS} within the Galilean framework. If the clock form is closed (see Eq. \eqref{dthetat}), locally $\uptheta^{\hat t}=\text{d}\tau$ and one may alternatively choose the integration hypersurface $\upSigma_\tau$ as $\tau(t,\mathbf{x})$ kept constant. In this instance, we obtain $Q_N=\int_{\upSigma_\tau}\text{d}^{d}x \sqrt{a}\varrho$. Nevertheless, all choices of space-like hypersurface $\upSigma_d$ lead to the same charge.}
  \begin{equation}
Q_N=\int_{\mathscr{S}}\text{d}^{d}x \sqrt{a}\left(\varrho+b_iN^i \right).
  \label{carconch}
 \end{equation}
Following the Galilean steps, time-independence reveals by replacing $\mathscr{S}$ in \eqref{carconch} with
$\mathscr{V}\subset \mathscr{S}$, where the boundary $\partial\mathscr{V}$ does not depend on $t$, for convenience. Using \eqref{delScarcon}, the time evolution of the matter/charge content of  $\mathscr{V}$ is 
 the following:
  \begin{equation}
\frac{\text{d}}{\text{d}t}\int_{\mathscr{V}}\text{d}^{d}x \sqrt{a} \left(\varrho+b_iN^i \right)=-\int_{\mathscr{V}}\text{d}^{d}x \, 
\partial_i\left(\sqrt{a} \Omega N^i\right)
=-\int_{\partial\mathscr{V}}\Omega\star\mathbf{N} .
  \label{carcontime}
 \end{equation}
If $\mathscr{V}$ is extended to the whole $\mathscr{S}$ the time dependence fades and we find that $Q_N$ is 
  conserved.

 \subsubsection*{Isometries, conservation and non-conservation}

Isometries of Carrollian spacetimes are generated by  Killing fields of the Carrollian type \eqref{carkil}, required to obey  
\begin{equation}
\label{carkill}
\mathscr{L}_\upxi a_{ij}=0, \quad\mathscr{L}_\upxi \text{e}_{\hat t}=0,
\end{equation}
because the metric \eqref{cardegmet} and the field of observers \eqref{kert}
are the fundamental geometric 
data in the spacetimes at hand (see Refs. \cite{Duval:2014uoa,Duval:2014uva, Duval:2014lpa, Ciambelli:2019lap}).
For Carrollian diffeomorphisms ($\xi^{i}\equiv \xi^{\hat \imath}$ is only $\mathbf{x}$-dependent), Eqs. \eqref{Liedaijcar} and  \eqref{Liedcfoocar} lead to
\begin{equation}   
\label{carkilleq}
\boxed{\hat\nabla_{(i}\xi^ka_{j)k}+\xi^{\hat t}  \hat \gamma_{ij}
 =0
,\quad
\frac{1}{\Omega} \partial_t
\xi^{\hat t }+
\varphi_i \xi^i =0.}
 \end{equation}
These equations refer to the invariance of a \emph{weak Carroll structure} \cite{Duval:2014uva} and possess an infinite set of solutions. As for the Newton--Cartan case discussed in Sec. \ref{FLUIDS}, \emph{strong Carroll structures} are further equipped with a torsionless metric-compatible connection, which is also required to be invariant under Carrollian isometries. This constricts the solution space of \eqref{carkilleq}. Observe however that one does not demand the Ehresmann be invariant, hence for a Carrollian Killing field $\upxi$, using \eqref{carkilleq} inside \eqref{Liedcehrecar} we obtain:
\begin{equation}   
\label{Liedcehrecarkil}
\mathscr{L}_\upxi \uptheta^{\hat t}=\left(\left(\hat\partial_i-\varphi_i\right)\xi^{\hat t }-
2\xi^j \varpi_{ji}\right)\text{d}x^i.
 \end{equation}

The case of a Carroll spacetime with  $a^{ij}=\delta^{ij}$, $\Omega = 1$ and constant $b_i$ (standard flat Carroll spacetime with our connection) provides a nice illustration of the above. Equations \eqref{carkilleq} possess an infinite number of solutions: 
\begin{equation}   
\label{carkillflatsol}
\upxi= \left(\Omega_i^{\hphantom{i}j}x^i +X^j\right)\partial_j + f(\mathbf{x}) \partial_t
 \end{equation}
with constant and antisymmetric $\Omega_{ij}= \Omega_i^{\hphantom{i}k}\delta_{kj}$ generating the rotations in $\mathfrak{so}(d)$, constant $X^j$ for the space translations, and an arbitrary function of space $f(\mathbf{x}) $. The latter is only linear if the connection of the strong Carroll structure is required to remain invariant under $\upxi$: $f= T - B_i x^i $, $T$ generating time translations and $B_i$ being the Carroll boosts. The total number of solutions is now 
$\nicefrac{(d+2)(d+1)}{2}$,
which is the dimension of the Carroll algebra $\mathfrak{carr}(d+1)$. Besides, we find that 
\begin{equation}   
\label{carkillflatsol-lie}
\mathscr{L}_\upxi \uptheta^{\hat t}= - \left(B_i+ \Omega_i^{\hphantom{i}j}b_j\right)\text{d}x^i\neq 0, 
 \end{equation}
exhibiting a constant shift in the Ehresmann connection. 

We can now handle the conservation law that would take the Carrollian form \eqref{delScarcon} with a Carrollian scalar $\kappa$ and a Carrollian vector $K^i$ determined from the components $\xi^{\hat t}$ and $\xi^{\hat\imath}$ of a Carrollian Killing, and from the Carrollian  momenta, i.e. the energy density $\Pi$, the energy current $\Pi^i$ and the energy--stress tensor $\Pi^{ij}$ defined in Eqs. \eqref{carvarstren}, \eqref{carvarencur}, \eqref{carvarenden}, as well as the momentum $P^i$, and satisfying the conservation equations \eqref{delScarmomeq} and \eqref{delScareneq}. If such a conservation exists, the  Carrollian scalar
 \begin{equation}
\mathcal{K}= \left(
\frac{1}{\Omega}\partial_t+
\theta \right)\kappa
+\left(\hat \nabla_i +\varphi_i\right)K^i
 \label{kilcarcon}
 \end{equation}
shall vanish on-shell. In fine, $\kappa$ and $K^i$ are disclosed in the on-shell boundary terms of  $\delta_\upxi S$ (see \eqref{carvarS}) -- or likewise,  inherited from the relativistic-current components \eqref{restIinvcar}, \eqref{inviotacar}, \eqref{invicar}:\footnote{We anticipate here Sec. \ref{carfluids2}, where Eqs. \eqref{jexpCextinv} are obtained as a small-$c$ expansions of
 \eqref{inviotacar} and \eqref{invicar}, leading to \eqref{carkappaK}, which includes  \eqref{invkappacar} and \eqref{invKcar}.\label{car-con-curr}}
\begin{eqnarray}
\label{invkappacar}
\kappa&=& \xi^{i} P_i-\xi^{\hat t}  \Pi,\\
\label{invKcar}
K^i&=&\xi^{j}\Pi_{j}^{\hphantom{j}i}-
 \xi^{\hat t} \Pi^i.
\end{eqnarray}
The scalar  $\mathcal{K}$ can be determined using the conservation equations \eqref{delScarmomeq} and \eqref{delScareneq}:
 \begin{eqnarray}
\label{carcon}
\mathcal{K}&=&- \Pi\left(
\frac{1}{\Omega} \partial_t
\xi^{\hat t }+
\varphi_i \xi^i 
\right)
-\Pi^i 
\left(
\left(\hat\partial_i-\varphi_i\right)\xi^{\hat t }-
2\xi^j \varpi_{ji}
\right)
+\Pi^{i}_{\hphantom{i}j}\left(
\hat\nabla_{i}\xi^j+\xi^{\hat t}  \hat \gamma_{i}^{\hphantom{i}j}
\right)
\\
\label{carnoncon}
&=&-\Pi^i 
\left(
\left(\hat\partial_i-\varphi_i\right)\xi^{\hat t }-
2\xi^j \varpi_{ji}
\right)
.
\end{eqnarray}
We have obtained \eqref{carnoncon} from \eqref{carcon} thanks to  the Killing equations \eqref{carkilleq}. This latter result shows that \emph{in Carroll structures, a Killing field does not guarantee an on-shell conservation law for Carrollian dynamics}.\footnote{Credit should be given to the authors of  \cite{CM1} for observing this phenomenon in a quite general framework.}

The above result is expected as in the Galilean case, where a similar non-conservation was proven: the energy current $\Pi^i$ is conjugate to $b_i$  \eqref{carvarencur} and $b_i$ does transform under diffeomorphisms (see \eqref{Liedwiomcar}), \emph{even when this diffeomorphism is an isometry}. Nonetheless, Eq.  \eqref{carnoncon} infers that a conservation law exists for the subalgebra of Killing vectors such that $\mathscr{L}_\upxi \uptheta^{\hat t}= 0$,\footnote{The Jacobi identity is used to show that  the commutator of two $\upxi $s obeying $\mathscr{L}_\upxi \uptheta^{\hat t}= 0$, satisfies the same condition.}  in agreement with general N\oe ther's theorem, for which isometry seems insufficient and a stronger symmetry required. Incidentally, one cannot exclude that the right-hand side of \eqref{carnoncon} originates from a boundary term
(as for the corresponding Galilean equation \eqref{galnoncon}). Conservation would then occur -- put differently  \eqref{kilcarcon} would  vanish with effective Carrollian curent $\kappa'$ and $K^{i\prime}$, amended by the boundary-term contributions. 

As we have emphasized slightly above, \emph{ordinary Carrollian boosts in flat Carroll structures ($a_{ij}=\delta_{ij}$, $\Omega=1$, $b_i$ constants)  do not satisfy the  extra condition $\mathscr{L}_\upxi \uptheta^{\hat t}= 0$ (see \eqref{carkillflatsol-lie}), and thus no conservation law is necessarily associated with them}. This property was disregarded in Refs. \cite{dutch, dutch2, newdutch}, where the authors took for granted that such a conservation should exist in the primitive sense i.e. with $\mathcal{K}=0$ in \eqref{carnoncon}.\footnote{Notice that conservation might be valid with non-vanishing $\mathcal{K}$, if the latter is a generalized divergence. In flat space we find $\mathcal{K}=\Pi^i\left(B_i+ \Omega_i^{\hphantom{i}j}b_j\right)$ using  Eqs. \eqref{carkillflatsol} and \eqref{carkillflatsol-lie}. Under the assumption of \emph{potential flow} (we borrow the Galilean language) $\Pi_i = \hat\partial_i \phi + \partial_t \phi_i$ with $\phi(t,\mathbf{x})$ and $\phi_i(t,\mathbf{x})$ the potential  functions so that $\mathcal{K}=\hat \partial_i \phi U^i + \partial_t \phi_i U^i$ with $U^i = B^i+ \Omega^{ij} b_j$. In this very specific instance, there is a conserved Carrollian current associated with boosts and rotations, and components $\kappa -\phi_iU^i $ and $K^i - \phi U^i $ making \eqref{kilcarcon} vanish. \label{57}}  This assumption led to the conclusion that $\Pi^i$ should always vanish, as we have already pointed out. 
A similar reasoning  in the Galilean case (see \eqref{galnoncon})  amounts to stating that the fluid momentum $P_i $ ought to vanish. This is a notoriously degenerate state, where either the fluid is absent, or motion is absent -- global equilibrium is reached. Having no  
intuition for Carrollian fluids, we leave open a teensy possibility for a state with vanishing energy flux to make sense. Such a state is by no means a consequence of any spacetime symmetry though.\footnote{Situations of this sort are not forbidden but are not demanded a priori. They may occur outside the realm of fluids. For instance, a Carrollian scalar field has always vanishing energy flux in its electric edition, and can be set to zero in magnetic some configurations \cite{newdutch, CM1, RBV22}. However, this flux is generically non-zero in the magnetic version, although configurations do exist for which it vanishes.} This feature will be recast at the end of Sec. \ref{carfluids2}, where besides the Carrollian current \eqref{invkappacar},  \eqref{invKcar}, more currents of the same sort appear, which may or may not be conserved as a consequence of the vanishing of a vector, be it the energy flux $\Pi^i$ met here, or another vector emerging in the small-$c$ expansion of the relativistic heat current.

\subsubsection*{Weyl invariance, conformal isometries, conservation and non-conservation}

As for the Newton--Cartan spacetimes, Weyl transformations can be investigated in Carrollian manifolds. They
act on their basic geometric data as 
\begin{equation}
\label{weyl-geometry}
a_{ij}\to \frac{1}{\mathcal{B}^2}a_{ij},\quad \Omega\to \frac{1}{\mathcal{B}}\Omega,\quad b_{i}\to \frac{1}{\mathcal{B}}b_{i},
\end{equation} 
where $\mathcal{B}=\mathcal{B}(t,\mathbf{x})$ is an arbitrary function. The Carrollian momenta $\Pi^{ij}$, $\Pi^i$ and $\Pi$ defined in 
\eqref{carvarstren}, \eqref{carvarencur} and  \eqref{carvarenden} inherit conformal weights $d+3$, $d+2$ and $d+1$ when the effective action is presumed Weyl-invariant. The momentum $P^i$ appearing in \eqref{delScarmomeq} has also weight $d+2$, and in the matter sector, assuming the gauge field $B$ and $B_i$ be weight-zero, we conclude from \eqref{carmomchi} and \eqref{carmomphii} that the density $\varrho$ and the matter current $N^i$ have weights $d$ and $d+1$.

Requiring Weyl invariance for the effective action  $\delta_{\mathcal{B}}S=0$, expression \eqref{delSgravRP} 
implies that 
\begin{equation}
\label{car-conf-cond}
\Pi_i^{\hphantom{i}i}=\Pi.
\end{equation} 
In order to implement elegantly Weyl covariance,  the appropriate covariant derivatives will be introduced in Sec. \eqref{carfluids} for time and space, dubbed Weyl--Carroll.  For the moment, we wish to circumscribe our discussion and adapt to Carroll 
the pattern discussed in Sec. \ref{FLUIDS} for Newton--Cartan spacetimes, Weyl-invariant dynamics and conformal isometries. This will lead to the same conclusion as above:  \emph{a conformal Killing field does not always provide a conservation law in Weyl-invariant Carrollian  dynamics}.

Following \cite{Duval:1990hj, Duval:2014uoa,  Duval:2014uva,  Duval:2014lpa, Ciambelli:2019lap} a conformal isometry is generated by a vector field $\upxi$ satisfying
\begin{equation}
\label{PRkillconf}
\mathscr{L}_\upxi a_{ij}=\lambda a_{ij},
\end{equation}
where
\begin{equation}
\label{carkilleqconf}
\lambda(t,\mathbf{x}) =\frac{2}{d}
\left(\hat  \nabla_{i}\xi^{i}
+\theta \xi^{\hat t}  
\right).
 \end{equation}
The extra condition imposed for reaching an operational definition of conformal Killing vectors is again  \eqref{extra-conf-cond} i.e. 
$2\mu + \lambda=0$ with (see \eqref{Liedcfoocar})
\begin{equation}
\label{PRkillconfmu}
\mu(t,\mathbf{x})= -\left(\frac{1}{\Omega} \partial_t
\xi^{\hat t }+
\varphi_i \xi^i \right) .
 \end{equation}
A dynamical exponent $z$, here equal to $1$, can also be defined (see footnote \ref{dynexp}), and the distinction of weak versus strong Carroll structures supplements the discussion on the web of conformal Killing fields. 

Assuming the existence of a conformal isometry,  the conservation equations \eqref{delScareneq}  and \eqref{delScarmomeq} can be used for computing the Carrolian scalar $\mathcal{K}$  \eqref{kilcarcon} with \eqref{invkappacar},  \eqref{invKcar} and \eqref{car-conf-cond}:
\begin{equation}
\label{carconconf}
\mathcal{K}= \Pi\left(\frac{\lambda}{2}+\mu\right)
-\Pi^i\left(\left(
\hat\partial_i-\varphi_i\right)\xi^{\hat t }-
2\xi^j \varpi_{ji}
\right).
\end{equation}
The defining equation \eqref{extra-conf-cond} for conformal Killing vectors on Carrollian spacetimes expectedly arises in \eqref{carconconf}, but is insufficient to ensure $\mathcal{K}=0$. As anticipated, \emph{a plain conformal Killing field does not generically provide a conservation law in Weyl-invariant Carrollian dynamics}.

The same conclusion has been reached for Galilean spacetimes in Sec. \ref{FLUIDS}. It will be further investigated from the small-$c$ viewpoint in the following paragraphs, and finally in App. \ref{varprinccar}.

As an example, let us again consider the standard flat Carroll spacetime ($a^{ij}=\delta^{ij}$, $\Omega = 1$ and constant $b_i$). Equations \eqref{PRkillconf} and \eqref{extra-conf-cond} possess an infinite number of solutions, which for a strong Carroll structure read \cite{Duval:2014uva, Duval:2014lpa, Ciambelli:2019lap}: 
\begin{equation}   
\label{carkillflatsolcon}
\upxi= Y^j(\mathbf{x})\partial_j + \left(T(\mathbf{x})+\frac{t}{d}\partial_iY^i\right)\partial_t
 \end{equation}
with $T(\mathbf{x})$ an arbitrary function generating the \emph{supertranslations} and $Y^i(\mathbf{x})\partial_j $ being  conformal Killing fields of Euclidean $d$-dimensional space, generating $\mathfrak{so}(d+1, 1)$. This is the conformal Carroll algebra $\mathfrak{ccarr}(d+1)\equiv\mathfrak{so}(d+1,1)\loplus \text{supertranslations}$, also known as $\text{BMS}_{d+2}$ for Bondi--van der Burg--Metzner--Sachs) \cite{Duval:2014uva,Duval:2014lpa}.\footnote{In the presence of a dynamical exponent defined via $2\mu +z \lambda=0$ (in this case the Carrollian structure is not inherited from the Carrollian limit of a pseudo-Riemannian spacetime), the algebra exhibits a level $N=\nicefrac{2}{z}$: $\mathfrak{ccarr}_{N}(d+1)$. Strictly speaking $\mathfrak{ccarr}(d+1)\equiv \mathfrak{ccarr}_{2}(d+1)$ is $\text{BMS}_{d+2}$ for $d=1,2$ only, because for higher $d$ the BMS algebra is finite-dimensional, whereas $\mathfrak{ccarr}_{N}(d+1)$ is not. Infinite-dimensional extensions of the BMS algebra have been nevertheless presented in the literature  (see e.g. \cite{Campoleoni:2020ejn} for a recent account and further references).} We also find how the clock form behaves:
\begin{equation}   
\label{carkillflatsol-lie-con}
\mathscr{L}_\upxi \uptheta^{\hat t}= \left(\partial_i\left(T-Y^jb_j\right)
+\frac{b_i}{d}\partial_jY^j +\frac{t}{d}\partial_i\partial_jY^j
\right)\text{d}x^i.
 \end{equation}
The associated current is not conserved since $\mathcal{K}$ in \eqref{carconconf} does not generically vanish,\footnote{For general Carrollian fluids it has no a priori reason to vanish, and no N\oe ther current exists. Nonetheless, one finds explicit dynamics where these currents are enforced (see e.g.  \cite{Bagchi:2019clu} for the scalar electrodynamics), and where the full conformal Carroll group is realized  in terms of the associated conserved charges.}  
 unless $\partial_jY^j= C_0$ and $T=T_0+Y^jb_j -\frac{C_0}{d}b_i x^i$, thus linear in $x^i$ ($C_0$ and $T_0$ are constants). This excludes the $d$ special conformal transformations of $\mathfrak{so}(d+1,1)$ and leaves the supertranslations with the time translation as  unique freedom, leading to a symmetry subgroup of finite dimension $\frac{d^2}{2}+\frac{d}{2}+2$.

One should stress again that generally, $\Pi^i\left(\left(\hat\partial_i-\varphi_i\right)\xi^{\hat t }-2\xi^j \varpi_{ji} \right)$ with $\upxi$ a conformal Killing field of a Carrollian manifold $\mathscr{M}$, might be a boundary term possibly leading to a conserved Carrollian current. This needs however to be appreciated case-by-case and not premised.

\subsection{Carrollian hydrodynamics from relativistic fluids -- I}  \label{carfluids}

\boldmath
\subsubsection*{The small-$c$ expansion}
\unboldmath

Following the pattern introduced in \cite{CMPPS1}, we will now study the vanishing-$c$ limit of relativistic hydrodynamics on a pseudo-Riemannian manifold, in Papapetrou--Randers coordinates.  As for the Galilean case, all $c$-dependence in the geometry is explicit.  In particular, the fluid velocity is parameterized with $\beta^i$ introduced in \eqref{vbetacar}, and for small $c$ we obtain:
 \begin{equation}
\label{carbetav}
v^i=c^2 \Omega \beta^i + \text{O}\left(c^4\right).
\end{equation}
The fluid velocity vanishes at zero $c$ -- this is not a surprise -- but a kinematical parameter with dimensions of an inverse velocity is bound to remain in as a Carrollian-fluid variable. The full fluid congruence then reads:
\begin{equation}
\label{carvel}
u_0=-c\Omega+\text{O}\left(c^3\right),
\quad
u^i=c^2\beta^i+\text{O}\left(c^4\right),
\end{equation}
whereas the expansion and the shear behave as
\begin{eqnarray}
\label{caracclim} 
\label{carexp} 
&&\Theta=
\frac{1}{\Omega}
\partial_t \ln\sqrt{a}+\text{O}\left(c^2\right)
=\theta+\text{O}\left(c^2\right),
\\
\label{carsiglim} 
&&\sigma^{ij}=-\frac{1}{\Omega}\left(\frac{1}{2} \partial_t a^{ij}+
\frac{1}{d} a^{ij} \partial_t \ln\sqrt{a}\right)+\text{O}\left(c^2\right)
=
\xi^{ij}+\text{O}\left(c^2\right)
\end{eqnarray}
with $\xi^{ij}$ and $\theta$ defined for a Carrollian manifold in \eqref{dgammaCartime} and \eqref{carexp-tempcon}.

In order to go on with the fluid equations, we should handle the behaviour of the energy--momentum tensor at small $c$. This includes $\varepsilon$, $p$ $q^i$ and $\tau^{ij}$. The matter current $j^i$, if present, plays no role since no relationship amongst $\varepsilon$ and $\varrho $ survives in the Carrollian limit. In the absence of thermodynamic or transport hints for the behaviour of these quantities, we will consider an ansatz motivated by the Carrollian fluids emerging in flat holography \cite{CMPPS2}. The simplest is 
\begin{eqnarray}
\varepsilon &=&\eta + \text{O}\left(c^2\right),
\label{eexpC}
\\
p &=&  \varpi+ \text{O}\left(c^2\right),
\label{pexpC}
\\
\label{QexpC}
q^i&=&Q^{i}+{c^2}\pi^i + \text{O}\left(c^4\right),
\\
\label{sigexpC}
\tau^{ij}&=&-\Xi^{ij} + \text{O}\left(c^2\right).
\end{eqnarray}
This follows the pattern of the Galilean counterpart \eqref{sigexpG} and \eqref{qgalun}, except that the energy is now of order $1$, as for the case of a massless-carrier Galilean fluid. In Sec. \ref{carfluids2} we will consider a Laurent expansion with order-$\nicefrac{1}{c^2}$ terms,  as required in flat-holography fluids.

Although a conserved current is not essential in the discussion, we present it for completeness with the following ansatz 
\begin{equation}
\label{jexpC}
\varrho_0=\chi + \text{O}\left({c^2}\right),\quad
j^k=n^k+ \text{O}\left({c^2}\right).
\end{equation}
We recall that $\varrho_0$ is the proper density i.e. the density measured by an observer with velocity $u^\mu$. We could consider a fiducial observer, who would play here the role of $\text{u}_{\text{Z}}$ in Zermelo frame:
\begin{equation}
\label{fidRP}
\text{u}_{\text{PR}}
=\frac{1}{\Omega}\partial_t.
\end{equation}
This observer is not geodesic (unless $\partial_t$ it is a Killing field), but this is of secondary importance since inertial frames play no role in the Carrollian limit. For this observer, the fluid density is $ -\frac{1}{c^2} J_\mu u^\mu_{\text{PR}}=\frac{c}{ \Omega}J_0$, which coincides with   $\varrho_{0\text{r}} $ given in \eqref{invrhoRP}.

\subsubsection*{Carrollian momenta and hydrodynamic equations}

With the data \eqref{eexpC}, 
\eqref{pexpC}, 
\eqref{QexpC}, \eqref{sigexpC},
the invariant pieces of the relativistic energy--momentum tensor defined in \eqref{restTinvRP}, 
\eqref{invenRP}, \eqref{invqRP} and \eqref{invtauRP} read:
\begin{eqnarray}
\label{limcarT0i}
q^i_{\text{r}}&=&
\Pi^{i}+c^2  P^{i}+\text{O}\left(c^4\right),
\\
\label{limcarT00}
\varepsilon_{\text{r}}&=&
\Pi +\text{O}\left(c^2\right),
\\
\label{limcarTij}
p_{\text{r}} a^{ij} +\tau_{\text{r}}^{ij} &=&
\Pi^{ij}
+ \text{O}\left(c^2\right)
\end{eqnarray}
with
\begin{equation}
\Pi=
\eta
+2\beta_iQ^{i}, \quad 
\Pi^{i}
=Q^{i},\quad
\Pi^{ij}
=Q^{i}\beta^j+\beta^iQ^{j}+\varpi a^{ij}-\Xi^{ij},
\label{carflux}
\end{equation}
and 
\begin{equation}
 P^{i}
=\pi^{i}
+\beta^i \left(\eta+\varpi+\beta_kQ^{k}\right)-\beta_k\Xi^{ki}+\frac{\pmb{\beta}^2}{2}Q^{i}.
\label{carheatflux}
\end{equation}

Equations \eqref{conT} with the energy--momentum tensor at hand demand the following expressions
 \begin{eqnarray}
 \label{conTcarexp0} 
\frac{c}{\Omega}\nabla_\mu T^{\mu}_{\hphantom{\mu}0}&=&
{\mathcal{E}}
+\text{O}\left(c^2\right),
\\
\label{conTcarexpi} 
 \nabla_\mu T^{\mu i}&=&\frac{1}{c^2} \left\{
\left(\frac{1}{\Omega}\hat D_t +\theta
\right)\Pi^{i}+\Pi^j\hat \gamma^{\hphantom{j}i}_j
\right\}
 +\mathcal{G}^i
 +\text{O}\left(c^2\right) 
\end{eqnarray}
be zero with 
 \begin{eqnarray}
\mathcal{E}&=&-  \left(\frac{1}{\Omega}\hat D_t +\theta\right) \Pi
 -\left(\hat\nabla_i +2\varphi_i \right)\Pi^{i}-\Pi^{ij}\hat \gamma_{ij} ,
  \label{carEbis} 
  \\
  \mathcal{G}_j&=&
\left(\hat \nabla_i+\varphi_i\right)\Pi^{i}_{\hphantom{i}j}
+2\Pi^i\varpi_{ij}
+\Pi \varphi_j +\left(\frac{1}{\Omega}\hat D_t +\theta
\right) P_{j}+ P^i \hat \gamma_{ij}.
  \label{carG} 
\end{eqnarray}
Hence, we recover the Carrollian momenta conservation equations \eqref{delScareneq} and \eqref{delScarmomeq}
(notice the use of  the Carrollian time covariant derivative \eqref{Cartimecovdervecform})
\emph{augmented with an extra equation on the energy current}
\begin{equation}
\left(\frac{1}{\Omega}\hat D_t +\theta
\right)\Pi^{i}+\Pi^j\hat \gamma^{\hphantom{j}i}_j=0.
\label{heatfluxextraeq}
\end{equation}

The whole Carrollian scheme of the present chapter, and Eq. \eqref{heatfluxextraeq} in particular, resonate with the discussion made on the Galilean side, Sec. \ref{comgal}. This equation is indeed absent when working directly in the framework of a Carrollian manifold with Carrollian diffeomorphisms, as in Sec. \ref{cardifcons}. It is in fact a boundary term that could not be retrieved by a variational principle -- as the Carrollian momentum $ P^i$ appeared to be necessary but undetermined. Getting the Carrollian dynamics as a zero-$c$ limit of relativistic hydrodynamics is richer. 

On the one hand, equation  \eqref{heatfluxextraeq} emerges as a vestige  of the original full-diffeomorphism relativistic invariance,\footnote{Precise statements on this reminiscence of full diffeomorphisms are illusive. In particular, no central extension can accompany the Carrollian contraction. According to Ref. \cite{bacryLLinfmas}, the prerequisite for this to occur is the existence of an absolute time i.e. Galilean or Aristotelian frameworks. \label{centrext}} and is the dual of the Galilean constraint equation \eqref{extra-co-gal} -- remember that time and space play dual roles in the two limits considered here, interchanging, among others,  momentum and energy current. On the other hand, the momentum $ P^i$ is no longer undetermined, and stands for the subleading term of the relativistic heat current \eqref{limcarT0i}, expressed explicitly in terms of the kinematical and ``thermodynamic--transport'' observables -- $\beta^i$ and $\eta$, $\varpi$, $Q^i$, $\pi^i$, $\Xi^{ij}$ in \eqref{carheatflux}.

This last observation calls for a comparison with the existing literature. At the first place it should be accepted that the very concept of spacetime energy--momentum tensor is loose in Carrollian (and Galilean) physics. This was clearly emphasized in the early work \cite{CM1}, where the role of Carrollian momenta as a necessary replacement to the energy--momentum tensor was proposed. It is stated again here, and one should moreover avoid the latent confusion amongst momenta (or energy--momentum tensor, if any) as a response of the system to geometry disturbances, 
and N\oe therian  currents, which are generically absent as no isometries have been assumed. Next, it is clear that nothing requires the vanishing of neither the energy flux $\Pi^i $, nor the momentum $ P^i$, irrespective of the approach -- Carrollian conservation or zero-$c$ limit. This has been undermined in \cite{dutch, dutch2, newdutch}. Finally, the momenta are expressed in terms of \emph{an ``inverse velocity'' $\beta^i$ and not a velocity $v^i$}, since no velocity is compatible with Carrollian physics due to the shrinking of the light cone. It should be added that no matter density enters the momenta because no relationship exists any longer between energy and mass. Equation \eqref{epsrho} is obsolete in the Carrollian limit, and it is fair to admit that Carrollian thermodynamics remains in limbo -- as pointed out in App. \ref{thermo}. This is sometimes overlooked.

From a more abstract viewpoint, a relation between energy and some other charge (possibly, but not necessarily the mass) might appear only if a conservation exists that defines this charge, independently of the conservation involving the energy. Such a conservation may or may not be present for the relativistic fluids, and is emergent in the Galilean limit. For the Carrollian case, if a $U(1)$ conservation law of the type \eqref{conJ} is available in the ascendent theory, we find, after inserting \eqref{jexpC} inside \eqref{invrhoRP} and \eqref{invjRP}:
\begin{eqnarray}
\label{invrhoRPexp}
\varrho_{0\text{r}}&=&\varrho +\text{O}\left(c^2\right) ,\\
\label{invjRPexp}
j^{i}_{\text{r}}&=&N^{i}+\text{O}\left(c^2\right),
\end{eqnarray}
with
\begin{equation}
 \label{Careff} 
\varrho=\chi +\beta_i n^i,\quad
N^i=n^i,
\end{equation}
the matter Carrollian momenta explicitly determined in terms of $\beta^i$. We can now compute the 
divergence of \eqref{curdec} in the Papapetrou--Randers background \eqref{carrp}. The result is
 \begin{equation}
 \label{conJcarexp} 
\nabla_\mu J^{\mu}= {\mathcal{J}}
+\text{O}\left(c^2\right)
\end{equation}
with 
 \begin{equation}
\mathcal{J}=
 \left(\frac{1}{\Omega}\partial_t +\theta\right)\varrho+\left(\hat\nabla_j +\varphi_j \right)N^j,
 \label{carJ} 
\end{equation}
and demanding the conservation, we recover the  Carrollian continuity equation \eqref{delScarcon}.

\subsubsection*{Hydrodynamic-frame invariance}

We discussed hydrodynamic-frame invariance in the framework of relativistic fluids, where it states that 
the relativistic fluid equations remain invariant under arbitrary unimodular transformations of the velocity field $\text{u}$, performed together with transformations of the energy density, pressure, heat current and stress tensor -- see e.g.
\eqref{delepsRP}, \eqref{delqRP} and \eqref{deltau2RP}.\footnote{In the Papapetrou--Randers frame, the local unimodular transformations \eqref{loclor} are captured by $\beta^i\to \beta^i + \delta \beta^i(t,\mathbf{x})$ (see Eqs. \eqref{vel}), \eqref{vbetacar}, \eqref{gammabeta}) parameterized as $ \delta \beta^i= B^i -c^2B^j \beta_j \beta^i + \Omega^{ij}\beta_j$. Infinitesimal Lorentz boosts are associated with $B^i(t,\mathbf{x})$, while infinitesimal rotations go along with  the antisymmetric $\Omega^{ij}(t,\mathbf{x})$. In the Carrollian limit, the general transformation, which captures Carrollian boosts and rotations, reads: $ \delta \beta^i= B^i + \Omega^{ij}\beta_j$.} The question we want now to answer is again whether this invariance survives the Carrollian limit.

From the experience we have acquired in the Galilean chapter, answering requires a careful analysis, and the output depends on several options. Assumptions are made about the small-$c$ behaviour of the various observables (see \eqref{eexpC}, 
\eqref{pexpC}, 
\eqref{QexpC}, 
and \eqref{sigexpC}) and this behaviour may not be stable under Carrollian hydrodynamic-frame transformations. As opposed to the Galilean situation, there is no physical intuition that can argue in favour or against. There are however concrete results from flat holography \cite{CMPR, CMPRpos} suggesting that Carrollian hydrodynamic-frame invariance should hold as a local boundary symmetry, translating in a bulk diffeomorphism transformation.

As for the Galilean case,  the operators entering \eqref{carEbis}, \eqref{carG} and \eqref{heatfluxextraeq} are velocity-independent, and the momenta  $ P_i$, $\Pi$, $\Pi_i$ and $\Pi_{ij}$ appear as coefficients in the expansion of the hydrodynamic-frame-invariant relativistic momenta \eqref{limcarT0i}, \eqref{limcarT00} and \eqref{limcarTij}. In order to conclude about hydrodynamic-frame invariance, we must investigate the stability of the scaling properties encoded in \eqref{eexpC}, 
\eqref{pexpC}, 
\eqref{QexpC} and \eqref{sigexpC},  using the transformation rules set in the Papapetrou--Randers frame \eqref{delepsRP},  \eqref{delqRP} and \eqref{deltau2RP}. We find the following:
\begin{eqnarray}
\label{delepsRPC}
\delta \eta&=& -2 \delta\beta_iQ^{i} ,
\\
\label{delQRPC}
\delta Q^{i}&=&0,
\\
\label{delpiRPC}
\delta \pi^i&=&\delta\beta_j\left(\Xi^{ij}-(\eta+\varpi) a^{ij}+ \beta^i Q^{j}\right),
\\
\delta \left(\Xi^{ij}-\varpi a^{ij}\right)&=&\delta\beta_k\left(
Q^{i} a^{jk}+Q^{j} a^{ik}\right).
\label{delxitauRPC}
\end{eqnarray}
Remarkably, under the frame transformations at hand, the Carrollian densities and fluxes defined in \eqref{carflux} and \eqref{carheatflux}  are invariant:
\begin{equation}
\delta\Pi=
0, \quad 
\delta\Pi^{i}
=0,\quad
\delta\Pi^{ij}
=0,\quad 
\delta P^{i}
=0.
\label{delcarflux}
\end{equation}

Contrary to the Galilean massive case (Sec. \ref{galdyn}), but similarly to the Galilean massless case (Sec. \ref{conFLUIDS}),  the energy--momentum-tensor dynamics and the current dynamics are decoupled here. This decoupling holds in particular for hydrodynamic-frame invariance, and we are invited to iterate the above course for matter dynamics. For matter, the transformation rules in Papapetrou--Randers frame are 
\eqref{delrho0RP} and \eqref{deljRP},  whereas the invariant relativistic momenta \eqref{invrhoRPexp}, \eqref{invjRPexp} should be used in conjunction with the small-$c$ behaviour \eqref{jexpC}. The output is now
\begin{eqnarray}
\label{delrho0RPC}
\delta \chi&=& -\delta \beta_i n^i,
\\\label{deljRPC}
\delta  n^{i}&=& 0
.
\end{eqnarray}
Using \eqref{Careff}, one shows that
\begin{equation}
\label{delCareff}
\delta\varrho= 0,
\quad  
\delta N^i=0
.
\end{equation}
This result demonstrates the invariance of  \eqref{carJ}. In conclusion, \emph{matter dynamics is hydrodynamic-frame-invariant}, establishing thereby that \emph{Carrollian fluid dynamics is hydrodynamic-frame invariant}.

\subsubsection*{Weyl-invariant Carrollian fluids}

Carrollian fluids are fundamental ingredients of flat holography, where they appear in their Weyl-invariant version \cite{CMPPS2}.  On a Papapetrou--Randers frame,  Weyl transformations generated by $\mathcal{B}(t,\mathbf{x})$ act as they do on the fundamental data of a Carrollian spacetime, Eqs. \eqref{weyl-geometry}. The Weyl-invariance condition \eqref{car-conf-cond} is here reached as the zero-$c$ limit of the relativistic Weyl-invariance condition $T^\mu_{\hphantom{\mu}\mu}=0$ discussed at the end of Sec. \ref{rreleq}, using \eqref{restraceinvRP}, 
\eqref{limcarT0i}, \eqref{limcarT00} and \eqref{limcarTij}. For Carrollian fluids, the various Carrollian momenta \eqref{carflux}, \eqref{carheatflux}, \eqref{Careff} are expressed in terms of fluid variables such as the inverse velocity $\beta^i$, the energy density $\eta$, the pressure $\varpi$ as well as $Q_i $, $\pi_i$ and $\Xi_{ij}$. Their conformal weights are\footnote{We mentioned in Sec. \ref{cardifcons} that $\Pi^{ij}$, $\Pi^i$, $\Pi^i$ and $\Pi$ have conformal weights $d+3$, $d+2$ , $d+2$ and $d+1$, whereas the density $\varrho$ and the matter current $N^i$ have weights $d$ and $d+1$.} $1$, $d+1$, $d+1$, $d$, $d$ and $d-1$. Similarly the weights of $\chi $ and $n^i$ are $d$ and $d+1$. Condition \eqref{car-conf-cond} reads $\eta=d\varpi -\Xi^i_{\hphantom{i}i}$, which is split as usual:
\begin{equation}\label{conn}
\eta= d\varpi,\quad \Xi^i_{\hphantom{i}i}. 
\end{equation}

The geometric tools necessary for handling Carrollian Weyl covariance were introduced in App. A.2 of \cite{CMPPS1} and we summarize them here.
We define \emph{Weyl--Carroll} covariant time and space derivatives using $\theta$ and  $\varphi_i$ defined in \eqref{carexp-tempcon}
and
\eqref{caromacc}, 
which transform as connections
 \begin{equation}
 \label{weyl-geometry-2-abs}
\theta\to \mathcal{B}\theta-\frac{d}{\Omega}\partial_t \mathcal{B}
  ,\quad 
\varphi_{i}\to \varphi_{i}-\hat\partial_i\ln \mathcal{B},
\end{equation} 
as opposed to the Carrollian shear $\xi_{ij}$ \eqref{dgammaCartime} and Carrollian vorticity $\varpi_{ij}$ \eqref{carconcomderf}, which are Weyl-covariant of weight $-1$. The action of the Carrollian Weyl-covariant time  derivative on a weight-$w$ function $\Phi$ is 
\begin{equation}
\label{CWtimecovdersc}
\frac{1}{\Omega}\hat{\mathscr{D}}_t \Phi=\frac{1}{\Omega}\hat D_t \Phi +\frac{w}{d} \theta \Phi=
\frac{1}{\Omega}\partial_t \Phi +\frac{w}{d} \theta \Phi,
\end{equation}
and this is a scalar of weight $w+1$. On a weight-$w$ vector, the action is 
\begin{equation}
\label{CWtimecovdervecform}
\frac{1}{\Omega}\hat{\mathscr{D}}_t V^l=\frac{1}{\Omega}\hat D_t V^l +\frac{w-1}{d} \theta V^l
=\frac{1}{\Omega}\partial_t V^l +\frac{w}{d} \theta V^l+\xi^{l}_{\hphantom{l}i} V^i .
\end{equation}
These are the components of a Carrollian vector of weight $w+1$. Similarly for any tensor by Leibniz rule and in particular we find:
\begin{equation}
\label{CWt-met}
\hat{\mathscr{D}}_t a_{kl}=0.
\end{equation}

For a weight-$w$ scalar function  $\Phi$,
we introduce the space Weyl-covariant Carrollian derivative
\begin{equation}
\label{CWs-Phi}
\hat{\mathscr{D}}_j \Phi=\hat\partial_j \Phi +w \varphi_j \Phi,
\end{equation}
which has the same conformal weight. Similarly, for a vector with weight-$w$ components $V^l$:
\begin{equation}
\hat{\mathscr{D}}_j V^l=\hat\nabla_j V^l +(w-1) \varphi_j V^l +\varphi^l V_j -\delta^l_j V^i\varphi_i.
\end{equation}
The Weyl--Carroll spatial derivative does not modify the weight of the tensor it acts on.
The action on any other tensor is obtained using the Leibniz rule, as in example for rank-two tensors:
\begin{equation}
\hat{\mathscr{D}}_j t_{kl}=\hat\nabla_j t_{kl} +(w+2) \varphi_j t_{kl} +\varphi_k t_{jl}+ \varphi_l t_{kj} -
a_{jl}t_{ki}\varphi^i
-
a_{jk}t_{il}\varphi^i.
\end{equation}
Moreover, it is metric-compatible:
\begin{equation}
\hat{\mathscr{D}}_j a_{kl}=0.
\end{equation}

Time and space Weyl--Carroll covariant derivatives do not commute. Their commutators allow to define further geometric tensors such as the Weyl--Carroll curvature (spatial and mixed space--time), which do also emerge in the small-$c$ expansion of the relativistic Weyl curvature tensors introduced in \eqref{relweylcurv}, \eqref{curlRic}, \eqref{curlRc}, and evaluated in a Papapetrou--Randers background. More information is available in the already quoted reference \cite{CMPPS1}.

With these derivatives, Carrollian equations \eqref{delScareneq}, \eqref{delScarmomeq} and equation  \eqref{heatfluxextraeq} 
read for a conformal fluid:
\begin{eqnarray}
\frac{1}{\Omega}\hat{\mathscr{D}}_t\Pi
+\hat{\mathscr{D}}_i \Pi^{i}
+\Pi^{ij}\xi_{ij}&=&0,
 \label{carEcon} 
\\
\hat{\mathscr{D}}_i \Pi^{i}_{\hphantom{i}j}+2\Pi^{i}\varpi_{ij}+ \left(\frac{1}{\Omega}\hat{\mathscr{D}}_t \delta^i_j +\xi^{i}_{\hphantom{i}j}\right) P_i &=&0,
  \label{carGcon0}\\
\frac{1}{\Omega}\hat{\mathscr{D}}_t 
\Pi_{j}
+ \Pi_{i}\xi^{i}_{\hphantom{i}j}&=&0.
 \label{heatfluxextraeqcon} 
\end{eqnarray}
These equations are Weyl-covariant of weights $d+2$, $d+1$ and $d+1$ ($P_i$ is weight-$d$). They are also manifestly hydrodynamic-frame invariant. Similarly, for the matter sector, \eqref{delScarcon} reads:
 \begin{equation}
 \frac{1}{\Omega}\hat{\mathscr{D}}_t \varrho+\hat{\mathscr{D}}_j N^j = 0,
 \label{Weyl-carJ} 
\end{equation}
and is Weyl-covariant of weight $d+1$ and hydrodynamic-frame invariant.

\subsection{Carrollian hydrodynamics from relativistic fluids -- II}  \label{carfluids2}

\subsubsection*{More degrees of freedom}

The behaviours \eqref{eexpC}, 
\eqref{pexpC}, 
\eqref{QexpC}
and \eqref{sigexpC}
of energy, pressure, heat current and stress tensor are required for matching the zero-$c$ limit of hydrodynamic equations with the Carrollian momenta conservation, while preserving hydrodynamic-frame invariance. At the end of  Sec. \ref{comgal}, we contemplated exotic Galilean situations involving more degrees of freedom and obeying extended systems of fluid equations reached at infinite $c$, such as \eqref{gal-no-eq-extra}. Although of limited use in the non-relativistic framework, this sort of extensions play a pivotal role when studying Carrollian fluids in flat holography, were more divergent terms appear to be needed \cite{CMPPS2}. It is worth elaborating in this direction in the spirit of \cite{CMPPS1}, and consider in particular\footnote{In holographic Carrollian fluids, one keeps terms up to order $\nicefrac{1}{c^4}$. The pattern is the same.}
\begin{eqnarray}
\varepsilon &=&\frac{\zeta}{c^2}+\eta + \text{O}\left(c^2\right),
\label{eexpCext}
\\
p &=&\frac{\phi}{c^2}+\  \varpi+ \text{O}\left(c^2\right),
\label{pexpCext}
\\
\label{QexpCext}
q^i&=&\frac{\psi^i}{c^2}+Q^{i}+{c^2}\pi^i + \text{O}\left(c^4\right),
\\
\label{sigexpCext}
\tau^{ij}&=&-\frac{\Sigma^{ij} }{c^2}-\Xi^{ij} + \text{O}\left(c^2\right).
\end{eqnarray}
We can at the same time extend the matter sector with 
\begin{equation}
\label{jexpCext}
\varrho_0=\frac{\omega}{c^2}+\chi + \text{O}\left({c^2}\right),\quad
j^k=\frac{m^i}{c^2}+n^k+ \text{O}\left({c^2}\right).
\end{equation}

With the new scalings, the expansion of the energy--momentum tensor components  \eqref{invenRP}, \eqref{invqRP}, \eqref{invtauRP} is now
\begin{equation}
\label{limcaremext}
\begin{cases}
\varepsilon_{\text{r}}=\frac{\tilde \Pi}{c^2}+
\Pi +\text{O}\left(c^2\right),
\\
q^i_{\text{r}}=\frac{\tilde \Pi^i}{c^2}+
\Pi^{i}+c^2  P^{i}+\text{O}\left(c^4\right),
\\
p_{\text{r}} a^{ij} +\tau_{\text{r}}^{ij} =\frac{\tilde \Pi^{ij}}{c^2}+
\Pi^{ij}
+ \text{O}\left(c^2\right).
\end{cases}
\end{equation}
The Carrollian momenta are displayed in \eqref{momcaremext}.
Similarly, the matter current (Eqs. \eqref{invrhoRP} and \eqref{invjRP}) exhibits the following:
\begin{equation}
\label{jexpCextinv0}
\varrho_{0\text{r}}=\frac{\tilde \varrho}{c^2} +\varrho + \text{O}\left({c^2}\right),\quad
j^k_{\text{r}}=\frac{ \tilde N^k}{c^2}+N^k+ \text{O}\left({c^2}\right)
\end{equation}
with $\tilde \varrho$, $\varrho$, $ \tilde N^k$ and $N^k$ given in \eqref{matcaremext}.

Using the above expansions in the relativistic divergence of the energy--momentum tensor on Papapetrou--Randers background \eqref{carrp} one obtains 
 \begin{eqnarray}
 \label{conTcarexp0ext} 
\frac{c}{\Omega}\nabla_\mu T^{\mu}_{\hphantom{\mu}0}&=&
 \frac{\mathcal{F}}{c^2}+ 
{\mathcal{E}}
+\text{O}\left(c^2\right),
\\
\label{conTcarexpiext} 
 \nabla_\mu T^{\mu i}&=&\frac{\mathcal{X}^i}{c^4} 
 +\frac{\mathcal{H}^i}{c^2}  +\mathcal{G}^i
 +\text{O}\left(c^2\right), 
\end{eqnarray}
while the divergence of the matter current reads:
\begin{equation}
 \label{conJcarexpn} 
\nabla_\mu J^{\mu}= \frac{\mathcal{N}}{c^2}
+ {\mathcal{J}}
+\text{O}\left(c^2\right).
\end{equation}
In these expressions, $\mathcal{E}$, $ \mathcal{G}_j$ and $ \mathcal{J}$ are given in \eqref{carEbis}, \eqref{carG} and \eqref{carJ}, whereas the new expressions are 
 \begin{eqnarray}
\mathcal{F}&=&-  \left(\frac{1}{\Omega}\hat D_t +\theta\right) \tilde \Pi
 -\left(\hat\nabla_i +2\varphi_i \right)\tilde \Pi^{i}-\tilde \Pi^{ij}\hat \gamma_{ij} ,
  \label{carF} 
  \\
  \mathcal{H}_j&=&
\left(\hat \nabla_i+\varphi_i\right)\tilde \Pi^{i}_{\hphantom{i}j}
+2\tilde \Pi^i\varpi_{ij}
+\tilde \Pi \varphi_j +\left(\frac{1}{\Omega}\hat D_t +\theta
\right) \Pi_{j}+ \Pi^i \hat \gamma_{ij}
,
  \label{carH}
  \\
  \mathcal{X}_j&=&
\left(\frac{1}{\Omega}\hat D_t +\theta
\right) \tilde \Pi_{j}+ \tilde \Pi^i \hat \gamma_{ij},
  \label{carX}
\end{eqnarray}
and 
 \begin{equation}
\mathcal{N}=
 \left(\frac{1}{\Omega}\partial_t +\theta\right)\tilde\varrho+\left(\hat\nabla_j +\varphi_j \right)\tilde N^j.
 \label{carN} 
\end{equation}
At zero $c$, the Carrollian energy and momenta equations are thus $\mathcal{E}=\mathcal{F}= \mathcal{G}_j= \mathcal{H}_j= \mathcal{X}_j=0$, and similarly  $\mathcal{J}=\mathcal{N}=0$ describe the matter sector.  All these equations are invariant under hydrodynamic-frame transformations because the differential operators are geometric and thus invariant, and because the momenta 
$\tilde\Pi$, $ \Pi$, 
$\tilde\Pi^{i}$,
$ \Pi^{i}$,
$P^i$,  
$\tilde\Pi^{ij}$,
$ \Pi^{ij}$, $\tilde\varrho$, $\varrho$, $\tilde N^i$,  $N^i$
also are, as shown in App. \ref{compl}.

\subsubsection*{Weyl-invariant Carrollian fluids}

For the system under investigation, the use of an effective action is not convenient, as it would require a complete set of variables conjugate to the momenta $\Pi$, $\Pi^i$, $\Pi^{ij}$ and $\tilde \Pi$, $\tilde\Pi^i$, $\tilde\Pi^{ij}$, which is  bigger than $a_{ij}$, $b_i$ and $\Omega$. Weyl invariance is here easier to impose as a zero-$c$ limit of  $T^\mu_{\hphantom{\mu}\mu}=0$ 
 discussed at the end of Sec. \ref{rreleq}, using \eqref{restraceinvRP} with \eqref{limcaremext}:
\begin{equation}
\label{restraceinvRPcon}
T^\mu_{\hphantom{\mu}\mu}=\frac{1}{c^2}\left(\tilde \Pi_{i}^{\hphantom{i}i}
-\tilde \Pi
\right) + \Pi_{i}^{\hphantom{i}i}
- \Pi
+ \text{O}\left(c^2\right)=0
,
\end{equation}
leading to
\begin{equation}
\label{car-conf-cond-more}
\tilde\Pi_i^{\hphantom{i}i}=\tilde\Pi,\quad
\Pi_i^{\hphantom{i}i}=\Pi.
\end{equation} 
These conditions can be recast in terms of Carrollian-fluid observables using the explicit expressions of the momenta \eqref{momcaremext}. Splitting them again \`a la \eqref{con}, into global-equilibrium equations of state plus conditions for the dynamical irreversible components, we find the following:
\begin{equation}\label{con-more}
\zeta= d\phi,\quad \Sigma^i_{\hphantom{i}i}=0,\quad \eta= d\varpi,\quad \Xi^i_{\hphantom{i}i}=\beta_i\beta_j \Sigma^{ij}. 
\end{equation} 
The conformal weights of  $\tilde \Pi$, $\tilde\Pi^i$, $\tilde\Pi^{ij}$ match those of $\Pi$, $\Pi^i$, $\Pi^{ij}$; those of the extra variables $\zeta$, $\phi$, $\psi_i$ and $\Sigma_{ij}$ are $d+1$, $d+1$, $d$ and $d-1$, while for $\omega $ and $m^i$ we find $d$ and $d+1$.

Finally, using \eqref{car-conf-cond-more}, the equations $\mathcal{E}=\mathcal{F}= \mathcal{G}_j= \mathcal{H}_j= \mathcal{X}_j=0$ and $\mathcal{J}=\mathcal{N}=0$ become Weyl-covariant with
\begin{eqnarray}
\mathcal{J}&=&
 \frac{1}{\Omega}\hat{\mathscr{D}}_t \varrho+\hat{\mathscr{D}}_j N^j,
 \label{Weyl-carJncon} 
\\
 \label{Weyl-carRcon} 
\mathcal{N}&=& 
\frac{1}{\Omega}\hat{\mathscr{D}}_t \tilde\varrho+\hat{\mathscr{D}}_j  \tilde N^j,
\end{eqnarray}
  and
\begin{eqnarray}
\mathcal{E}&=&-  \frac{1}{\Omega}\hat{\mathscr{D}}_t\Pi
-\hat{\mathscr{D}}_i \Pi^{i}
-\Pi^{ij}\xi_{ij} ,
  \label{carEbiscon} 
  \\
  \mathcal{F}&=&-  \frac{1}{\Omega}\hat{\mathscr{D}}_t \tilde \Pi
-\hat{\mathscr{D}}_i \tilde\Pi^{i}
-\tilde\Pi^{ij}\xi_{ij} ,
  \label{carFcon}  \\
  \mathcal{G}_j&=&
\hat{\mathscr{D}}_i \Pi^{i}_{\hphantom{i}j}+2\Pi^{i}\varpi_{ij}+ \left(\frac{1}{\Omega}\hat{\mathscr{D}}_t \delta^i_j +\xi^{i}_{\hphantom{i}j}\right) P_i ,
  \label{carGcon} 
  \\
  \mathcal{H}_j&=&
\hat{\mathscr{D}}_i \tilde \Pi^{i}_{\hphantom{i}j}+2\tilde\Pi^{i}\varpi_{ij}+ \left(\frac{1}{\Omega}\hat{\mathscr{D}}_t \delta^i_j +\xi^{i}_{\hphantom{i}j}\right) \Pi_i 
,
  \label{carHcon}
  \\
  \mathcal{X}_j&=&
\frac{1}{\Omega}\hat{\mathscr{D}}_t 
\tilde\Pi_{j}
+ \tilde\Pi_{i}\xi^{i}_{\hphantom{i}j}.
  \label{carXcon}
\end{eqnarray}
These equations are the seed for flat holography, see \cite{Campoleoni:2018ltl, CMPR, CMPRpos,CMPPS2}. They are covariant under Carrollian diffeomorphisms, covariant under Weyl rescalings and invariant under hydrodynamic-frame transformations, which are local Carroll transformations (Carroll boosts and rotations). 

\subsubsection*{Isometries and conformal isometries}

As a final application of the above results on multiplication of degrees of freedom, we can insert the behaviour \eqref{limcaremext} inside the components \eqref{restIinvcar}, \eqref{inviotacar} and \eqref{invicar} of a relativistic conserved current resulting from the combination of the energy--momentum tensor with a Killing or a conformal Killing field. We obtain
\begin{equation}
\label{jexpCextinv}
-\frac{1}{c\Omega}I_0=\iota_{0\text{r}}= \frac{\tilde{\tilde \kappa}}{c^4} + \frac{\tilde \kappa}{c^2} + \kappa + \text{O}\left(c^2\right),\quad
I^k=i_{\text{r}}^k= \frac{\tilde K^k}{c^2}+K^k+ \text{O}\left(c^2\right)
\end{equation}
with (remember that $\xi^i\equiv \xi^{\hat \imath}$ is a function of $\mathbf{x}$ only for Carrollian diffeomorphisms)
\begin{equation}
\label{carkappaK}
 \begin{cases}
\kappa= \xi^{i} P_i-\xi^{\hat t}  \Pi
\\
\tilde\kappa=\xi^{i} \Pi_i-\xi^{\hat t}  \tilde\Pi
\\
\tilde{\tilde \kappa}=\xi^{i} \tilde\Pi_i
\\
K^i=\xi^{j}\Pi_{j}^{\hphantom{j}i}-\xi^{\hat t} \Pi^i
\\
\tilde K^i =\xi^{j}\tilde \Pi_{j}^{\hphantom{j}i}-
 \xi^{\hat t} \tilde\Pi^i
,
\end{cases}
\end{equation}
where $\kappa$ and $K_i$ are precisely as anticipated in \eqref{invkappacar} and \eqref{invKcar}, and described in footnote \ref{car-con-curr}.
Inserting \eqref{jexpCextinv} in the relativistic divergence of the matter current $I^\mu$ in Papapetrou--Randers background we recover a multiplication of \eqref{kilcarcon} in the form
\begin{equation}
 \label{conJcarrexpn} 
\nabla_\mu I^{\mu}= \frac{\tilde{\tilde{\mathcal{K}}}}{c^4}+  \frac{\tilde{\mathcal{K}}}{c^2}+ \mathcal{K} + \text{O}\left(c^2\right)
\end{equation}
with
\begin{equation}
\label{carKN}
 \begin{cases}
\tilde{\tilde{\mathcal{K}}}=
\left(
\frac{1}{\Omega}\partial_t+
\theta \right)\tilde{\tilde\kappa}=
0
 \\\tilde{\mathcal{K}}=
\left(
\frac{1}{\Omega}\partial_t+
\theta \right)\tilde\kappa
+\left(\hat \nabla_i +\varphi_i\right)\tilde K^i=
-\tilde\Pi^i\left(\left(
\hat\partial_i-\varphi_i\right)\xi^{\hat t }-
2\xi^j \varpi_{ji}
\right)
 \\ \mathcal{K}=
\left(
\frac{1}{\Omega}\partial_t+
\theta \right)\kappa
+\left(\hat \nabla_i +\varphi_i\right)K^i=
-\Pi^i\left(\left(
\hat\partial_i-\varphi_i\right)\xi^{\hat t }-
2\xi^j \varpi_{ji}
\right).
\end{cases}
\end{equation}
The last equalities are obtained
owing to the Carrollian Killing \eqref{carkilleq} (or conformal Killing \eqref{extra-conf-cond}, \eqref{PRkillconf}, \eqref{carkilleqconf},   \eqref{PRkillconfmu})  conditions and the equations of motion 
$\mathcal{E}=\mathcal{F}= \mathcal{G}_j= \mathcal{H}_j= \mathcal{X}_j=0$ (see Eqs. \eqref{carEbiscon},  \eqref{carFcon},  \eqref{carGcon},  \eqref{carHcon},  \eqref{carXcon} -- and \eqref{car-conf-cond-more} for the Weyl-covariant situation).
\begin{comment}
\begin{equation}
\label{carKNonshell}
 \begin{cases}
\tilde{\tilde{\mathcal{K}}}=
0 \\\tilde{\mathcal{K}}=
-\tilde\Pi^i\left(\left(
\hat\partial_i-\varphi_i\right)\xi^{\hat t }-
2\xi^j \varpi_{ji}
\right)
 \\ \mathcal{K}=
-\Pi^i\left(\left(
\hat\partial_i-\varphi_i\right)\xi^{\hat t }-
2\xi^j \varpi_{ji}
\right).
\end{cases}
\end{equation}
\end{comment}
One of the three currents is conserved as a consequence of the (conformal) isometry, whereas the other two are not. 

The multiplication of degrees of freedom induced by the behaviour of the energy--momentum tensor \eqref{limcaremext}, triggers a multiplication of currents in the presence of an isometry (or a conformal isometry if the dynamics is Weyl-invariant) -- here three, but possibly more if more terms are present in \eqref{limcaremext}.  These currents are generically non-conserved, as we already observed in Sec. \ref{cardifcons} for a single current (Eqs. \eqref{carnoncon} or \eqref{carconconf}),  unless the symmetry is stronger than a Carrollian (conformal) isometry i.e. if 
$\mathscr{L}_\upxi \uptheta^{\hat t}=\left(\left(\hat\partial_i-\varphi_i\right)\xi^{\hat t }-
2\xi^j \varpi_{ji}\right)\text{d}x^i=0$. Alternatively 
each of those currents may or may not be conserved, irrespective of the others, when an appropriate vector in the expansion of $q_{\text{r}}^i$ vanishes, be it $\Pi^i $ or $\tilde\Pi^i$, and this explains why  $\tilde{\tilde{\mathcal{K}}}=0$ in the above paradigm, whereas $\tilde{\mathcal{K}}\neq0$ and $\mathcal{K}\neq0$. 

As we emphasized at the end of Sec. \ref{comgal} in the Galilean framework, i.e. for infinite $c$, it is puzzling that  the conservation law $\nabla_\mu I^{\mu}= 0$ of a relativistic current $I^\mu = \xi_\nu T^{\mu\nu}$ produced by a Killing vector $\upxi$ of a pseudo-Riemannian spacetime, similarly fails when the limit $c$-to-zero is taken. This is once again due to the nature of the Carrollian Killing fields; details are available in App. \ref{varprinccar}.

\section{Aristotelian dynamics}

\subsubsection*{Aristotelian spacetimes}

Aristotelian spacetimes were introduced in \cite{Penrose-Battelle}. They are part of a rich web of geometric structures, which incorporate Newton--Cartan and Carroll, among others. Their defining feature is a manifold $\mathscr{M}= \mathbb{R} \times \mathscr{S}$ of dimension $d+1$,  endowed with a degenerate metric and a degenerate cometric. The first implies that there is a symmetric rank-two tensor acting on tangent-space elements
\begin{equation}   
\label{arisdegmet}
\text{d}\ell^2=\ell_{\mu\nu}( t, \mathbf{x})  \text{d}x^\mu \,\text{d}x^\nu,\quad \mu,\nu\ldots \in \{0,1,\ldots,d\}
\end{equation}
with one-dimensional kernel, the field of observers
\begin{equation}   
\label{ariskert}
\upupsilon = \upsilon^\mu( t, \mathbf{x})\,  \partial_\mu:\quad \ell_{\mu\nu} \upsilon^\mu=0.
\end{equation}
The second  translates into the existence of symmetric rank-two tensor acting on cotangent-space vectors:
\begin{equation}   
\label{dmetinvaris}
\partial_m^2=m^{\mu\nu}( t, \mathbf{x})\, \partial_\mu\, \partial_\nu,
\end{equation}
where 
\begin{equation}   
\label{ariskerinv}
\upmu = \mu_\nu( t, \mathbf{x}) \, \text{d}x^\nu:\quad m^{\mu\nu} \mu_\nu=0,
\end{equation}
defines a clock form (one should say ``anti-clock'' because of the sign), such that 
\begin{equation}   
\label{arisclos}
- \upsilon^\mu \mu_\nu+m^{\mu\lambda}\ell_{\lambda\nu}=\delta^\mu_\nu, \quad  \upsilon^\nu \mu_\nu=-1, \quad m^{\mu\nu}\ell_{\mu\nu}=d.
\end{equation}
The topological structure $\mathscr{M}= \mathbb{R} \times \mathscr{S}$ provides here a genuine, vertical and horizontal foliation, as opposed to Carrollian manifolds, where it generally supports a fiber bundle over a $d$-dimensional basis, and to Newton--Cartan were the fibration is defined over a line. In other words, Aristotelian spacetimes lie in the intersection of Carrollian and Newton--Cartan geometries with trivial Ehresmann connection for the Carrollian and trivial field of observers for the Newton--Cartan.\footnote{Our definition of Aristotelian manifolds is that of \cite{Penrose-Battelle}, also used in \cite{dutch, dutch2}. In Refs. \cite{Bekaert:2014bwa, Bekaert:2015xua} ``Aristotelian'' is somewhat less restrictive. It is meant to be a Leibnizian structure and a Fr\"obenius-integrable distribution associated with the absolute clock form ($\upmu\wedge \text{d}\upmu=0$ obvious for \eqref{ariskert-invk}). The necessary extra ingredient to match the more conventional picture we use is the field of observers $\upupsilon$ \eqref{ariskert} (or \eqref{ariskert-invk} in the concrete realization), as described in Props. A.5 and A.6 of   \cite{Bekaert:2015xua}.}

The vector $\upupsilon$ and the covector $\upmu$ are intrinsic geometric objects. Aristotelian transformations are meant to leave them invariant and respect the associated foliation. This forbids boosts, which is a characteristic feature of Aristotelian spacetimes translating into the absolute nature of time \emph{and} space. In practice it allows to adopt a convenient choice:\footnote{This is not the most general choice, as \eqref{dmetinv}--\eqref{clock} and \eqref{cardegmet}-\eqref{kert} were not the most general either for Newton--Cartan and Carrollian manifolds.}
\begin{equation}   
\label{arisdegmet-invmet}
\text{d}\ell^2=a_{ij}\,  \text{d}x^i \,\text{d}x^j,\quad \partial_m^2=a^{ij}\, \partial_i\, \partial_j
\end{equation}
with $a^{ik}a_{kj}=\delta^i_j$ and 
\begin{equation}   
\label{ariskert-invk}
\upupsilon =\frac{1}{\Omega} \partial_t, \quad \upmu = -\Omega \text{d}t,
\end{equation}
where $a_{ij}$ and $\Omega$ are functions of $t$ and $\mathbf{x}$. Aristotelian diffeomorphisms act as 
\begin{equation}
\label{arisdifs} 
t'=t'(t),\quad \mathbf{x}^{\prime}=\mathbf{x}^{\prime}(\mathbf{x})
\end{equation}
with Jacobian 
\begin{equation}
 \label{arisj}
J(t)=\frac{\partial t'}{\partial t},\quad 
J^i_j(\mathbf{x}) = \frac{\partial x^{i\prime}}{\partial x^{j}}.
\end{equation}
The transformations of the metric and kernel components
are thus
\begin{equation}
\label{arisdifawom}
a^{\prime}_{ij} =a_{kl} J^{-1k}_{\hphantom{-1}i} J^{-1l}_{\hphantom{-1}j} ,
\quad
\Omega^{\prime }=\frac{\Omega}{J}.
\end{equation}

Aristotelian manifolds can be equipped with covariant derivatives, which allow to obtain genuine tensors under Aristotelian diffeomorphisms \eqref{arisdifs}. As for the various spacetimes met earlier, we will focus here on the simplest torsionless and metric-compatible space and time connections, which naturally appear in the conservation equations. 

The spatial connection is the same as the one introduced for the Galilean manifolds in \eqref{dgamma}:
\begin{equation}
\label{dgamma-aris}
\gamma^i_{jk}=\frac{a^{il}}{2}\left(\partial_j a_{lk}+\partial_k a_{lj}-\partial_l a_{jk}\right).
\end{equation}
The associated covariant derivative is spelled $\nabla_i$ since it needs not be distinguished from the ordinary Levi--Civita derivative on a Riemannian $d$-dimensional manifold -- it just depends on time here. It is torsionless, because $\gamma^i_{jk}$ are symmetric and also obeys 
$\nabla_ia_{jk}=0$.\footnote{For a more general discussion on compatible connections, see Props. A.17 and A.18 of   \cite{Bekaert:2015xua}.}
Its Riemann, Ricci and scalar curvature tensors are defined as usual $d$-dimensional Levi--Civita curvature tensors would be, except that they are $t$-dependent.

The time derivative operator $\frac{1}{\Omega}\partial_t$ is also promoted to an Aristotelian temporal metric-compatible covariant derivative 
\begin{equation}
\label{ariscartimemet}
  D_ta_{jk}=0,
\end{equation}
using
\begin{equation}
\label{arisdgammaCartime}
 \gamma_{ij}=\frac{1}{2\Omega}\partial_t a_{ij}
=\xi_{ij} + \frac{1}{d}a_{ij}\theta.
\end{equation}
We have introduced the familiar by now traceless shear tensor $\xi_{ij} $ and expansion scalar $\theta=\frac{1}{\Omega}              
\partial_t \ln\sqrt{a}$, which can be completed with the acceleration form
\begin{equation}
\label{ariscaromacc}
\varphi_i=\partial_i \ln \Omega.
\end{equation}
The action of $  D_t$ is as in the Carrollian case, Eqs. \eqref{Cartimecovdersc} and \eqref{Cartimecovdervecform}:
$D_t \Phi=\partial_t  \Phi$ and 
\begin{equation}
\label{aristimecovdervecform}
\frac{1}{\Omega} D_tV^i=\frac{1}{\Omega} \partial_tV^i+\gamma^i_{\hphantom{i}j} V^j,\quad
\frac{1}{\Omega} D_tV_i=\frac{1}{\Omega} \partial_tV_i-\gamma_i^{\hphantom{i}j} V_j
\end{equation}
generalized using the Leibniz rule. 

Time and space derivatives do not commute: 
\begin{equation}
\left[\frac{1}{\Omega}  D_t,\nabla_i\right]V^j= 
\varphi_{i}\left(\left(\frac{1}{\Omega}  D_t
+
\theta
\right)V^j -   \gamma^{j}_{\hphantom{j}k}V^k\right)
- \gamma_{i}^{\hphantom{i}k} \nabla_k V^j
-d   r^j_{\hphantom{j}ik}V^k
\label{aris3carriemanntimetilde}
\end{equation}
with\footnote{Again $\frac{1}{\Omega} \partial_t   \gamma^j_{ik}
=\left(  \nabla_i + \varphi_i\right)  \gamma^{j}_{\hphantom{j}k}
+\left(  \nabla_k + \varphi_k\right)  \gamma^{j}_{\hphantom{j}i}
-
\left(  \nabla^j + \varphi^j \right)  \gamma_{ik}
$
is an Aristotelian tensor, even though $   \gamma^j_{ik}$ is not.}
\begin{equation}
  r^j_{\hphantom{j}ik}=\frac{1}{d}
\left(\theta \varphi_i \delta^j_k 
+  \nabla_i  \gamma^{j}_{\hphantom{j}k}
-\frac{1}{\Omega} \partial_t   \gamma^j_{ik}
\right)
\label{aris3carriemanntime}
\end{equation}
another Aristotelian curvature tensor, similar to the one introduced for our Carrollian connection.

\subsubsection*{Dynamics from diffeomorphism invariance}

Aristotelian fluids were introduced in \cite{dutch}, and mentioned as ``self-dual'' in \cite{CMPPS1}. They received further attention \cite{dutch2} amid a soaring interest for non-boost-invariant dynamics \cite{Armas:2020mpr, Pinzani-Fokeeva:2021klb, Rajagopal:2021swe}. It is abusive though to call this ``hydrodynamics'' for two reasons. The first is shared with the Carrollian instance: a fluid is meant to assimilate a phenomenological description of a many-body system under local-equilibrium and slow-variation assumptions, and those assumptions are non controllable in the Carrollian or Aristotelian framework due the absence of a thermodynamic or kinetic theory -- the elementary particle motion is forbidden in both cases. Additionally, as opposed to Galilean or Carrollian manifolds, Aristotelian cannot be obtained naturally as limits of pseudo-Riemannian spacetimes. Thus, no guide for the would-be Aristotelian fluid equations exists that could be based on a limiting procedure, and this is the second reason. The only safe way to reach a set of equations covariant under Aristotelian diffeomorphisms \eqref{arisdifs} is to require this diffeomorphism invariance at the level of an effective action, and study its conservation consequences for the conjugate Aristotelian momenta. The latter remain however agnostic on a decomposition in terms of kinematical (as $v_i$ for Galilean, $\beta^i$ for Carrollian) and kinetic (energy, pressure, heat and stress) variables. 

Starting again with an action $S=\int \text{d}t\,  \text{d}^{d}x \sqrt{a}\Omega\mathcal{L}$ on $\mathscr{M}=\mathbb{R}\times \mathscr{S}$, now functional of $a_{ij}$ and $\Omega$, the Aristotelian momenta  conjugate to those variables are
\begin{eqnarray}
\label{arisvarstren} 
\Pi^{ij}&=&\frac{2}{\sqrt{a} \Omega}\frac{\delta S}{\delta a_{ij}},\\
\label{arisvarenden}
\Pi&=&-\frac{1}{\sqrt{a}}\frac{\delta S}{\delta \Omega}.
 \end{eqnarray}
These are the \emph{energy--stress tensor} and the \emph{energy density}, and with those the variation of the action in the gravitational sector is
\begin{equation}
\label{delSgravaris}
\delta S=-\int\text{d}t \, \Omega\int \text{d}^dx \sqrt{a}  \left(\frac{1}{2}\Pi_{ij}\delta a^{ij} 
+\Pi \delta \ln \Omega\right).
\end{equation}

Aristotelian diffeomorphisms \eqref{arisdifs} are generated by vector fields 
\begin{equation}\label{arsdiffeos}   
\upxi=\xi^t \partial_t +\xi^i \partial_i=
\xi^{\hat t}\frac{1}{\Omega} \partial_t+\xi^i \partial_i ,
 \end{equation}
where  $\xi^t =\xi^t (t)$ and $\xi^i =\xi^i (\mathbf{x})$.
The variation under diffeomorphisms is implemented as usual with
\begin{equation}   
-\delta_\upxi a_{ij}=\mathscr{L}_\upxi a_{ij}= 2\nabla_{(i}\xi^ka_{j)k}+2\xi^{\hat t}    \gamma_{ij}
.
\label{Liedaijaris}
 \end{equation}
Now
\begin{equation}   
\mathscr{L}_\upxi \upupsilon=\mu \upupsilon
, \quad
\mathscr{L}_\upxi \upmu=-\mu \upmu  
 \end{equation}
 with
 \begin{equation}
\label{PRkillconfmuaris}
\mu(t,\mathbf{x})= -\left(\frac{1}{\Omega} \partial_t
\xi^{\hat t }+
\varphi_i \xi^i \right) .
 \end{equation}
Thus
 \begin{equation}   
-\delta_\upxi \ln \Omega=\frac{1}{\Omega}\mathscr{L}_\upxi \Omega =-\mu . 
\label{Liedwiomaris}
\end{equation}
Using \eqref{Liedaijaris} and  \eqref{Liedwiomaris} in \eqref{delSgravaris} with $\xi^i=\xi^i(\mathbf{x})$, we obtain (indices are here lowered with $a_{ij}$)
 \begin{eqnarray}
 \delta_\upxi S&=&\int \text{d}t \text{d}^dx  \sqrt{a}\Omega 
\left\{-\xi^{\hat t}\left[
\left(\frac{1}{\Omega}\partial_t +\theta\right)\Pi
+\Pi^{ij}   \gamma_{ij}
\right]+\xi^i\left[\left(  \nabla_j+\varphi_j\right)\Pi^{j}_{\hphantom{j}i}
+\Pi \varphi_i
\right]\right\} 
 \nonumber\\
&&+\int \text{d}t \text{d}^dx  
 \left\{\partial_t\left[\sqrt{a} \xi^{\hat t}\Pi \right] -\partial_i \left[\sqrt{a} \Omega\xi^j \Pi^{i}_{\hphantom{i}j}  \right]
 \right\}.
\label{arisvarS}
 \end{eqnarray}
The boundary terms (last line of \eqref{arisvarS}) are ignored and  $\delta_\upxi S=0$ implies that the momenta defined earlier in \eqref{arisvarstren}, \eqref{arisvarenden}  are Aristotelian-covariant. It leads to two equations: the energy equation 
 \begin{equation}   
 \label{delSariseneq}
\boxed{\left(\frac{1}{\Omega}\partial_t +\theta\right)\Pi
+\Pi^{ij}   \gamma_{ij}
=-\left(\nabla_i+2\varphi_i\right)\Pi^i,}
\end{equation}
and the momentum equation 
  \begin{equation}   
\label{delSarismomeq}
\boxed{\left(\nabla_j+\varphi_j\right)\Pi^{j}_{\hphantom{j}i}
+\Pi \varphi_i = -\left(\frac{1}{\Omega}\partial_t +\theta\right) P_i .}
 \end{equation}
 In these equations the \emph{energy current} $\Pi^i $ and the \emph{momentum} $P^i$ are undetermined. They arise because $\xi^t$ depends on time and $\xi^i$ on space, exclusively. As a consequence, 
 $
 \sqrt{a}\Omega \xi^{\hat t}\left(\nabla_i+2\varphi_i\right)\Pi^i
 =\partial_i\left(\sqrt{a}\Omega \xi^{\hat t}\Pi^i\right)
 $
 and 
 $ \sqrt{a}\Omega \xi^{i }\left(\frac{1}{\Omega}\partial_t +\theta\right) P_i = \partial_t\left(\sqrt{a}\xi^i  P_i \right)$,  
which  are boundary terms and vanish inside the integral. 

Aristotelian dynamics stands at the intersection of Galilean and Carrollian. This is expected from first principles
and can be verified by comparing Eqs. \eqref{delSariseneq} and \eqref{delSarismomeq} with \eqref{delSgaleneq}\footnote{ Notice that a term of the type $2\varphi_i \Pi^i$ would also have been present in this Galilean energy equation if we had considered a torsionfull Newton--Cartan spacetime, with $\Omega=\Omega(t,\mathbf{x})$, as we assume in the present Aristotelian case.} and \eqref{delSgalmomeq}
in the Galilean at $w^i=0$, as well as \eqref{delScareneq} and \eqref{delScarmomeq} in the Carrollian $b_i=0$ instances. However, contrary to these, Aristotelian dynamics cannot be reached as a limit of relativistic hydrodynamics. No kinematical parameter such as $v_i$ or $\beta^i$ can be foreseen and no decomposition of the momenta \`a la \
\eqref{galmomP},
\eqref{stressgalgeni}, 
\eqref{endengalgeni}, 
\eqref{encurgalgeni}
or 
\eqref{carflux}, 
\eqref{carheatflux}. 
Finally, hydrodynamic-frame invariance cannot be argued or disputed. 

The presence of a matter/charge sector can be treated as in the previous families of dynamics. We assume the existence 
of a gauge field  $\text{B}=B (t, \mathbf{x}) \text{d}t +B_i (t, \mathbf{x}) \text{d}x^i$ associated with a local $U(1)$ 
transformation as in \eqref{delB}, and the conjugate momenta 
\begin{eqnarray}
\label{arismomchi}
\varrho&=&\frac{1}{\sqrt{a}}\frac{\delta S}{\delta B},\\
\label{arismomphii}
N^{i}&=&\frac{1}{\Omega\sqrt{a}}
\frac{\delta S}{\delta B_i}
\end{eqnarray}
are the \emph{charge density} and the \emph{charge current}.
 The gauge variation
 \begin{eqnarray}
  \delta_\Lambda S&=&
 \int \text{d}t \text{d}^{d}x \sqrt{a} \left(
\varrho\delta_\Lambda  B+
\Omega
N^i \delta_\Lambda  B_i \right) 
\nonumber
 \\
 &=&
 \int \text{d}t \text{d}^dx \left\{  \sqrt{a}\Omega 
\Lambda\left(\frac{1}{\Omega}\partial_t\varrho+
\theta \varrho+\left(\nabla_i +\varphi_i\right)N^i
\right)-
\partial_t\left(\sqrt{a} \Lambda \varrho\right) -\partial_i \left(\sqrt{a}\Lambda \Omega N^i\right)
 \right\}
 \label{U1aris}
 \end{eqnarray}
 is  assumed to vanish, and this leads to
  \begin{equation}
\boxed{\left(
\frac{1}{\Omega}\partial_t+
\theta \right)\varrho
+\left(\nabla_i +\varphi_i\right)N^i=0.
}
  \label{delSariscon}
 \end{equation}
In integrated form, thanks to Stokes and Gauss theorems, this conservation reads ($\mathscr{V}\subset \mathscr{S}$, a constant-time section,
and  $\partial \mathscr{V}$ does not depend on time):
 \begin{equation}
\frac{\text{d}}{\text{d}t}\int_{\mathscr{V}}\text{d}^{d}x \sqrt{a} \varrho =-\int_{\mathscr{V}}\text{d}^{d}x \, 
\partial_i\left(\sqrt{a} \Omega N^i\right)
=-\int_{\partial\mathscr{V}}\Omega\star\mathbf{N} 
  \label{ariscontime}
 \end{equation}
 with the conserved charge defined as an integral over the entire space $\mathscr{S}$:
   \begin{equation}
Q_N=\int_{\mathscr{S}}\text{d}^{d}x \sqrt{a}\varrho.
  \label{arisconch}
 \end{equation}
 As for the Galilean or the Carrollian situations discussed in Secs. \ref{FLUIDS} and \ref{cardifcons}, one can chose any space-like hypersurface $\upSigma_d$ for the determination of the charge:
\begin{equation}
Q_N= \int_{\upSigma_d}
\sqrt{a}\varrho\,
\text{d}x^1\wedge\ldots\wedge\text{d}x^d
+ \int_{\upSigma_d} \sqrt{a}\sum_{i=1}^d \text{d}x^1\wedge\ldots\wedge N^i \upmu\wedge\ldots\wedge\text{d}x^d,
\label{arisconch-gen}
\end{equation}
with the clock form $\upmu$ given in \eqref{ariskert-invk}, and $N^i \upmu$ at the $i$th position in the last exterior product.

\subsubsection*{Weyl invariance}

Weyl invariance can also be present in Aristotelian dynamics. The reasoning is familiar, starting from  $ \delta_{\mathcal{B}} S=0$, we are lead to the condition
\begin{equation}
\label{aris-conf-cond}
\Pi_i^{\hphantom{i}i}=\Pi.
\end{equation} 
The geometric tools for handling Weyl covariance are similar to those introduced in the previous chapters. The connections are $\theta $ and $\varphi_i$ (see Eqs. \eqref{arisdgammaCartime} and \eqref{ariscaromacc})
with
\begin{equation}
 \label{weyl-geometry-aris}
\theta\to \mathcal{B}\theta-\frac{d}{\Omega}\partial_t \mathcal{B}
  ,\quad 
\varphi_{i}\to \varphi_{i}-\partial_i\ln \mathcal{B}.
\end{equation} 
The action of the Aristotelian Weyl-covariant time  derivative on any tensor increases its weight by one unit. 
On a weight-$w$ 
function $\Phi$ it is
\begin{equation}
\label{AWtimecovdersc}
\frac{1}{\Omega}\mathscr{D}_t \Phi=\frac{1}{\Omega}D_t \Phi +\frac{w}{d} \theta \Phi=
\frac{1}{\Omega}\partial_t \Phi +\frac{w}{d} \theta \Phi,
\end{equation}
while on a weight-$w$ vector, we find using \eqref{aristimecovdervecform}
\begin{equation}
\label{AWtimecovdervecform}
\frac{1}{\Omega}\mathscr{D}_t V^l=\frac{1}{\Omega} D_t V^l +\frac{w-1}{d} \theta V^l
=\frac{1}{\Omega}\partial_t V^l +\frac{w}{d} \theta V^l+\xi^{l}_{\hphantom{l}i} V^i .
\end{equation}
Similarly for any tensor by Leibniz rule and in particular we find $\mathscr{D}_t a_{kl}=0$.

A spatial Aristotelian-Weyl-covariant derivative can also be introduced, and does not alter the conformal weight. 
For a weight-$w$ scalar function  $\Phi$ it acts as
\begin{equation}
\label{AWs-Phi}
\mathscr{D}_j \Phi=\partial_j \Phi +w \varphi_j \Phi;
\end{equation}
for a vector we find
\begin{equation}
\mathscr{D}_j V^l=\nabla_j V^l +(w-1) \varphi_j V^l +\varphi^l V_j -\delta^l_j V^i\varphi_i.
\end{equation}
The action on any other tensor is obtained using the Leibniz rule, and in particular $\mathscr{D}_j a_{kl}=0$.

Time and space Aristotelian--Weyl covariant derivatives do not commute and curvature tensors follow, which resemble those already quoted for the Carrollian or Galilean cases. We will not elaborate on this subject here, in particular because all these connections can be generalized and studied in a more abstract geometric framework, as for example 
in \cite{Bekaert:2014bwa, Bekaert:2015xua}.

With the above tools, Aristotelian equations \eqref{delSariseneq} and \eqref{delSarismomeq} read, under the assumption of conformal dynamics \eqref{aris-conf-cond}:
\begin{eqnarray}
\frac{1}{\Omega}\mathscr{D}_t\Pi
+\mathscr{D}_i \Pi^{i}
+\Pi^{ij}\xi_{ij}&=&0,
 \label{arisEcon} 
\\
\mathscr{D}_i \Pi^{i}_{\hphantom{i}j}+ \left(\frac{1}{\Omega}\mathscr{D}_t \delta^i_j +\xi^{i}_{\hphantom{i}j}\right) P_i &=&0.
  \label{arisGcon}
  \end{eqnarray}
These equations are Weyl-covariant of weights $d+2$ and $d+1$ ($\Pi $ and $\Pi^{i}_{\hphantom{i}j}$
have weight
$d+1$, whereas  $\Pi_i$ and 
$P_i$ are weight-$d$). Similarly, for the matter sector, \eqref{delSariscon} reads:
 \begin{equation}
 \frac{1}{\Omega}\mathscr{D}_t \varrho+\mathscr{D}_j N^j = 0,
 \label{Weyl-carJ} 
\end{equation}
and is Weyl-covariant of weight $d+1$.

\subsubsection*{Isometries, conformal isometries and conservation laws}

Isometries of Aristotelian spacetimes are generated by vectors fields \eqref{arsdiffeos} required to satisfy 
\begin{equation}
\label{ArsK}
\mathscr{L}_\upxi a_{ij}=0\,, \quad\mathscr{L}_\upxi \upmu=0\,.
\end{equation}
The latter lead to a set of Killing equations for $\xi^{\hat{t}}(t)$ and $\xi^{i}(\mathbf{x})$:
\begin{equation}   
\label{Arskilleq}
\boxed{\nabla_{(i}\xi^ka_{j)k}+\xi^{\hat t} \gamma_{ij}
 =0
,\quad
\frac{1}{\Omega} \partial_t
\xi^{\hat t }+
\varphi_i \xi^i =0.}
 \end{equation}
 For conformal isometries, the generators must satisfy 
 \begin{equation}
\label{ArsKcon}
\mathscr{L}_\upxi a_{ij}= \lambda a_{ij} 
\end{equation}
where $ \lambda $ is obtained by tracing the latter:
 \begin{equation}
\label{ariskilleqconf}
\lambda(t,\mathbf{x}) =\frac{2}{d}
\left(\nabla_{i}\xi^{i}
+\theta \xi^{\hat t}  
\right).
\end{equation}
As for the previous Galilean and Carrollian cases, the requirement \eqref{ArsKcon} must be supplemented with the usual extra condition \eqref{extra-conf-cond} $ 2\mu + \lambda=0$, in order to reach a well-defined set of conformal generators -- $\mu(t,\mathbf{x})$ is displayed in \eqref{PRkillconfmuaris}.

In the presence of isometries or conformal isometries, we can define an Aristotelian current $\kappa$ and $K^i$
(as $\varrho$ and $N^i$ in \eqref{arismomchi}, \eqref{arismomphii}), following the previous definitions in Newton--Cartan (\eqref{inviotagal} and \eqref{invigal}) or Carrollian spacetimes (\eqref{invkappacar} and \eqref{invKcar})
-- see also the boundary terms in $\delta_\upxi S$ \eqref{arisvarS} evaluated on-shell:
\begin{equation}
\label{invcuraris}
\kappa= \xi^{i} P_i-\xi^{\hat t}  \Pi
,\quad
K^i=\xi^{j}\Pi_{j}^{\hphantom{j}i}-
 \xi^{\hat t} \Pi^i
.
\end{equation}
Going on-shell and using the Killing equations, or the conformal Killing equations when the system is Weyl-invariant i.e. when \eqref{aris-conf-cond} is satisfied, we find:
\begin{equation}
\label{Aconservation}
\mathcal{K}=\left(\frac{1}{\Omega}\partial_t+\theta\right)\kappa+(\nabla_i+\varphi_i)K^i=0.
\end{equation}
Consequently, \emph{on Aristotelian structures, a (conformal) Killing field always supports an on-shell conservation law for (Weyl-invariant) Aristotelian dynamics}. 

\section{Conclusions} 

Hydrodynamic equations for Galilean and Carrollian fluids on arbitrary backgrounds were displayed in Ref. \cite{CMPPS1}. The present work augments these achievements and points towards an extensive set of features: (\romannumeral1) the fluid momenta conjugate to the geometric variables that define the spacetime and their role in the dynamics; (\romannumeral2) the inclusion of a matter conserved current and its potential interplay with the energy and momentum through thermodynamics; (\romannumeral3) the hydrodynamic-frame invariance; (\romannumeral4) the consequence of isometries in terms of conservation. For this purpose, two distinct and complementary approaches have been pursued. 

The first relies on local symmetries, primarily diffeomorphism invariance, but also local Weyl and $U(1)$ gauge invariance. The energy and momenta are defined through geometry and are conserved as a consequence of diffeomorphism invariance. They are different for the relativistic, Galilean or Carrollian theories, and it should be stressed that \emph{only the relativistic theory has an energy--momentum tensor $T_{\mu\nu}$} with zero general-covariant divergence. In Galilean dynamics there is an energy--stress tensor $\Pi_{ij}$, a momentum $P_i$ and an energy density $\Pi$, obeying a set of conservation equations, necessarily involving an extra variable, the energy flux $\Pi_i$, which is not conjugate to any available  geometric piece of data. Similarly, Carrollian dynamics goes through with an energy--stress tensor $\Pi^{ij}$, an energy flux $\Pi^i$ and an energy density $\Pi$, whereas the momentum $P^i$ comes now aside. 

The second approach consists in working within an appropriate coordinate system of the relativistic theory, and reach the Galilean or Carrollian dynamics in the infinite or zero speed-of-light limit. This method does not allow to retrieve Aristotelian dynamics, which is neither a large-$c$ nor a small-$c$ limit, but offers the possibility to capture more degrees of freedom by keeping overleading terms in the Laurent expansions. It is also better suited for unravelling the subtleties in the contraction of the relativistic group of invariance, as the persistence in the limit of a supplementary equation, absent in the direct approach. This is for instance  the continuity equation in the Galilean framework or a similar conservation equation in the Carrollian, which betray that the relativistic diffeomorphism group valid before the limit is bigger than the actual Galilean or Carrollian groups. Finally, only when considering a limit from a relativistic system, can we express the various momenta in terms of the relevant kinematical and thermodynamic variables. This is how the velocity appears in the Galilean limit, whereas the inverse velocity arises as the relevant observable for Carrollian fluids --  particularly important for the latter, where the we have no handle or hint from thermodynamics. The kinematical observables play a pivotal role in the further analysis of hydrodynamic-frame invariance.

The energy--momentum variables entering the dynamical equations are variations of some effective action with respect to geometric data. These \emph{are not} any sort of N\oe therian currents, which have no reason to exist since no isometry is assumed. Furthermore, even when isometries are present, N\oe therian conservations appear only in relativistic dynamics. In Galilean and Carrollian systems, conserved currents cannot be designed on the basis of an isometry, unless the latter obeys an extra condition. This unexpected result should be viewed  as one of our main accomplishments. It jeopardizes N\oe therian  descriptions of Galilean or Carrollian hydrodynamics on isometric backgrounds, and calls for extra attention when studying charges in the framework of flat holography. 

Focusing specifically on Galilean hydrodynamics, a few words should be devoted to the extra conserved matter current  introduced in the relativistic theory.  
As shown in \cite{CMPPS1}, this current is not necessary for recovering the continuity equation in the Galilean limit.  Additionally, its associated matter density defines the rest contribution to the energy density, and their relationship is dictated by thermodynamics. This is how the chemical potential, absent in  \cite{CMPPS1} enters the dynamics. Most importantly, in the limiting process, this current contributes to the Galilean heat current, as much as the relativistic heat current does, confirming that both Eckart and Landau--Lifshitz hydrodynamic-frame choices are viable. This brings us to a central theme of the present work: the hydrodynamic-frame invariance in the Galilean limit.

The fluid velocity in the relativistic system is merely a book-keeping device. The effect of modifying it by local Lorentz transformations propagates on the various observables (heat current, stress tensor, thermodynamic variables) so as to keep the energy--momentum tensor, the matter current and the entropy current invariant. When considering a limit on the speed of light, the expansion of the observables matters on the possible preservation of the hydrodynamic-frame invariance. Using the standard behaviour dictated by the physical transport coefficients, this invariance does not survive in the Navier--Stokes equations. This is not a surprise because the velocity and the matter density of a Galilean fluid are measurable observables, protected by a symmetry supported by the conservation of mass -- itself a consequence of a central extension of the Galilean group emerging in a Poincar\'e-group contraction. Adopting an alternative behaviour, with overleading terms in the heat current saves the hydrodynamic-frame invariance at the expense of altering the continuity equation. The resulting dynamics is reminiscent of diffusion processes or superfluids, where indeed all species are not simultaneously conserved. 

Fluids based on massless energy carriers are revealed as an exception to this Galilean scheme. For these fluids, with or without a charge conservation, the behaviour of the heat current, possibly but not necessarily with exotic terms, is compatible with Galilean hydrodynamic-frame invariance. It should be emphasized that this holds irrespective of Weyl invariance, which is not even a priori assumed. The deep reason of this feature is rather the absence of a rest term in the energy density, behaving like $c^2$ -- and thereupon, the inexistence of mass conservation, usually in conflict with hydrodynamic-frame invariance.

Although to some extent dual to Galilean, Carrollian fluids exhibit different features. As opposed to the Galilean case, we have no inkling on what thermodynamics for these systems is, due to the absence of motion and thus of kinetic theory. For the same reasons, the energy density cannot be decomposed into rest plus kinetic contributions. Therefore, if a conserved charge exists, it decouples from  the hydrodynamic equations. More importantly, no transport theory is available that would serve as a guide for the behaviour of the physical observables (heat current and the stress tensor) when the speed of light gets extinct. Instead, Carrollian fluids have been recognized as holographic duals of asymptotically flat spacetimes -- their unique successful application so far -- and this gives a handle on the terms to keep in the small-$c$ expansion. With those, hydrodynamic-frame invariance is maintained in the Carrollian limit, reflecting holographically residual bulk diffeomorphisms.

Aristotelian dynamics was the last subject treated in this context. As we stressed earlier, only the geometric method is applicable, based on Aristotelian diffeomorphisms. The set of equations reached in this way appears as the intersection of Galilean (with torsion though) and Carrollian dynamics. The reduction of the light cone onto points, the degenerate nature of motion and the impossibility to bridge this theory with the relativistic theory through a limit,\footnote{Notice that Aristotelian spacetimes are related to the infinite-mass limit introduced by Henri Bacry and Jean-Marc L\' evy-Leblond in \cite{bacryLLinfmas}, as a mean to reach the static group by contraction of the Galilean group.}
leaves little room for a more in-depth discussion in terms of fluids.

All this summarizes our achievements, their relationships and the general context. This activity, mostly based on classical physics and differential geometry, is part of a palette  of timely developments.  The subject of hydrodynamic-frame invariance has been treated here in a kinematic fashion, ignoring the entropy current and the constitutive relations. This latter facet has come back in the forefront \cite{Bemfica:2017wps, Grozdanov:2019kge, Grozdanov:2019uhi, Kovtun:2019hdm, Bemfica:2019knx, Hoult:2020, Bemfica:2020zjp, Dore:2021xqq, Hoult:2021gnb} (see also \cite{Jensen:2014ama, Karch}), and further investigation of these physical features is certainly desirable in light of our more formal results. In particular, certain aspects of the photon fluid deserve some attention, even at an anecdotic level \cite{preEckart} with the notorious Planck--Ott paradox: \emph{does a moving body appear cool?} According to Isra\"el  about this paradox \cite{Israel1981} ``\emph{it is not yet quite dead},'' and one indeed finds articles  where it is still debated \cite{Gavassino:2020lgo}. On a less frivolous tone, relativistic and Galilean hydrodynamics can be studied using Boltzmann equation, and this should also apply to Carrollian fluids. More ambitiously, one may even try to root Boltzmann equation inside field theory, which in turn would require mastering Galilean or Carrollian fields on general curved spacetimes. At the classical level, some results are available \cite{Jensen:2014ama, Karch, Geracie:2015xfa,  CM1, Lebellac, Souriau, Bagchi:2019xfx, Bagchi:2019clu, Gupta:2020dtl,  henneaux2021carroll,Chen:2021xkw,RBV22}, but the quantum theory remains elusive -- see e.g. \cite{LLQ}. 

Besides the caveats plaguing hydrodynamic-frame invariance in relation with causality or stability and rooted in the constitutive relations, this local gauge symmetry does also disclose global issues.  Local velocity transformations may leave the system with global distinctnesses. To handle them, one should be able to design charges associated with the energy and momenta, the matter current or the entropy current, possibly sensitive to global properties. For relativistic or Carrollian fluids, those charges can be handled holographically via a gravity dual, asymptotically anti-de Sitter or flat. They actually obey algebras, which depend on the chosen hydrodynamic frame \cite{Campoleoni:2018ltl, CMPR, CMPRpos}. Extensions and refinements to this analysis would be expedient.

\section*{Acknowledgements}

We thank our colleagues  G. Barnich, X. Bekaert, G. Bossard, A. Campoleoni, L. Ciambelli, L.~Donnay, A.~Fiorucci, L. Freidel, P. Gauduchon, M. Geiller, J. Gomis, E. Gourgoulhon, D. Grumiller, M.~Henneaux, Y.~Herfray, P.~Horvathy, A.~Kleinschmidt, C. Kopper, P. Kovtun, C. Lorc\'e, C. Markou, C.~Marteau, G.~Montambaux, N. Obers, B. Oblak, G. Oling, J. Rossell, R. Ruzziconi, S. Vandoren, M. Vilatte and C. Zwikel for rich scientific discussions. We are also grateful to the \emph{IP2I de l'Universit\'e de Lyon}, the \emph{Universit\' e libre de Bruxelles} and the \emph{Universit\' e de Mons}, where presentations were made by us  on these topics, triggering fecund exchanges. Tasos Petkou and Marios Petropoulos have also often visited the \emph{Ecole Polytechnique} and the \emph{Aristotelian University of Thessaloniki}, where parts of the present work were carried on. Finally, many thanks to the \emph{TU Wien} and to the organizers of the \emph{Carroll Workshop} held in Vienna in February 2022, where our results were previewed and refined, taking advantage of the inimitable emulation created during this event. The work of A. C. Petkou  was supported by the Hellenic Foundation for Research and Innovation (H.F.R.I.) under the \textsl{First Call for H.F.R.I. Research Projects to support Faculty members and Researchers and the procurement of high-cost research equipment grant} (MIS 1524, Project Number: 96048). The work of D. Rivera-Betancour was funded by Becas Chile (ANID) Scholarship No. 72200301.

\appendix 

\section{A primer on thermodynamics}\label{thermo}

\subsubsection*{Relativistic thermodynamics}

We remind here the usual observables of global-equilibrium thermodynamics. These are supposed to make sense also in local-equilibrium thermodynamics, as for fluids where the absence of short wave-length modes is assumed. In this case they depend on time and space and refer to measurements performed by an observer comoving with respect to the fluid. Matter conservation is generically (but not necessarily) akin to the existence of massive carriers in conserved number. 

\begin{itemize}

\item The temperature $T$.

\item The mass density  $\varrho_0$ per unit proper volume.

\item The entropy per unit proper volume $\sigma$, and the entropy per unit mass $s$ (specific entropy):
\begin{equation}
\label{sigrho}
\sigma=s \varrho_0.
\end{equation}

\item The relativistic internal energy density per unit proper volume $\varepsilon$, which contains the rest mass,  and the specific energy per unit mass $e$:
\begin{equation}
\label{epsrho}
\varepsilon=\left(e+c^2\right) \varrho_0.
\end{equation}

\item The pressure $p$ and the relativistic enthalpy ${\cal w}$ per unit proper volume:
\begin{equation}
\label{renth}
{\cal w}=\varepsilon+p.
\end{equation}

\item The relativistic chemical potential per unit mass (specific chemical potential) $\mu_0$. This contains the rest-mass contribution, as opposed to $\mu$:
\begin{equation}
\label{mumu0}
\mu_0=\mu +c^2.
\end{equation}

\end{itemize}
These quantities obey the grand-potential equation, sometimes referred to as 
the Gibbs--Duhem equation:
\begin{equation}
\label{GD}
{\cal w}=T\sigma + \mu_0 \varrho_0\Leftrightarrow p=Ts \varrho_0 + \mu \varrho_0-e \varrho_0.
\end{equation}

The energy density is a functional of two thermodynamic variables: $\varepsilon=\varepsilon\left(\sigma,\varrho_0\right)$. The first law of thermodynamics reads:
\begin{equation}
\label{1stl}
\text{d}\varepsilon=T\text{d}\sigma + \mu_0 \text{d}\varrho_0 \Leftrightarrow
\text{d}e=T\text{d}s -p  \text{d}\left(\frac{1}{\varrho_0}\right) .
\end{equation}
The Gibbs--Duhem equation allows to  exhibit  the dependence 
of the enthalpy per unit proper volume ${\cal w}={\cal w}\left(\sigma,p,\varrho_0\right)$
\begin{equation}
\label{1stlL}
\text{d}{\cal w}= T\text{d}\sigma +\text{d}p+ \mu_0 \text{d}\varrho_0,
\end{equation}
whereas a double Legendre transformation on $\varepsilon$ infers the dependence 
of the grand potential $p=p\left(T,\mu_0\right)$
\begin{equation}
\label{1stlLL}
\text{d}p=\sigma\text{d}T + \varrho_0 \text{d}\mu_0=s \varrho_0\text{d}T + \varrho_0 \text{d}\mu.
\end{equation}

We would like to mention the situations where no massive degrees of freedom are present in the microscopic theory.\footnote{This happens effectively in the usual ultra-relativistic limit, meant to be relevant microscopically at high temperature or high pressure.} A gas of photons is the prime example but other instances exist in condensed matter, in particular when fermions are involved, as in graphene (see e.g. \cite{GM}). In the latter case, as opposed to the gas of photons, there is a conserved quantity. So $\varrho_0$ is non-vanishing, but it\emph{ is not} a mass density; $\varepsilon=e \varrho_0$ and  $\mu_0=\mu$,  without rest-mass contribution. These systems can be conformally invariant, and in that case
 the dependence $p=p(T,\mu)$ is 
\begin{equation}
\label{pTmu-confinv}
p = T^D f\left(\nicefrac{\mu}{T}\right)
\end{equation}
in $D=d+1$ spacetime dimensions.\footnote{The precise bearing between conformal invariance, absence of mass and existence of conserved currents is subtle and tight to the microscopic theory.\label{masless2}}

Coming back to a system with massless carriers and no conserved charge, as for the gas of photons, the above thermodynamic relationships simplify by setting $\mu=0$ and dropping the rest-mass terms. Specific quantities are no longer significant in this instance. Fluid dynamics of such systems does not involve any conserved current.\footnote{This instance was discussed in the precise framework of relativistic fluid dynamics in \cite{LandauF2} \textsection 134, footnote 1 and exercise 2. The general thermodynamic aspects are presented in greater detail in  \cite{LandauMS} \textsection 60\label{masless}.} The basic laws are summarized as follows:
\begin{equation}
\label{1stlmassless}
\begin{cases}
&{\cal w}=T\sigma
\\&\text{d}{\cal w}= T\text{d}\sigma +\text{d}p
\\& \text{d}\varepsilon=T\text{d}\sigma
\\& \text{d}p=\sigma\text{d}T,
\end{cases}
\end{equation}
and when the system is furthermore conformal,  $p \propto  T^D$.

Several conserved charges might exist simultaneously in a thermodynamic system. They would each be associated  with a density and a chemical potential. Only one, if any, would however enter the energy density \eqref{epsrho}.

\subsubsection*{Non-relativistic limit}

The thermodynamic variables introduced earlier in the relativistic theory such as $\varrho_0$, $\varepsilon$, $\mu$ etc. are referring to a comoving observer. Measurements performed by another observer, be this an inertial observer in special relativity or some fiducial observer in a general gravitational background, are more relevant for the Galilean framework, but are not equal and should be spelled stricto sensu with some distinctive index. Their differences, however, are of order $\nicefrac{1}{c^2}$ and vanish in the infinite-$c$ limit. In order to avoid inflation in notation, we will keep the same  symbols, $e$, $T$, $p$, $s$, $\mu$, with the exception of $\varrho_0$, which becomes $\varrho$ for the fiducial observer. The $\nicefrac{1}{c^2}$ corrections amongst $\varrho_0$ and $\varrho$ (see \eqref{newvarrho}) play no role in thermodynamics, but are indispensable in recovering Navier--Stokes equations as the Galilean limit of the relativistic hydrodynamic equations.

In non-relativistic thermodynamics, it is customary to introduce the specific volume   (not to be confused with the velocity)
\begin{equation}
\label{specvol}
v=\frac{1}{\varrho},
\end{equation}
as well as specific enthalpy $h=h\left(s,p\right)$ as\footnote{This is spelled $w_{\text{nr}}$ in footnote 1, \textsection 134 of \cite{LandauF2}.} 
\begin{equation}
\label{hgal}
h=e+ p v,
\end{equation}
which also enters in
\begin{equation}
\label{mugal}
\mu=h-Ts.
\end{equation}
Using these definitions and the various relativistic laws mentioned above, we find the standard expressions:
\begin{eqnarray}
\label{dh}
\text{d}h&=&T\text{d}s + v \text{d}p,\\
\label{derhoh}
\text{d}\left(e \varrho\right)&=&\varrho T\text{d}s  + h \text{d}\varrho,\\
\text{d}e&=&T \text{d}s -p \text{d}v,
\label{1stlg}\\
\text{d}\mu&=&-s\text{d}T + v \text{d} p.
\label{dmu}
\end{eqnarray}

Before closing this chapter, let us quote that Galilean thermodynamics can accommodate fluids with massless energy carriers, as long as the macroscopic velocity is small compared to $c$ -- although at the microscopic level the dynamics is ultra-relativistic. Again, a conserved current may or may not exist. In case such a current is available, $\varrho$ is the charge density with\footnote{Notice the distinction: 
$\upvarepsilon=\lim\limits_{c\to \infty} \varepsilon =e \varrho$,  $\upsigma=\lim\limits_{c\to \infty} \sigma = s \varrho$. In order to avoid multiplication of symbols, we keep ${\cal w}=h \varrho$, $p$ and $\mu$ both for the relativistic quantities and for their Galilean limits.} $\upvarepsilon= e\varrho$ the internal energy density and $\mu$ the chemical potential. The basic relationships are now 
\begin{equation}
\label{GDGM}
{\cal w}=p+\upvarepsilon=T\upsigma + \mu \varrho \Leftrightarrow p=Ts \varrho + \mu \varrho-e \varrho,
\end{equation}
and
\begin{equation}
\label{2stlmassless}
\begin{cases}
& \text{d}{\cal w}= T\text{d}\upsigma +\text{d}p +\mu\text{d}\varrho
\\& \text{d}\upvarepsilon=T\text{d}\upsigma +\mu\text{d}\varrho
\\& \text{d}p=\upsigma\text{d}T +\varrho\text{d}\mu.
\end{cases}
\end{equation}
Equations \eqref{specvol}, \eqref{hgal}, \eqref{mugal}, \eqref{dh}, \eqref{derhoh}, \eqref{1stlg}, \eqref{dmu} remain also valid, together with \eqref{pTmu-confinv} in case of conformal invariance. 

If no conserved charge is present, the chemical potential vanishes (as does $\text{d}\mu$) and the relevant equations are expressed with 
${\cal w}$, 
 $\upvarepsilon$ and $\upsigma$ rather that $h\varrho$, $e\varrho$ or $s\varrho$.

\subsubsection*{Carrollian thermodynamics}

Carrollian thermodynamics is poorly understood. In most parts of this work dealing with the fluid equations, we have kept the energy density $\varepsilon$ and the pressure $p$  unaltered in the limit of vanishing velocity of light. Neither have we introduced any temperature, nor discussed an entropy equation, and when a conserved current was assumed (as eluded in \cite{Ciambelli:2017wou}), no relationship was established or set among energy and conserved-charge densities. This is minimalistic by default. Indeed, the shrinking of the light cone and the absence of particle motion or signal propagation, raise fundamental questions regarding the origin -- and even the definition --  of energy, pressure, entropy, temperature and thermalization processes. Even the kinematic parameter of the fluid is an inverse velocity, which could point towards the dynamics of instantonic space-filling branes, as mentioned in \cite{CMPPS1}. Obviously, this sort of objects are tachyonic -- like those introduced later in \cite{newdutch} -- and we feel uneasy advocating any sort of kinetic theory for setting up thermodynamic laws and deviations from equilibrium.

\section{Carrollian momenta and hydrodynamic-frame invariance} \label{compl}

In Sec. \ref{carfluids2} we obtained the Carrollian fluid equations under $c^2$-scaling assumptions involving more degrees of freedom than the standard ones treated in Sec. \ref{carfluids}: 
\eqref{eexpCext}, 
\eqref{pexpCext}, 
\eqref{QexpCext} and \eqref{sigexpCext}  for the relativistic energy $\varepsilon$, pressure $p$, heat current $q^i$ and stress tensor $\tau^{ij}$, and similarly for the matter sector with the matter density $\varrho_0$ and the non-perfect current $j^i$ in \eqref{jexpCext}. These equations involve the Carrollian momenta    
$\tilde\Pi$, $ \Pi$, 
$\tilde\Pi^{i}$,
$ \Pi^{i}$,
$P^i$,  
$\tilde\Pi^{ij}$,
$ \Pi^{ij}$, $\tilde\varrho$, $\varrho$, $\tilde N^i$,  $N^i$, which have the following expressions in terms of the observables entering the already quoted $c$-expansions: 
\begin{equation}
\label{momcaremext}
\begin{cases}
\tilde \Pi^i =\psi^i
\\
\Pi^i = Q^i-\beta_j \left(\Sigma^{ij}-\phi a^{ij}\right)+\beta^i\left(\zeta+\beta_j \psi^j
\right)+\frac{\pmb{\beta}^2 }{2}\psi^i
\\
P^i = \pi^i -\beta_j \left(\Xi^{ij}-\varpi a^{ij}\right)+\beta^i\left(\eta+\beta_j Q^j
\right)+\frac{\pmb{\beta}^2 }{2}\left(Q^i+\frac{3\pmb{\beta}^2}{4} \psi^i\right)
+\beta^i\pmb{\beta}^2 \left(\zeta+
\phi+\frac{1}{2}\beta_j \psi^j
\right)
\\
\tilde \Pi = \zeta + 2\beta_i \psi^i
\\
 \Pi = \eta + 2\beta_i Q^i-\beta_i\beta_j \left(\Sigma^{ij}-\phi a^{ij}\right) +\pmb{\beta}^2 \left(\zeta+\beta_i \psi^i\right)
 \\
\tilde \Pi^{ij}=\psi^{i}\beta^j+\beta^i\psi^{j}+\phi a^{ij}-\Sigma^{ij}
\\
\Pi^{ij} = Q^{i}\beta^j+\beta^iQ^{j}+\varpi a^{ij}-\Xi^{ij} +\beta^i\beta^j\left(\zeta+\phi\right)+\frac{\pmb{\beta}^2 }{2}
\left(\psi^{i}\beta^j+\beta^i\psi^{j}
\right),
\end{cases}
\end{equation}
and
\begin{equation}
\label{matcaremext}
\begin{cases}
\tilde N^j =m^j
\\
N^j = n^j+\beta^j\omega
\\
 \tilde \varrho = \omega+\beta_k m^{k}
 \\
\varrho=\chi +\beta_k n^k+\frac{\pmb{\beta}^2}{2}\omega.
\end{cases}
\end{equation}
The aim of the present appendix is to show that these are hydrodynamic-frame-invariant.

Although hydrodynamic-frame invariance is built-in for the relativistic momenta  \eqref{invenRP}, \eqref{invqRP}, \eqref{invtauRP}, \eqref{invrhoRP} and \eqref{invjRP}, it is not guaranteed to persist in the vanishing-$c$ limit because it can be incompatible with the presumed small-$c$ behaviour of the physical observables. This happens in the  Galilean (infinite-$c$) limit, as we have witnessed in Sec. \ref{galdyn} for the ordinary i.e. with $n^i=0$ non-relativistic fluids, because $\delta n^i \propto \delta v^i$ (see Eq. \eqref{delgalnk}). Here it turns out to hold and in order to prove that we use the relativistic transformation rules  \eqref{delepsRP},  \eqref{delqRP},  \eqref{deltau2RP}, \eqref{delrho0RP}  and \eqref{deljRP} in the Papapetrou--Randers frame \eqref{carrp}. Using \eqref{eexpCext}, 
\eqref{pexpCext}, 
\eqref{QexpCext}, \eqref{sigexpCext} and  \eqref{jexpCext},
and expanding we find
\begin{equation}
\label{delmatcaremext}
\begin{cases}
\delta m^j =0
\\
\delta n^j = -\delta\beta^j\omega+\delta\beta_k m^{k} \beta^j
\\
\delta \omega= - \delta\beta_k m^{k}
 \\
\delta\chi=  -\delta\beta_k\left( n^k+\frac{\pmb{\beta}^2}{2}m^k\right),
\end{cases}
\end{equation}
and
\begin{equation}
\label{delmomcaremext}
\begin{cases}
\delta \psi^i=0
\\
\delta Q^i=\delta\beta_j \left(\Sigma^{ij}   -\phi a^{ij}\right)-\delta\beta^i\zeta
+\delta\beta_j \psi^j \beta^i
\\
\delta \pi^i= \delta\beta_j\left( \Xi^{ij}-(\eta+
\varpi)a^{ij}\right)+\frac{\pmb{\beta}^2 }{2} \delta\beta_j\left( \Sigma^{ij}-(\zeta+
\phi)a^{ij}\right)
+\delta\beta_j\beta^i\left(\beta^j (\zeta+
\phi)+Q^j\right)
+\pmb{\beta}^2 \beta^i\delta\beta^j\psi^j
\\
\delta \zeta =- 2\delta \beta_i \psi^i
\\
\delta \eta= - 2\delta\beta_i \left(Q^i+\frac{\pmb{\beta}^2 }{2}\psi^i\right) \\
\delta \Sigma^{ij}-\delta\phi a^{ij}= \psi^{i}\delta\beta^j+\psi^{j}\delta \beta^i
\\
\delta \Xi^{ij} -\delta\varpi a^{ij}= Q^{i}\delta\beta^j+Q^{j}\delta\beta^i
+
\delta \beta_k
\left(\Sigma^{ki} \beta^j
+\Sigma^{kj} \beta^i
\right)
+\frac{\pmb{\beta}^2 }{2}
\left(\psi^{i}\delta\beta^j+\psi^{j}\delta\beta^i
\right)
\\+ \delta \beta_k\beta^k\left(\psi^{i}\beta^j+\psi^{j}\beta^i\right)
+ \beta^i\beta^j
\delta\phi.
\end{cases}
\end{equation}
It is straightforward to show that the variations of all momenta \eqref{momcaremext} and \eqref{matcaremext} vanish. 

\section{Free motion} \label{frmotNC}

Our results on the failure of conservation laws associated with some Galilean or Carrollian (conformal) Killing vector fields are generic and rooted to the nature of the underlying geometries. The same phenomenon occurs when studying free-particle motion in Newton--Cartan spacetimes, or instantonic branes on Carrollian structures (see \cite{CMPPS1} for motivations on the latter paradigm). For concreteness we will illustrate here the former case. 

The stage is set with an action
\begin{equation}
\label{galactgal}
S[\mathbf{x}]=\int_{\mathscr{C}}\text{d}t\, \Omega(t) \, \mathcal{L}(t,\mathbf{x},\mathbf{v}),
\end{equation}
where $L= \Omega \mathcal{L}$ is the Lagrangian -- as opposed to the Lagrangian density. The generalized Lagrange momenta are 
\begin{equation}
\label{galmom}
p_i=\frac{\partial \mathcal{L}}{\partial\frac{ v^i}{\Omega}}
\end{equation}
and the energy $E =\Omega \mathcal{E} $ with
\begin{equation}
\label{galenmom}
\mathcal{E}
= \frac{p_iv^i}{\Omega}-
\mathcal{L}.
\end{equation}
The equations of motion are Euler--Langrange
\begin{equation}
\label{galeom}
\frac{1}{\Omega}\dot{p_i}-\frac{\partial\mathcal{L}}{\partial x^i}=0.
\end{equation}
The dot stands for the total derivative along the trajectory, which can act also as $\partial_t+v^i\partial_i$ on \emph{any} tensor, and should not be confused with $\nicefrac{\text{d}}{\text{d}t}$ defined in \eqref{galfder} unless they  act on scalars (cf. ordinary vs. covariant spatial derivative). 

Consider Galilean diffeomorphisms generated by 
\begin{equation}   
\label{genkil-app1}
\upxi=\xi^t\partial_t +\xi^i \partial_i
,
 \end{equation}
where $\xi^t=\xi^t(t)$ and
\begin{equation}   
\xi^{\hat t} = \xi^t \Omega, \quad \xi^{\hat\imath} = \xi^i - \xi^t w^i, \quad \xi_{\hat t} = -c^2  \xi^{\hat t},
\quad  \xi_{\hat\imath} = a_{ij} \xi^{\hat\jmath} =\xi_i.
 \end{equation}
Their effect on the dynamical variables is 
\begin{equation}   
\begin{cases}
t\to t+ \xi^t \\
x^i\to x^i+ \xi^i \\
v^i\to v^i+ \partial_t \xi^i+v^j\partial_j \xi^i -v^i  \partial_t \xi^t.
\end{cases}
\label{xi-effect}
 \end{equation}
 
 On the one hand, the invariance of the action is characterized as follows:
 \begin{equation}   
\label{invact0}
\delta S=0 \Leftrightarrow
\Omega \delta \mathcal{L}+ \mathcal{L} \partial_t\xi^{\hat t}  =\frac{\text{d}\phi}{\text{d}t},
 \end{equation}
where $\phi= \phi(t, \mathbf{x})$  is an arbitrary function, that needs not be zero. 
On the other hand, one can determine the on-shell  variation of the Lagrangian density:
\begin{equation}   
\label{varcurlL}
\delta \mathcal{L}=- \frac{\mathcal{L}}{\Omega} \partial_t\xi^{\hat t} 
+\frac{1}{\Omega} 
\frac{\text{d}}{\text{d}t}\left(p_i\xi^i-\mathcal{E}\xi^{\hat t} \right).
 \end{equation}
The simplest of N\oe ther's theorems states that 
 \begin{equation}   
\label{invact}
\delta S=0 \Leftrightarrow
p_i\xi^i-\mathcal{E}\xi^{\hat t} -\phi = \text{constant of motion}.
 \end{equation}

Suppose now that the motion is free on a Newton--Cartan spacetime featured by $a^{ij}$, $w^i$ and $\Omega$. The Lagrangian density is
\begin{equation}
\label{gallag}
\mathcal{L}=\frac{1}{2\Omega^2}a_{ij} \left(v^i-w^i \right)\left(v^j-w^j\right)
\end{equation}
with 
\begin{equation}
\label{galmomfree}
p_i=\frac{1}{\Omega}(v_i-w_i). 
\end{equation}
Euler--Lagrange equations read:
\begin{equation}
\label{galeomfree}
\left(\frac{1}{\Omega}\frac{\hat{\text{D}} }{\text{d}t}+p^j\hat\nabla_j
\right)p_i+ p_j\hat\gamma^{wj}_{\hphantom{wj}i} =0.
\end{equation}
As an aside remark, the latter equation is the infinite-$c$ limit of the spatial component of the geodesic equation $u^\mu\nabla_\mu u_i=0$, in a Zermelo background. The time component  $u^\mu\nabla_\mu u^0=0$ leads to
\begin{equation}
\label{galenergyfree}
\frac{1}{2\Omega}\frac{\text{d}p_ip^i} {\text{d}t} + p_i p_j\hat\gamma^{wij} =0,
\end{equation}
which is the energy equation, obtained by contracting \eqref{galeomfree} with $p^i$.

We can now compute the generic variation of \eqref{gallag} under Galilean diffeomorphisms acting as \eqref{xi-effect}. We obtain the following:
\begin{equation}   
\label{varcurlLfree}
\delta \mathcal{L}=p_i p_j\left(\hat \nabla^{(i}\xi^{\hat\jmath)} + \xi^{\hat t}  \hat\gamma^{wij}\right)
-p_i p^i \frac{1}{\Omega}\frac{\hat{\text{D}} \xi^{\hat t}}{\text{d}t}+p_ i\left(\frac{1}{\Omega}\frac{\hat{\text{D}}\xi^{\hat\imath}}{\text{d}t}-\hat\gamma^{wi}_{\hphantom{wi}j} \xi^{\hat\jmath} \right)
.
 \end{equation}
If $\upxi $ is a Killing field it satisfies  \eqref{galkilleq}, the first two terms drop and
\begin{equation}   
\label{varcurlLfreeKill}
\delta \mathcal{L}=p_ i\left(\frac{1}{\Omega}\frac{\hat{\text{D}}\xi^{\hat\imath}}{\text{d}t}-\hat\gamma^{wi}_{\hphantom{wi}j} \xi^{\hat\jmath} \right)
 \end{equation}
does not vanish, exactly as in the Galilean fluid dynamics in the presence of an isometry. This betrays the break down of conservation, unless the right-hand side of Eq. \eqref{varcurlLfreeKill} happens to be of the form \eqref{varcurlL}, in which case N\oe ther's theorem applies in its version \eqref{invact}. 

As already emphasized repeatedly, this pattern works the same way in all situations we have met, involving Galilean or Carrollian dynamics. In flat spacetimes (either Galilean or Carrollian) boosts belong invariably to the class of isometries with non-vanishing Lagrangian variation (see \eqref{galkillflatsol-lie} and \eqref{carkillflatsol-lie}). There is not much we could extract from this in fluid dynamics (except for the case of flat-space potential flows, see footnotes \ref{33} and \ref{57}), but for Galilean free-particle motion on flat spacetime ($a_{ij}=\delta_{ij}$, $\Omega = 1$, $w^i$ constants) the situation is simpler. We find indeed:
\begin{equation}   
\label{varcurlLfreeKillonshell}
\delta \mathcal{L}=  \left(\dot x^i-w^i\right) \left(V_i+ w^k \Omega_{ki}\right)=\frac{\text{d}}{\text{d}t}\left(
x^i V_i - w^i V_i  t + w^k x^i \Omega_{ki}
\right).
 \end{equation}
In this particular case,  \eqref{invact} applies and gives the general constant of motion as (see also \eqref{galkillflatsol})
\begin{equation}   
\label{varcurlLfreeKillonshell-con}
V_i \left( \dot x^i t-x^i\right)-\frac{T}{2}\left(
{\dot{\mathbf{x}}^2} - {\mathbf{w}^2} 
\right)+X_i \left( \dot x^i-w^i\right) +\Omega_{ij}x^i \dot x^j.
 \end{equation}
The boosts $V^i$ do not generate any useful first integral (the initial position $x_0^i$), as opposed to time translation $T$, space translations  $X^i$ and rotations $\Omega_{ij}$, which lead to energy, momentum and angular momentum conservations. 

\section{From conservation to potential non-conservation}
\label{garbage}

\subsection{Galilean law from infinite speed of light}
\label{galgarbage}

Our starting point is a pseudo-Riemannian spacetime in Zermelo frame \eqref{galzerm}
\begin{equation}
\label{galzerm-app}
\text{d}s^2 =-\Omega^2 c^2 \text{d}t^2+a_{ij} \left(\text{d}x^i -w^i  \text{d}t\right)\left( \text{d}x^j-w^j \text{d}t\right)
\end{equation}
with an energy--momentum tensor $T^{\mu\nu}$ obeying $\nabla_\mu T^{\mu\nu}=0$, and a vector field 
\begin{equation}   
\label{genkil-app}
\upxi=\xi^t\partial_t +\xi^i \partial_i=
\xi^{\hat t}\text{e}_{\hat t}+ \xi^{\hat\imath }\text{e}_{\hat\imath}
,
 \end{equation}
where the frame and coframe are defined as in \eqref{zerfrcofr},
and 
\begin{equation}   
\xi^{\hat t} = \xi^t \Omega, \quad \xi^{\hat\imath} = \xi^i - \xi^tw^i, \quad \xi_{\hat t} = -c^2  \xi^{\hat t},
\quad  \xi_{\hat\imath} = a_{ij} \xi^{\hat\jmath} =\xi_i.
 \end{equation}
We define a current as in \eqref{relconcur}, $I_\mu=T_{\mu\nu} \xi^\nu $, and compute its on-shell divergence,  using Eqs. \eqref{restTinvZer}: 
\begin{eqnarray}
\nabla_\mu I^\mu=-\frac12  T_{\mu\nu} \mathscr{L}_\upxi g^{\mu\nu}= - 
\frac{\varepsilon_{\text{r}}}{\Omega}\frac{\hat{\text{D}} \xi^{\hat t}}{\text{d}t}
+  \left(p_{\text{r}} a_{ij}+\tau_{\text{r}{ij}}
\right) \left( \hat \nabla^{i}\xi^{\hat\jmath} + \xi^{\hat t}  \hat\gamma^{wij}\right)
&&
\nonumber\\
+\frac{1}{c^2}q_{\text{r}i}
 \left(\frac{1}{\Omega}\frac{\hat{\text{D}}\xi^{\hat\imath}}{\text{d}t}-\hat\gamma^{wi}_{\hphantom{wi}j} \xi^{\hat\jmath} -c^2 a^{ij}\partial_j  \xi^{\hat t}
\right)
&.&
\label{relconcureq-app}
\end{eqnarray}
This result is relativistic, expressed with Galilean derivatives though. It vanishes iff
\begin{equation}   
\begin{cases}   
\frac{1}{\Omega}\frac{\hat{\text{D}} \xi^{\hat t}}{\text{d}t}=0
\\
 \hat \nabla^{(i}\xi^{\hat\jmath)} + \xi^{\hat t}  \hat\gamma^{wij}=0\\
\frac{1}{\Omega}\frac{\hat{\text{D}}\xi^{\hat\imath}}{\text{d}t}-\hat\gamma^{wi}_{\hphantom{wi}j} \xi^{\hat\jmath} -c^2 a^{ij}\partial_j  \xi^{\hat t}=0
,
\end{cases}
\end{equation}
which are simply the conditions for $\upxi$ be a Killing field of the pseudo-Riemannian manifold.

We would like now to consider the infinite-$c$ limit of \eqref{relconcureq-app}. At the first place, we must provide the bahaviour of  $\varepsilon_{\text{r}}$, $q_{\text{r}i}$ and $p_{\text{r}} a_{ij}+\tau_{\text{r}{ij}}$ for large $c$. This is typically of the form  \eqref{gal-no-curem}\footnote{More general behaviours have appeared in 
\eqref{nlimgalvrep}, 
\eqref{nlimgalqri}, 
\eqref{nlimgalij}, 
or in \eqref{gal-no-curem-extra}. These choices wouldn't  change our present argument though.}
\begin{equation}
\label{gal-no-curem-app}
 \begin{cases}
\varepsilon_{\text{r}}=  \Pi 
+ \text{O}\left(\nicefrac{1}{c^2}\right)
\\
q_{\text{r}i}=
c^2 P_i+\Pi_{i}+
\text{O}\left(\nicefrac{1}{c^2}\right)
\\
p_{\text{r}} a_{ij}+\tau_{\text{r}{ij}}=
\Pi_{ij}
+ \text{O}\left(\nicefrac{1}{c^2}\right),
\end{cases}
\end{equation}
and \eqref{relconcureq-app} becomes:
\begin{equation}
\nabla_\mu I^\mu = - 
\frac{\Pi}{\Omega}\frac{\hat{\text{D}} \xi^{\hat t}}{\text{d}t}+
\Pi_{ij} \left( \hat \nabla^{i}\xi^{\hat\jmath} + \xi^{\hat t}  \hat\gamma^{wij}\right)+ \left( P_i+\frac{\Pi_{i}}{c^2}\right)
 \left(\frac{1}{\Omega}\frac{\hat{\text{D}}\xi^{\hat\imath}}{\text{d}t}-\hat\gamma^{wi}_{\hphantom{wi}j} \xi^{\hat\jmath} -c^2 a^{ij}\partial_j  \xi^{\hat t}
\right)
+ \text{O}\left(\nicefrac{1}{c^2}\right).
\label{relconcureq-app-lim}
\end{equation}
For this expression to remain finite at infinite $c$, we must impose that\footnote{Following the discussion on Sec. \ref{comgal}, one may refine the limiting procedure for the Killing fields, and reach the Galilean diffeomorphisms as 
$\xi^{\hat t}(t,\mathbf{x}) =  \xi_{\text{G}}^{\hat t}(t) + \frac{1}{c^2}\nu(t,\mathbf{x}) +\text{O}\left( \frac{1}{c^4}\right)$.
This would alter Eq. \eqref{Galkli3} as $\frac{1}{\Omega}\frac{\hat{\text{D}}\xi_{i}}{\text{d}t}-\hat\gamma^{wj}_{\hphantom{wj}i} \xi_{j} -\partial_i \nu=0$. Similarly the arbitrary function $\nu(t,\mathbf{x}) $ would also appear in the large-$c$ expansions of Eqs. \eqref{inviotaZer} and \eqref{inviZer}, altering the Galilean currents \eqref{galkappaK}.
Ultimately, this would have no incidence on our conclusions about the interplay between Galilean isometries and conservation. It may nevertheless provide a complementary view on the large-$c$ contraction of the relativistic diffeomorphisms, possibly in line with the approach followed in Ref. \cite{Gomis:2022spp}, where a further duality relationship has been established among leading Galilean and subleading Carrollian contributions (see footnote~\ref{theca}), and vice-versa. 
\label{thenu}}
\begin{equation}
\partial_j  \xi^{\hat t} = 0,
\end{equation}
which is the requirement that $\upxi$ generates a Galilean diffeomorphism. Conservation holds in the limit if expression \eqref{relconcureq-app-lim}
vanishes, which is again a threefold condition:
\begin{eqnarray}   
\label{Galkli1}
\frac{1}{\Omega}\frac{\hat{\text{D}} \xi^{\hat t}}{\text{d}t}&=&0,
\\
\label{Galkli2}
 \hat \nabla^{(i}\xi^{\hat\jmath)} + \xi^{\hat t}  \hat\gamma^{wij}&=&0,
 \\
 \label{Galkli3}
\frac{1}{\Omega}\frac{\hat{\text{D}}\xi^{\hat\imath}}{\text{d}t}-\hat\gamma^{wi}_{\hphantom{wi}j} \xi^{\hat\jmath} &=&0.
\end{eqnarray}
Equations \eqref{Galkli1} and \eqref{Galkli2} are nothing but \eqref{galkilleq} i.e. the definition of a Galilean Killing field. Equation \eqref{Galkli3} is an extra  condition, absent for generic Galilean isometries. The latter \emph{do not guarantee the existence of a conserved Galilean current}. The break down of the conservation is read off in 
\begin{equation}
\lim_{c\to\infty} \nabla_\mu I^\mu= P_i \left( \frac{1}{\Omega}\frac{\hat{\text{D}}\xi^{\hat\imath}}{\text{d}t}-\hat\gamma^{wi}_{\hphantom{wi}j} \xi^{\hat\jmath} \right) = 
\frac{P_i}{\Omega} \left(\partial_t \xi^{\hat \imath}+ \mathscr{L}_{\mathbf{w}} \xi^{\hat \imath}\right),
\end{equation}
which agrees with \eqref{galnoncon} or \eqref{galNkill}. As stressed in Sec. \ref{FLUIDS}, the failure might be only apparent, if the term  $\frac{P_i}{\Omega} \left(\partial_t \xi^{\hat \imath}+ \mathscr{L}_{\mathbf{w}} \xi^{\hat \imath}\right)$ turns out to be a boundary term, that would then contribute the Galilean current.

\subsection{Carrollian law from zero speed of light}
\label{varprinccar}

We will here consider pseudo-Riemannian spacetime in Papapetrou--Randers frame \eqref{carrp}
\begin{equation}
\label{carrp-app}
\text{d}s^2 =- c^2\left(\Omega \text{d}t-b_i \text{d}x^i
\right)^2+a_{ij} \text{d}x^i \text{d}x^j.
\end{equation}
We assume
 a conserved energy--momentum tensor $T^{\mu\nu}$ and a vector field as in 
\eqref{genkil-app}
with
 \begin{equation}   
  \xi^{\hat t} = \xi^t \Omega-\xi^ib_i, \quad \xi^{\hat\imath} = \xi^i, \quad \xi_{\hat t} = -c^2  \xi^{\hat t}, \quad \xi_{\hat\imath} =a_{ij}\xi^{\hat\jmath} =  \xi_i+ \xi_{\hat t}  b_i.
 \end{equation}
 The frame and coframe are defined in \eqref{RPframe}.

We now compute the  on-shell divergence  of the current \eqref{relconcur} $I_\mu=T_{\mu\nu} \xi^\nu $,  using Eqs. \eqref{restTinvRP}: 
\begin{eqnarray}
\nabla_\mu I^\mu=\frac12  T^{\mu\nu} \mathscr{L}_\upxi g_{\mu\nu}= -\varepsilon_{\text{r}}
\left(\frac{1}{\Omega} \partial_t
 \xi^{\hat t}+
\varphi_i \xi^i 
\right)+  \left(p_{\text{r}} a^{ij}+\tau_{\text{r}}^{ij}
\right) \left( \hat\nabla_{i}\xi_{\hat\jmath} + \xi^{\hat t}  \hat \gamma_{ij}\right)&&
\nonumber\\
-q_{\text{r}}^i \left(\left(\hat\partial_i-\varphi_i\right) \xi^{\hat t}-
2\xi^j \varpi_{ji}
-\frac{1}{c^2\Omega}a_{ij}\partial_t \xi^j
\right)
&.&
\label{varTCar}
\end{eqnarray}
Although expressed with Carrollian derivatives, this is relativistic and vanishes iff
\begin{equation}   
\begin{cases}   
\frac{1}{\Omega} \partial_t
 \xi^{\hat t}+
\varphi_i \xi^i =0
\\
\hat\nabla_{i(}\xi_{\hat\jmath)} + \xi^{\hat t}  \hat \gamma_{ij}=0\\
\left(\hat\partial_i-\varphi_i\right) \xi^{\hat t}-
2\xi^j \varpi_{ji}
-\frac{1}{c^2\Omega}a_{ij}\partial_t \xi^j=0
.
\end{cases}
\end{equation}
These conditions  define a  Killing field $\upxi$ on a pseudo-Riemannian manifold.

We would like now to consider the zero-$c$ limit of \eqref{varTCar}. We must provide the bahaviour of  $\varepsilon_{\text{r}}$, $q_{\text{r}}^i$ and $p_{\text{r}} a^{ij}+\tau_{\text{r}}^{ij}$ for small $c$, which is typically of the form \eqref{limcarT0i}, 
\eqref{limcarT00}, 
\eqref{limcarTij}\footnote{More general behaviours have appeared in 
 \eqref{limcaremext}. The latter wouldn't  change our present conclusions though.}
\begin{equation}   
\begin{cases}   
\varepsilon_{\text{r}}=
\Pi +\text{O}\left(c^2\right)
\\
q^i_{\text{r}}=
\Pi^{i}+c^2  P^{i}+\text{O}\left(c^4\right)
\\
p_{\text{r}} a^{ij} +\tau_{\text{r}}^{ij} =
\Pi^{ij}
+ \text{O}\left(c^2\right).
\end{cases}
\end{equation}
Equation 
\eqref{varTCar} reads now:
\begin{eqnarray}
\nabla_\mu I^\mu&=& -\Pi 
\left(\frac{1}{\Omega} \partial_t
 \xi^{\hat t}+
\varphi_i \xi^i 
\right)+ \Pi^{ij}\left(\hat \nabla_{i}\xi_{\hat\jmath} + \xi^{\hat t}  \hat \gamma_{ij}\right)
\nonumber\\
&&-\left(\Pi^{i}+c^2  P^{i}\right)\left(\left(\hat\partial_i-\varphi_i\right) \xi^{\hat t}-
2\xi^j \varpi_{ji}-\frac{1}{c^2\Omega}a_{ij}\partial_t \xi^j
\right)
+ \text{O}\left(c^2\right).
\label{varTCar-lim}
\end{eqnarray}
Finiteness  at zero $c$, demands\footnote{\label{theca} Mirroring footnote \ref{thenu}, an option is to set 
$\xi^{i}(t,\mathbf{x}) =  \xi_{\text{C}}^{i}(\mathbf{x}) + c^2\nu^i(t,\mathbf{x}) +\text{O}\left(c^4\right)$. With this,
 Eq. \eqref{carkli3} becomes  $\left(\hat\partial_i-\varphi_i\right) \xi^{\hat t}-
2\xi_{\text{C}}^j \varpi_{ji}
-\frac{1}{\Omega}a_{ij}\partial_t \nu^j=0$, and further work would be necessary on  Eqs. \eqref{inviotacar}, \eqref{invicar} and \eqref{carkappaK}, that would not alter our final conclusions, but could shed light on the 
small-$c$ contraction of general diffeomorphisms.}
\begin{equation}
\partial_t \xi^{i} = 0,
\end{equation}
hence $\upxi$ generates a Carrollian diffeomorphism. Conservation holds in the limit if expression \eqref{varTCar-lim}
vanishes. This is occurs if
\begin{eqnarray}   
\label{carkli1}
\frac{1}{\Omega} \partial_t
 \xi^{\hat t}+
\varphi_i \xi^i 
&=&0,
\\
\label{carkli2}
\hat\nabla_{(i}\xi_{\hat\jmath)} + \xi^{\hat t}  \hat \gamma_{ij}
&=&0,
 \\
 \label{carkli3}
\left(\hat\partial_i-\varphi_i\right) \xi^{\hat t}-
2\xi^j \varpi_{ji}
&=&0.
\end{eqnarray}
Equations \eqref{carkli1} and \eqref{carkli2} are as in  \eqref{carkilleq} i.e. the definition of a Carrollian Killing field. Equation \eqref{carkli3} is an extra  condition, absent for generic Carrollian isometries, which therefore \emph{do not guarantee the existence of a conserved Carrollian current}. The disruption to the conservation is measured as 
\begin{equation}
\lim_{c\to0} \nabla_\mu I^\mu= -\Pi^i \left(\left(\hat\partial_i-\varphi_i\right) \xi^{\hat t}-
2\xi^j \varpi_{ji}
\right),
\end{equation}
in agreement with \eqref{carnoncon} or \eqref{carKN}.

\end{document}